\documentclass[11pt,a4paper]{article}
\pdfoutput=1
\usepackage{amssymb} \usepackage{amsmath} \usepackage{graphicx}
\usepackage{epsfig,latexsym,color,xcolor}
\usepackage{ulem}
\usepackage{amstext}    
\usepackage{array} 
\usepackage{slashed}
\usepackage[height=10in ,hmargin={1.6cm,0.8in}]{geometry}
\usepackage{braket}

\DeclareGraphicsExtensions{.png, .pdf, .eps,}

\newcommand{\pa}{\partial}

\makeatletter
\newcommand*{\rom}[1]{\expandafter\@slowromancap\romannumeral #1@}
\makeatother

\numberwithin{equation}{section}

\begin{document}

\large
 \begin{center}
 {\Large \bf  DDF operators, open string coherent states \\ and \\ their scattering amplitudes}

 \end{center}

 \vspace{0.1cm}
 \begin{center}
{\large Massimo Bianchi\footnote{E-mail: massimo.bianchi@roma2.infn.it}
 and Maurizio Firrotta\footnote{E-mail: maurizio.firrotta@roma2.infn.it}}

\vspace{0.5cm}
{\it Dipartimento di Fisica, Universit\`a di Roma Tor Vergata, \\
I.N.F.N. Sezione di Roma Tor Vergata, \\
Via della Ricerca Scientifica, 1 - 00133 Roma, ITALY}
\end{center}

\vspace{1cm}
\begin{abstract}
We study interactions of string coherent states in the DDF (after Di Vecchia, Del Giudice, Fubini) formalism. For simplicity we focus on open bosonic strings. After reviewing basic properties of DDF operators and of excited open strings, we present some classical profiles and show how they become more and more compact as the number of harmonics increases at fixed mass. 
We then compute various three- and four-point amplitudes with insertions of coherent states, tachyons and vector bosons on the boundary of the disk relying on a convenient choice of reference null momenta. We find that the amplitudes exponentiate in a rather subtle and interesting way. We then study the high-energy fixed-angle limit, dominated by a saddle-point when coherent states are present, and the soft behaviour as the momentum of a vector boson is taken to zero. We briefly comment on generalisation of our analysis to multiple intersecting and magnetised D-branes and to closed strings. 
\end{abstract}

\vspace{1cm}
\tableofcontents 
\newpage
\section*{Introduction}

Direct detection of gravitational waves (GW) by the LIGO/Virgo Collaboration from Black-Hole (BH)  
\cite{Abbott:2017oio, Abbott:2017vtc} and Neutron-Star (NS) \cite{TheLIGOScientific:2017qsa} 
mergers has triggered renewed interest in the dynamics of compact gravitating systems, in connection with gamma-ray bursts \cite{Monitor:2017mdv}, too.  

In string theory, the objects colloquially called `black holes' (BH's) can be represented as bound states of strings and branes \cite{Maldacena:1996ds, Maldacena:1996ky}. A lot of progress has been achieved in the micro-state counting for certain classes of BPS (charged) BH's in various dimensions \cite{Strominger:1996sh, Maldacena:1997de, Ooguri:2004zv}, leading to the `fuzzball' proposal \cite{Lunin:2001fv, Lunin:2001jy, Lunin:2002qf, Mathur:2005zp} that has been tested in various contexts \cite{Giusto:2011fy, Bianchi:2016bgx, Bena:2016ypk, Bianchi:2017bxl, Bena:2017xbt}. Yet very little is known at the dynamical level \cite{Bianchi:2017sds, Bianchi:2018kzy}. 

BH production and evaporation in high energy scattering \cite{Amati:1987wq, Muzinich:1987in, tHooft:1987vrq, Amati:1987uf, Gross:1987kza, Gross:1987ar} has been addressed by several groups 
\cite{Eardley:2002re,Kohlprath:2002yh, Veneziano:2004er, Amati:2007ak, Bianchi:2010dy, Dvali:2014ila, Addazi:2016ksu}, but a complete unitary S-matrix description is still missing that is expected to include gravitational bremsstrahlung. In particular it would be extremely interesting to derive string corrections to the GW signals predicted by General Relativity \cite{Smarr:1977fy, Amati:1990xe, Gruzinov:2014moa, Ciafaloni:2015xsr, Ciafaloni:2015vsa, Ciafaloni:2018uwe, Addazi:2019mjh, Damour18}. To this end one would like to find a reliable description of non-supersymmetric and non BPS BH's as bound states of strings, possibly refining and elaborating on the principles of Strings/Black-Hole complementarity \cite{Susskind:1993ki, Susskind:1993if, Lowe:1995ac, Horowitz:1996nw, Horowitz:1997jc, Damour:1999aw, Chialva:2009pf}. This approach has already been revived and improved in recent years \cite{Brustein:2016msz}, relying on the thermal-scalar picture \cite{Horowitz:1997jc} and the collapsed polymer description of highly-excited, long closed strings at and above the Hagedorn temperature, whose joining and splitting interactions were studied earlier on \cite{Lowe:1994nm}. The emerging picture suggests how `stringy' matter can resist gravitational collapse \cite{Brustein:2018web} and be probed in GW experiments \cite{Brustein:2017koc} but seems to require a more quantitative control on the highly excited string interactions.  

While BRST invariant vertex operators for massless states are well known, the identification of BRST invariant vertex operators for very massive states, possibly with high spin, is rather laborious. 
Covariant vertex operators for massive higher spin states have been studied in the past
\cite{Bianchi:2010dy, Bianchi:2010es, Bianchi:2011se, Black:2011ep} [ADD TAYLOR + TAIWAN] also in connection with the holographic AdS/CFT correspondence in the weak coupling / high curvature limit \cite{Bianchi:2003wx, Beisert:2004di, Bianchi:2004ww, Bianchi:2004xi, Bianchi:2005yh, Bianchi:2006gk, Bianchi:2010mg} and with scattering off D-branes in the eikonal regime \cite{DAppollonio:2010krb,Bianchi:2011se,Black:2011ep, DAppollonio:2013mgj, DAppollonio:2013okd}.

Interactions of higher spins in the first Regge trajectory have been studied at tree level for the open bosonic string \cite{Sagnotti:2010at}, for heterotic strings \cite{Bianchi:2010dy, Bianchi:2010es}  and for (open) superstrings \cite{Schlotterer:2010kk} but a systematic approach to the interactions of massive higher spin states is not yet available.

A possible way out is to rely on the time-honoured Del Giudice, Di Vecchia and 
Fubini (DDF) operators \cite{DelGiudice:1971yjh, DelGiudice:1972ru}, that produce BRST invariant states by construction. Three-point amplitudes have been studied and the three-Reggeon Vertex has been derived for bosonic strings \cite{Ademollo:1974kz, DiVecchia:1986jv}  as well as for the Neveu-Schwarz sector of the superstring \cite{Hornfeck:1987wt}.

In the past ten years or so, the DDF approach and the construction of macroscopic `coherent' states of strings has been revived in connection with cosmic strings\footnote{See also the talk \cite{SklirosOAC} for superstring coherent states.}, 
\cite{Skliros:2009cs, Hindmarsh:2010if, Skliros:2011si}, including applications to the decay of macroscopic strings \cite{Skliros:2013pka} and to the generating function for highly excited strings
\cite{Skliros:2016fqs}. The role of the DDF approach has also been recently emphasised in connection with causality and unitarity in string theory \cite{DAppollonio:2015fly} as well as with absorption in high-energy string-brane collisions \cite{DAppollonio:2015oag}. Other recent work on DDF operators has focussed on their role in the BCFW construction of Veneziano Amplitude \cite{Fotopoulos:2010jz} and on three-point amplitudes with massive legs \cite{Boels:2012if}. 

Aim of the present paper is to study interactions of string coherent states in the DDF approach. For simplicity we will focus on open bosonic strings, although we plan to generalise our results to closed strings in view of their role in the understanding of BH physics and GW emission. As in the familiar case of the harmonic oscillator, coherent states are the quantum version of classical (string) configurations that can capture the dynamics of long, very massive and highly excited strings. As mentioned above, our results or their generalization to closed strings may find applications in cosmic strings \cite{Skliros:2009cs, Hindmarsh:2010if, Skliros:2011si} and BH dynamics \cite{Brustein:2016msz, Lowe:1994nm, Brustein:2018web}. Moreover, coherent states can be used as an (over)complete set to built higher order perturbative string amplitudes by sewing procedures \cite{Skliros:2013pka, Skliros:2016fqs, SklirosOAC}. Let us finally stress that the DDF approach is not limited to coherent states but it can profitably be used to investigate the dynamics of arbitrary BRST invariant states in any string theory.  

The plan of the paper is as follows. 

In Section 1, we will review DDF operators and the construction of physical BRST invariant states for open bosonic strings, closely following \cite{Skliros:2009cs, Hindmarsh:2010if, Skliros:2011si} and \cite{Skliros:2016fqs, SklirosOAC}. In particular we will show how the first ($N=1$) and second ($N=2$) level are neatly reproduced and describe how to construct coherent states. 

In Section 2, we will recall some properties of coherent states that emerge from the computation of 2-point `amplitudes'. In particular we reproduce the results of \cite{Skliros:2009cs, Hindmarsh:2010if, Skliros:2011si} for the average mass, gyration radius and angular momentum. We then present some random open string profiles and their time evolution and show how they become more and more compct as the number of harmonics is increased at fixed mass. 

In Section 3, we will compute 3-point `amplitudes' involving coherent states as well as tachyons and vector bosons. Quite remarkably, we find that the amplitudes exponentiate in a subtle way that requires a decomposition over the levels. As a by-product of our computations we will find some interesting identities involving cycle index polynomials that are ubiquitous when working with coherent states. 3-point amplitudes encode the decay (production) rates of coherent states into (from) other coherent states or low-lying states and, in particular, can be used to extract the `spectrum' of radiation emitted from this kind of processes. We will mostly work in the case of a single D25-brane\footnote{In fact any Dp-brane, if the momenta are properly restricted.} but we will discuss how to generalise the results to multiple D25-branes with the inclusion of Chan-Paton (CP) factors. We will only mention but not discuss in any detail the case of intersecting and magnetised D-branes. 

In Section 4, we will compute 4-point `amplitudes' involving coherent states with tachyons and vector bosons. As in the previous section, we will find that the 4-point amplitudes exponentiate when decomposed over levels and new and  interesting identities involving cycle index polynomials are found. We will then consider the high-energy fixed-angle regime and find how the known saddle-point is modified by the presence of coherent states. Moreover, we will study the soft limit at the open string level as the momentum of a massless vector is taken to zero, along the lines of the old literature \cite{Gell-MannGoldberger, Low} for QED and \cite{BurnettKroll} for gauge theories, that has been later on extended to gravity \cite{Weinberg, GrossJackiw, Jackiw, Donoghue:1999qh, White1103.2981} and recently generalised, after \cite{Cachazo:2014fwa}, in various contexts \cite{Casali:2014xpa, Schwab:2014xua, Afkhami-Jeddi:2014fia, Bern:2014vva}, including string theory \cite{Bianchi:2014gla, DiVecchia:2015oba, Bianchi:2015lnw, Bianchi:2016tju}, effective theories \cite{DiVecchia:2015jaq, Bianchi:2016viy, Guerrieri:2017ujb} and loop corrections to the universal tree-level behaviour \cite{Bern:2014oka, Sen:2017xjn, Laddha:2018rle,Sahoo:2018lxl}.

We will conclude in Section 5, with preliminary considerations on how to generalise our analysis to the case of open strings ending on intersecting or magnetised D-branes and to closed strings. This is particularly easy for 3-point amplitudes thanks to the simplicity of KLT formula in this case.  


\section{DDF operators and the String Spectrum}

In this Section, closely following  DDF \cite{DelGiudice:1971yjh, DelGiudice:1972ru} as well as \cite{Skliros:2009cs, Hindmarsh:2010if, Skliros:2011si}, we review the definition of DDF operators and their crucial role in the construction of physical BRST invariant states for open bosonic strings. After recalling the DDF approach that combines the virtues of the light-cone and covariant approaches, we show how to construct the first ($N=1$) and second ($N=2$) excited level and then describe the construction of coherent states.
\subsection{DDF operators for open bosonic strings}
In the open bosonic string, the DDF operators  are defined 
as
\begin{equation}\label{DDFop}
A^i_n= \dfrac{i}{\sqrt{2\alpha'}}\oint {dz \over 2\pi i} \, \pa_{z}X^i(z)e^{inq{\cdot}  X(z)}
\end{equation}
where $i=1,..., D{-}2$ ($D=26$) and $q^2=0$. For convenience, we set $q^+=q^i=0$ and $q^-\neq 0$ from the start, so that 
$q{\cdot}  X = {-}q^+X^-{-}q^-X^+{+}q_iX^i = {-}q^-X^+$ with 
\begin{equation}
X^+={1\over \sqrt{2}}(X^0+X^{D{-}1})= x^+ + 2\alpha' p^+ \tau
\end{equation}
Computing the OPE and imposing 
 \begin{equation} \label{12pq}
 2\alpha'{{p}}{\cdot}  q=1
 \end{equation}
 with ${{p}}^\mu$ the zero-mode of the momentum `operator'  
  \begin{equation} \label{Poperator}
{{p}}^\mu = i \oint {dz \over 2\pi i} \pa X^\mu = \oint {dz \over 2\pi i} P^\mu
\end{equation}
one finds the commutators\footnote{Note that $2\alpha'{{p}}{\cdot}  q\approx 1$ is `central' in that it commutes with $A^i_n$.}
\begin{equation}\label{DDFcomm}
[A^i_{n},A^j_{m}]=n\,\delta^{ij}\delta_{m+n,0}
\end{equation}
 Despite appearance, DDF operators reproduce ordinary BRST invariant and covariant vertex operators for the open bosonic string as we will see momentarily. For closed strings, as usual, one has a doubling of the modes $A^i_{n}\rightarrow (A^i_{n,L},A^j_{n,R})$.

\subsection{Vertex operators for open bosonic strings}
Starting from the tachyon vertex operator
\begin{equation}\label{tachVert}
V_T(z,p)={:}e^{ip{\cdot}  X(z)}{:}
\end{equation}
with 
\begin{equation}\label{TachMS}
p^2=\dfrac{1}{\alpha'} = - M^2
\end{equation}
and imposing (\ref{12pq}) one can successively construct BRST invariant vertex operators with on-shell momenta 
\begin{equation}\label{TachMS}
p_{_N} = p - Nq \qquad {\rm such \ that} \qquad - {\alpha'}p_{_N}^2 = {\alpha'}M_N^2 = N-1
\end{equation}
We henceforth set ${\alpha'}=1/2$ for convenience.

In particular at the massless level ($N=1$) one has the vector boson vertex operator
\begin{equation}\label{VectVertNoCov}
{:}\lambda_{i} A^i_{-1} e^{ip{\cdot}  X}{:}= \lambda_{i}(\delta^{i}_{\mu} - p^i q_{\mu})\, i\pa X^{\mu} e^{i(p-q){\cdot}  X}
\end{equation}
which can be written in the conventional form
\begin{equation}\label{VectVert}
V_A(k, \varepsilon, z)=\varepsilon_{\mu} \,i\pa_{z}X^{\mu} e^{ik{\cdot}  X(z)}
\end{equation}
with $k=p-q$ and $\varepsilon_{\mu}=\lambda_{i}(\delta^{i}_{\mu} - p^i q_{\mu})$  such that $k^2=0$ and $k{{\cdot} }\varepsilon =0$. 

At the next ($N=2$)  level one has two possibilities.
The first one is 
\begin{equation}\label{A2VertOp}
{:}{{b}}_{i}A^i_{-2}\, e^{ip{\cdot}  X}{:}=\left[i{{b}}{\cdot}\pa^2 X + {{b}}{\cdot}p \,\left( \dfrac{{{u}}_{1}^{2}}{2} + {{{u}}_{2}\over 2} \right) +i {{b}}{\cdot}  \pa X \,{{u}}_{1}\right]  e^{i(p-2q)X}
 \end{equation} 
where ${{u}}_\ell ={\cal U}^{(n=2)}_\ell$ with
\begin{equation}\label{defUcal}
{\cal U}^{(n)}_\ell(z) = -i\dfrac{n}{(\ell{-}1)!} q{{\cdot} }{\pa_z^{\ell}X} = i\dfrac{n}{(\ell{-}1)!}  q^-{\pa_z^{\ell}X^+}
 \end{equation} 

The second one is 
\begin{equation}
{:}{{e}}_{ij}A^i_{-1}A^j_{-1} e^{ip{\cdot}  X}{:}={{e}}_{ij} \Big[{H}^i_1{H}^j_1\, - \delta^{ij}\Big( (iq{\cdot}\pa X)^2- iq{\cdot}  \pa^2 X \Big)  \Big]   e^{i(p-2q)X}
\end{equation}
where \begin{equation}\label{Hdef}
{H}^{i}_{n}(z)=p^{i} {\cal Z}_{n}({\cal U}^{(n)}_\ell ) + {\mathcal{P}}^{i}_{n}(z)
\end{equation}
with 
\begin{equation}\label{Pdef}
\boxed{{\mathcal{P}}^{i}_{n}(z)=\sum_{h=1}^{n}  i\,\dfrac{\pa^{h} X^{i}_{z}}{(h-1)!} \,{\cal Z}_{n-h}({\cal U}^{(n)}_\ell ) }
\end{equation}
${\cal Z}_{n}({{u}}_{\ell})$ are degree $n$ cycle index polynomials of the symmetric group $S_n$, in the $n$ variables $u_\ell$, with $\ell=1,...,n$. They are related to Schur polynomials by ${\cal S}_{n}({{u}}_{\ell}) = {\cal Z}_{n}(\ell {{u}}_{\ell})$. Some of their properies will be recalled momentarily. 

The two vertex operators can be combined and written in the form
\begin{equation}\label{VectVertCov}
V_{\cal E}(k, \varepsilon, z)={\cal E}_{\mu\nu} i\pa_{z}X^{\mu} i\pa_{z}X^{\nu}e^{iP{\cdot}  X} \quad , \quad V_{\cal B}(k, \varepsilon, z)={\cal B}_{\mu} i\pa^2_{z}X^{\mu} e^{iP{\cdot}  X} \quad , 
\end{equation}
with $P=p-2q$, such that $P^2=-2$,
\begin{equation}
{\cal E}_{\mu\nu}=e_{ij}(\delta^i_\mu - p^i q_\mu)(\delta^j_\nu - p^j q_\nu) +\delta^{ij} e_{ij} q_\mu q_\nu - b_i[(\delta^i_\mu - p^i q_\mu)q_\nu +(\delta^i_\nu - p^i q_\nu)q_\mu]\end{equation}
and 
\begin{equation}
{\cal B}_{\mu}=b_i(\delta^i_\mu - p^i q_\mu) - \delta^{ij} e_{ij} q_\mu
\end{equation}
so that 
\begin{equation}
P^\mu{\cal E}_{\mu\nu} = - {\cal B}_{\nu}\quad , \quad \eta^{\mu\nu} {\cal E}_{\mu\nu} = -P^\mu{\cal B}_{\mu}\end{equation}
In fact one can `gauge away' ${\cal B}_{\mu}$ and impose  
$P^\mu{\cal E}_{\mu\nu} = 0$ and $\eta^{\mu\nu} {\cal E}_{\mu\nu}=0$ \cite{Skliros:2009cs, Hindmarsh:2010if, Skliros:2011si, Bianchi:2010dy, Bianchi:2011se, Bianchi:2010es}.

The explicit form of the cycle index polynomials is given by 
\begin{equation}\label{Zdef}
\boxed{{\cal Z}_{n}(u_{\ell})= \oint {dz \over 2\pi i z^{n+1}} e^{\sum_{\ell=1}^{\infty}{1\over \ell}{u_{\ell}} z^{\ell}} =
\dfrac{1}{n!}\pa_{w}^{n} \left(  e^{\sum_{\ell=1}^{\infty}{1\over \ell} {u_{\ell}}(w-z)^{\ell}} \right) \Big|_{w=z} = \sum_{\ell_k: \Sigma_k k \ell_k=n} \prod_{k=1}^n \dfrac{u_{k}^{\ell_k}}{\ell_k ! k^{\ell_k}} }
\end{equation}
For low degree one finds
\begin{equation}\label{Z123}
{\cal Z}_{0}=1 \quad , \quad{\cal Z}_{1}={{u}}_{1}
\quad {\cal Z}_{2}=
 \dfrac{{{u}}_1^2}{2}+{{{u}}_2 \over {{2}}} \quad , \quad
{\cal Z}_{3}=
 \dfrac{{{u}}_1^3}{6}+{{{u}}_2 {{u}}_1 \over {{2}}} +{{{u}}_3\over {{3}}} 
 \end{equation}
 \begin{equation}\label{Z45}
 {\cal Z}_{4}{=}\dfrac{1}{24} \left({{u}}_1^4{+}6 {{u}}_2 {{u}}_1^2{+}3 {{u}}_2^2\right){+}{{{u}}_1 {{u}}_3\over {{3}}} {+}{{{u}}_4\over {{4}}}\,, \,\, {\cal Z}_{5}({{u}}_{\ell}){=}\dfrac{{{u}}_1^5}{120}{+}\dfrac{1}{12}{{u}}_2 {{u}}_1^3{+}\dfrac{1}{6} {{u}}_3
   {{u}}_1^2{+}\left(\dfrac{{{u}}_2^2}{8}{+}{{{u}}_4\over {{4}}}\right){{u}}_1{+}{{{u}}_2 {{u}}_3 \over {{6}}}{+}{{{u}}_5\over {{5}}} \end{equation}

Some relevant properties of cycle index polynomials are
 \begin{eqnarray}\label{Zproperties}
 \label{Zprop1} &&{\cal Z}_{n}(u_\ell + v_\ell) = \sum_{k=0}^{n} {\cal Z}_{n-k}(u_\ell){\cal Z}_{k}(v_\ell) \\
  \label{Zprop2} &&{\cal Z}_{n}(\lambda^\ell u_\ell)= \lambda^{n} {\cal Z}_{n}(u_\ell) \\
  \label{Zprop3} &&{\cal Z}_{n}(u) = {1\over n}  {\Gamma(u+n) \over \Gamma(u)\Gamma(n)} \\
  \label{Zprop4} &&{\cal Z}_{n}(u_\ell)= {1\over n} \sum_{k=1}^{n} u_k {\cal Z}_{n-k}(u_\ell) 
 \end{eqnarray}

In general, the action of two DDF operators produces expressions of the form
\begin{equation} \label{AAonTach}
{:}A_{-n}^i A_{-m}^{j} e^{i p{{\cdot} } X}{:}=[\delta^{ij} \mathcal{S}_{m,n}\, + H_{m}^{j} H_{n}^{i} - H_{m}^{j} m q^i {\cal Z}_{n}({\cal U}^{(n)}_\ell )]e^{i [p-(n{+}m)q]{\cdot}X}
\end{equation}
where
\begin{equation}\label{SmnDef}
\boxed{\mathcal{S}_{m,n}(z)=\sum_{{h}=1}^{n} h \, {\cal Z}_{m+{h}}({\cal U}^{(m)}_\ell ) \,{\cal Z}_{n-{h}}({\cal U}^{(n)}_\ell) = \mathcal{S}_{n,m}(z)} \end{equation}
which, as indicated, is symmetric under the exchange of $m$ and $n$, though not manifestly so.

Iterating (\ref{AAonTach}) one finds 
\begin{equation}\label{AAAonTach}
{:}A_{-l}^{i}A_{-m}^{j}A_{-n}^{k}\,e^{ip{\cdot}  X}{:}= [H^{i}_{l} H_{m}^{j} H^{k}_{n} + \delta^{ij} \mathcal{S}_{l,m} H^{k}_{n} + \delta^{ik} \mathcal{S}_{l,n} H^{j}_{m} + \delta^{jk} \mathcal{S}_{n,m} H^{i}_{l}] 
e^{i [p-(n{+}m{+}l)q]{\cdot}X}  
\end{equation}
and in general
\begin{equation}\label{AgtimesonTach}
{:}A^{i_{1}}_{-n_{1}} ... A^{i_{g}}_{-n_{g}}\,e^{ip{\cdot}  X}{:}=\sum_{a=0}^{[g/2]} \sum_{\pi \in S_{g}} \prod_{l=1}^{a} \delta^{i_{\pi(2l+1)} i_{\pi(2l)}} \mathcal{S}_{n_{\pi(2l+1)},n_{\pi(2l)}} \prod_{q=2a+1}^{g} H^{i_{\pi(q)}}_{n_{\pi(q)}} \,e^{-i (p - \sum_{r} n_{r}q){\cdot}  X}
\end{equation}

This allows to constructed BRST invariant vertex operators that correspond to physical states at arbitrary level $N$ as well as to coherent states, which we now turn our attention on.

\subsection{Coherent states of the open bosonic string} 
Taking $n_{1}= ... = n_{g}=n$ the sum over permutations in (\ref{AgtimesonTach}) can be easily computed and reduces to
\begin{equation}
{:}\dfrac{1}{g!}( \lambda_n{{\cdot} } A_{-n} )^{g}\, e^{ip{\cdot}  X}{:}= \sum_{h=0}^{[g/2]} \dfrac{1}{h! (g-2h)!} \Big( \dfrac{\lambda_n{\cdot}  \lambda_n}{2} \mathcal{S}_{n,n} \Big)^{h} \Big( \lambda_n{{\cdot} } H_{n} \Big)^{g-2h} \,e^{i(p-gnq)X}
\end{equation}

The vertex operator for a coherent state can be contructed as an exponential of DDF creation operators $A^i_{-n}$ (with $n>0$) acting on the tachyonic ground-state. It reads 
\begin{equation}\label{CohVertOp}
{{V}}_{\mathcal{C}}(z)= {:}e^{\sum_{n=1}^{\infty} {1\over n}{\lambda_n}{\cdot}  A_{-n}}\, e ^{i p{{\cdot} } X }{:}= \sum_{g=0}^{\infty} \dfrac{1}{g!}  {:}\left( \sum_{n=1}^{\infty} {1\over n}\lambda_n{{\cdot} } A_{-n}\right)^{g} e^{ip{\cdot}  X}{:}
\end{equation}
It is coherent in that it is an eigenstate of the annihilation operators $A^i_{n}$ (with $n>0$) {\it viz.}
\begin{equation}\label{EigenvEq}
A^i_m e^{\sum_{n=1}^{\infty} {1\over n}{\lambda_n}{\cdot}  A_{-n}}\, e ^{i p{{\cdot} } X }|0\rangle =
\lambda^i_m e^{\sum_{n=1}^{\infty} {1\over n}{\lambda_n}{\cdot}  A_{-n}}\, e ^{i p{{\cdot} } X }|0\rangle
\end{equation}

After normal ordering one finds
\begin{equation}\label{CohVertOpFin}
\boxed{{{V}}_{\mathcal{C}}(z)=\exp\Bigg\{\sum_{r,s}^{1,\infty}\dfrac{\zeta_{r}{{\cdot} } \zeta_s }{2rs}\, \mathcal{S}_{r,s} \,e^{-i(r+s)q{\cdot}  X} + \sum_{n}^{1,\infty}\dfrac{1}{n}{\zeta_n}{\cdot}  \mathcal{P}_{n} \,e^{-inq{\cdot}  X}  \Bigg\} \,e^{ip{\cdot}  X}}
\end{equation}
with $\zeta_{n}^{\mu}=\lambda^{i}_n( \delta^{i \mu} - [2\alpha'] p^i q^{\mu} )$ such that $\zeta_n{\cdot}p=0= \zeta_n{\cdot}q$ and $\zeta_m{\cdot}  \zeta_n = \lambda_m{{\cdot} } \lambda_n$, while
\begin{equation}\label{HtoPdef}
\boxed{\zeta_{n}{\cdot}  \mathcal{P}_{n} = \lambda_n{{\cdot} } H_{n} = \sum_{h=1}^{n} \dfrac{i}{(h-1)!}\,{\cal Z}_{n-h}({{u}}_{\ell})\,\zeta_{n}{{\cdot} } \pa^{h}X }
\end{equation}
In order to derive  (\ref{CohVertOpFin}) use has been made of property (\ref{Zprop4}) of the cycle index polynomials.
 
For later purposes, it proves useful to expand the vertex for a coherent state as a superposition of vertex operators for states at fixed level $N$ 
\begin{equation}\label{CohVertExpan}
{{V}}_{\mathcal{C}}(z)=\sum_{N=0}^\infty \sum_{I=1}^{d_N} \mathcal{V}^{(I)}_{N}(z) e^{i(p-Nq){\cdot}  X} = \sum_{N} \sum_{h_n: \Sigma nh_n=N} \mathcal{V}_{\{n,h_n\}}(z) e^{i(p-q\sum nh_n){\cdot}  X}= V_T + V_A + V_H + ...\end{equation}
\begin{equation}
= e^{ip{\cdot}  X}\vert_{N=0} + iA(\zeta_1){\cdot}\partial X e^{i(p-q){\cdot}  X}\vert_{N=1}+
[\partial X{\cdot} H(\zeta_1,\zeta_2){\cdot}\partial X+i B(\zeta_1,\zeta_2){\cdot}\partial^2 X  ]e^{i(p-2q){\cdot}X}\vert_{N=2}+ ... \nonumber
\end{equation} 
and get the explicit expressions of the polarisation tensors $A^\mu,H^{\mu\nu},B^\mu,...$ in terms of $\zeta^i_n$ from 
\begin{equation}\label{CohVertExpan}
{{V}}_{\mathcal{C}}(z)=\sum_{N=0}^\infty e^{i(p-Nq){\cdot}  X} \sum_{(\ell_n,k_{r,s}):\sum_n n\ell_n + \sum_{r,s} k_{r,s} (r+s) = N} \prod_{n=1}^N {1\over \ell_n!} \left(\dfrac{1}{n}{\zeta_n}{{\cdot} }\mathcal{P}_{n}\right)^{\ell_n} \prod_{r,s}^{1,N{-}1} {1\over k_{r,s}!} \left(\dfrac{\zeta_{r}{{\cdot} }\zeta_s }{2\,rs}\, \mathcal{S}_{r,s}\right)^{k_{r,s}}
\end{equation}
where
\begin{equation}\label{PdefZexp}
\zeta_{n}{{\cdot} }\mathcal{P}_{n}
= \sum_{\ell=1}^{n} \dfrac{i}{(\ell-1)!} \oint {du\over 2\pi i u^{n-\ell +1}} e^{i n \sum_{s=1}^{n-\ell} \frac{1}{s!}{u^s} q^-{\pa^{s}X^+}}
\zeta_{n}{{\cdot} }\pa^{\ell}X
\end{equation}
and 
\begin{equation}\label{SmnExpDef}
\mathcal{S}_{m,n}=\sum_{{h}=1}^{n} h \oint {du\over 2\pi i u^{m+h +1}} e^{i m \sum_{s=1}^{m+h} \frac{1}{s!}{u^s }  q^-{\pa^{s}X^+}} \oint {dv\over 2\pi i v^{n-h +1}} e^{i n \sum_{r=1}^{n-h} \frac{1}{r!}v^r q^-{\pa^{r}X^+}}
\end{equation}


\section{2-point `functions' of Coherent states}

In this Section, we recall some properties of coherent states that emerge from the computation of 2-point `amplitudes'. In particular we reproduce the results of \cite{Skliros:2009cs, Hindmarsh:2010if, Skliros:2011si, SklirosOAC} for the average mass, gyration radius and spin. 

\subsection{Classical profiles, Normalization}

A classical string configuration $X^\mu(\sigma,t)$ is physically acceptable when it satisfies the Virasoro constraints $T_{\alpha\beta} = 0$. 
For open strings one should impose boundary conditions at $\sigma=0,\pi$. For simplicity we will only consider Neumann b.c. that correspond to D25-branes\footnote{As already mentioned, switching to Dp-branes is straightforward.}. In this case the mode expansion reads
\begin{equation}
X^\mu(\sigma,\tau) = x^\mu + p^\mu {{\tau}} + \sum_{n=1}^\infty ({{a}}_n^\mu e^{-in\tau} + {{a}}_n^{\mu *} e^{+in\tau})\cos n\sigma = x^\mu + p^\mu \tau + \sum_{n\neq 0}^\infty {{a}}_n^\mu e^{-in\tau} \cos n\sigma
\end{equation}
with ${{a}}_{-n}^\mu = {{a}}_n^{\mu *}$.

For illustrative purposes, we display in Fig. 1 and Fig. 2 some classical open string profiles in $d=3$ space dimensions with different numbers of harmonics, generated randomly,  together with their time evolution.

\begin{figure}[!h]
\begin{center}
\includegraphics[scale=0.5]{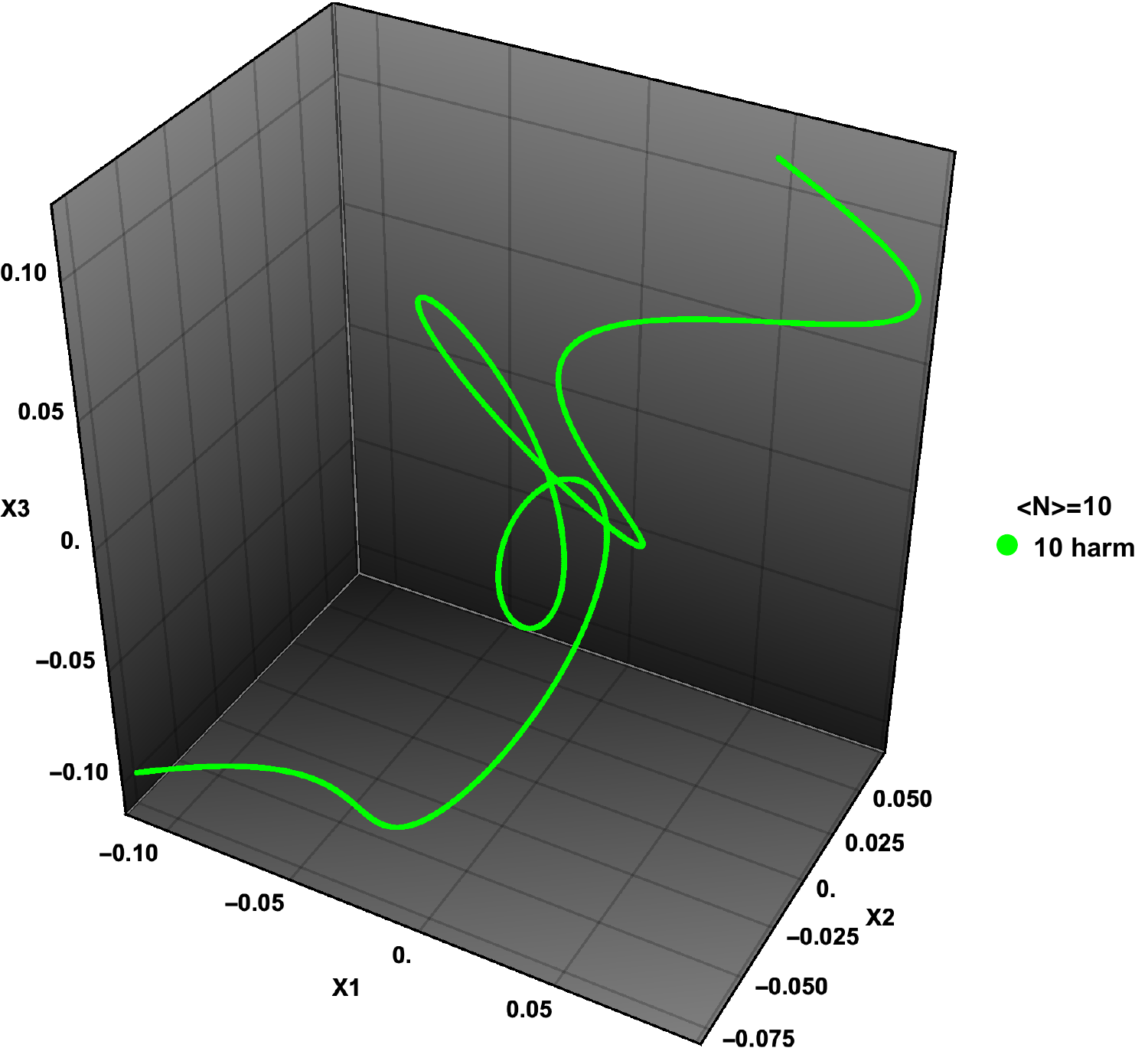} \includegraphics[scale=0.5]{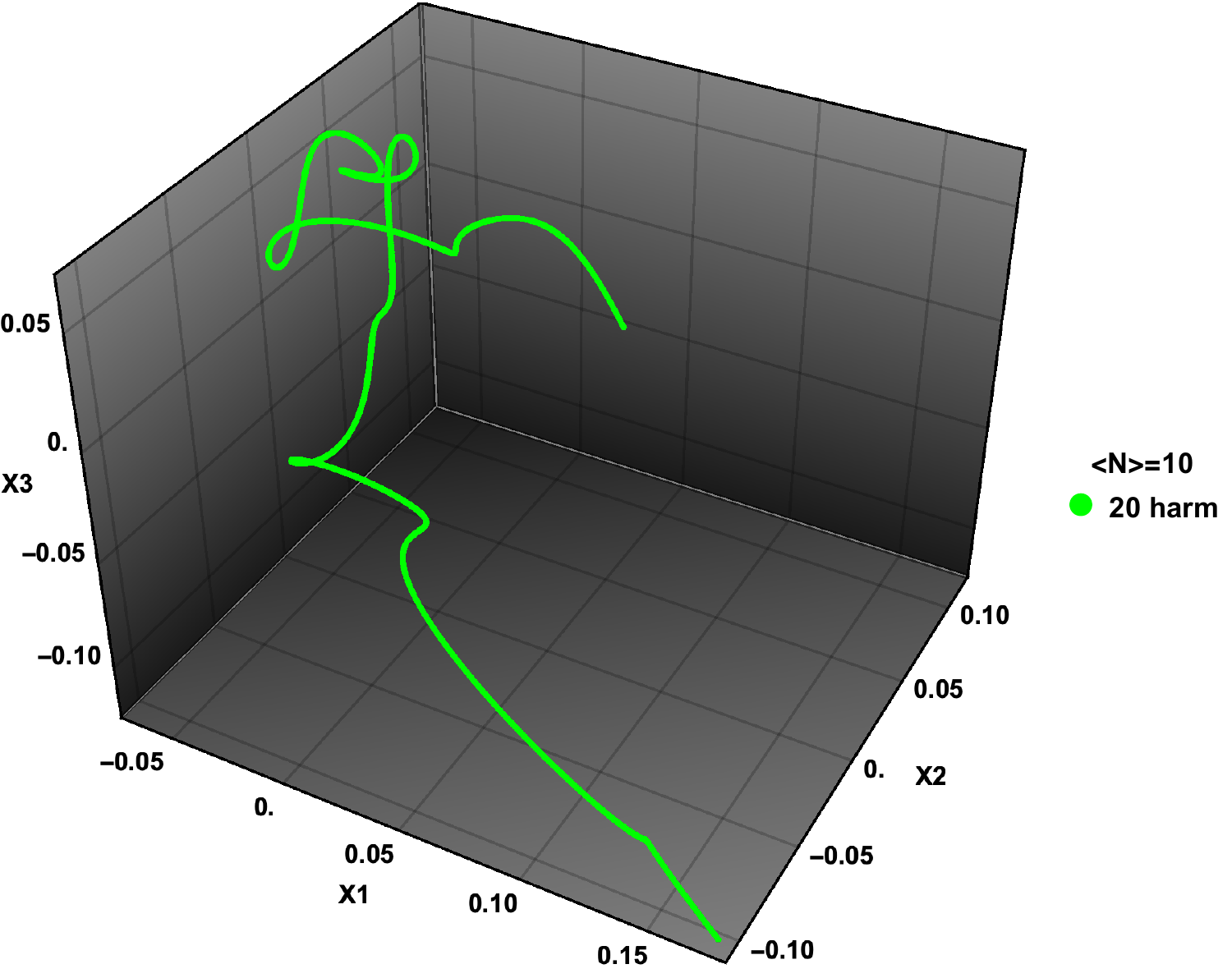}

\includegraphics[scale=0.5]{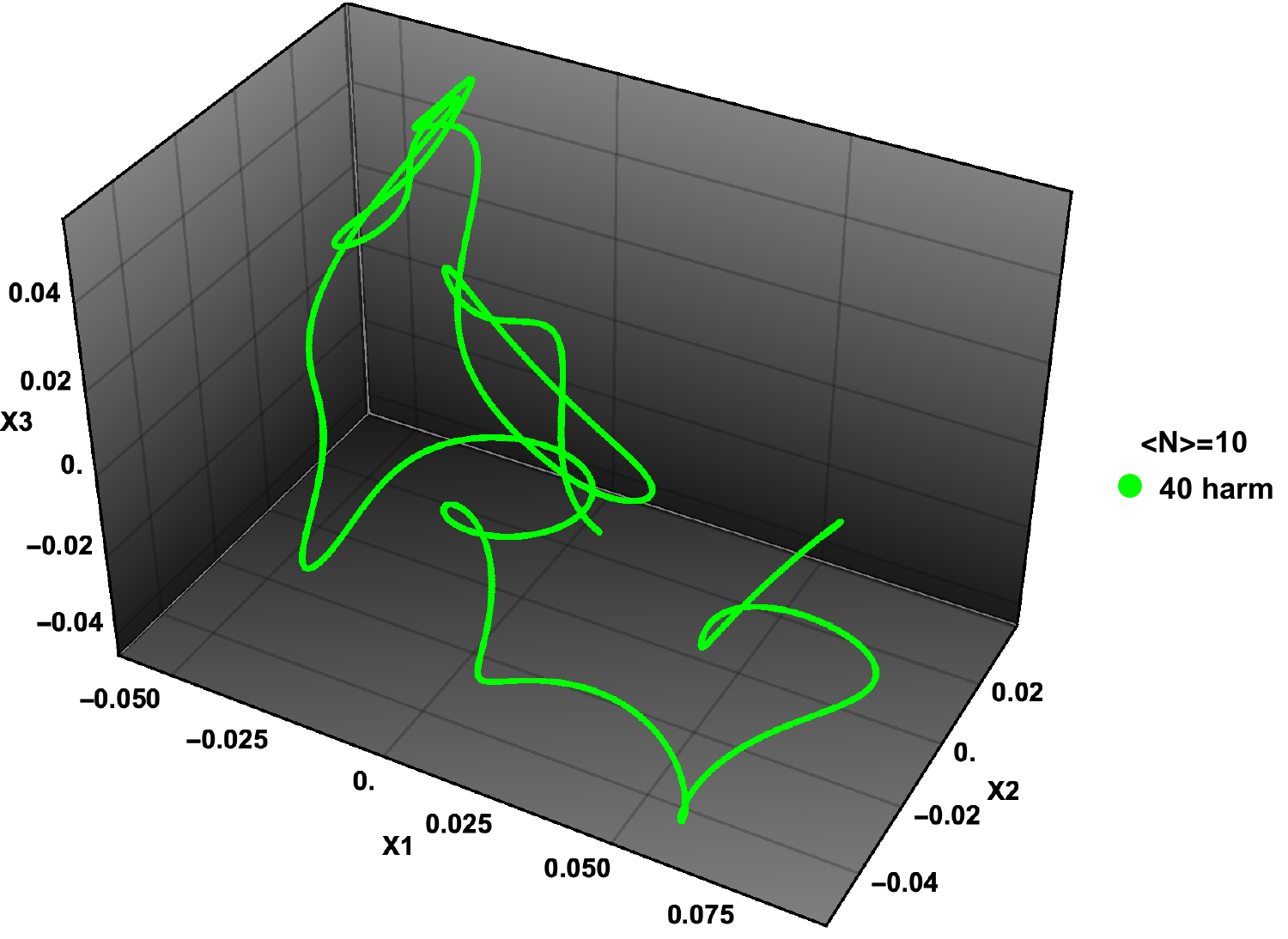}\includegraphics[scale=0.5]{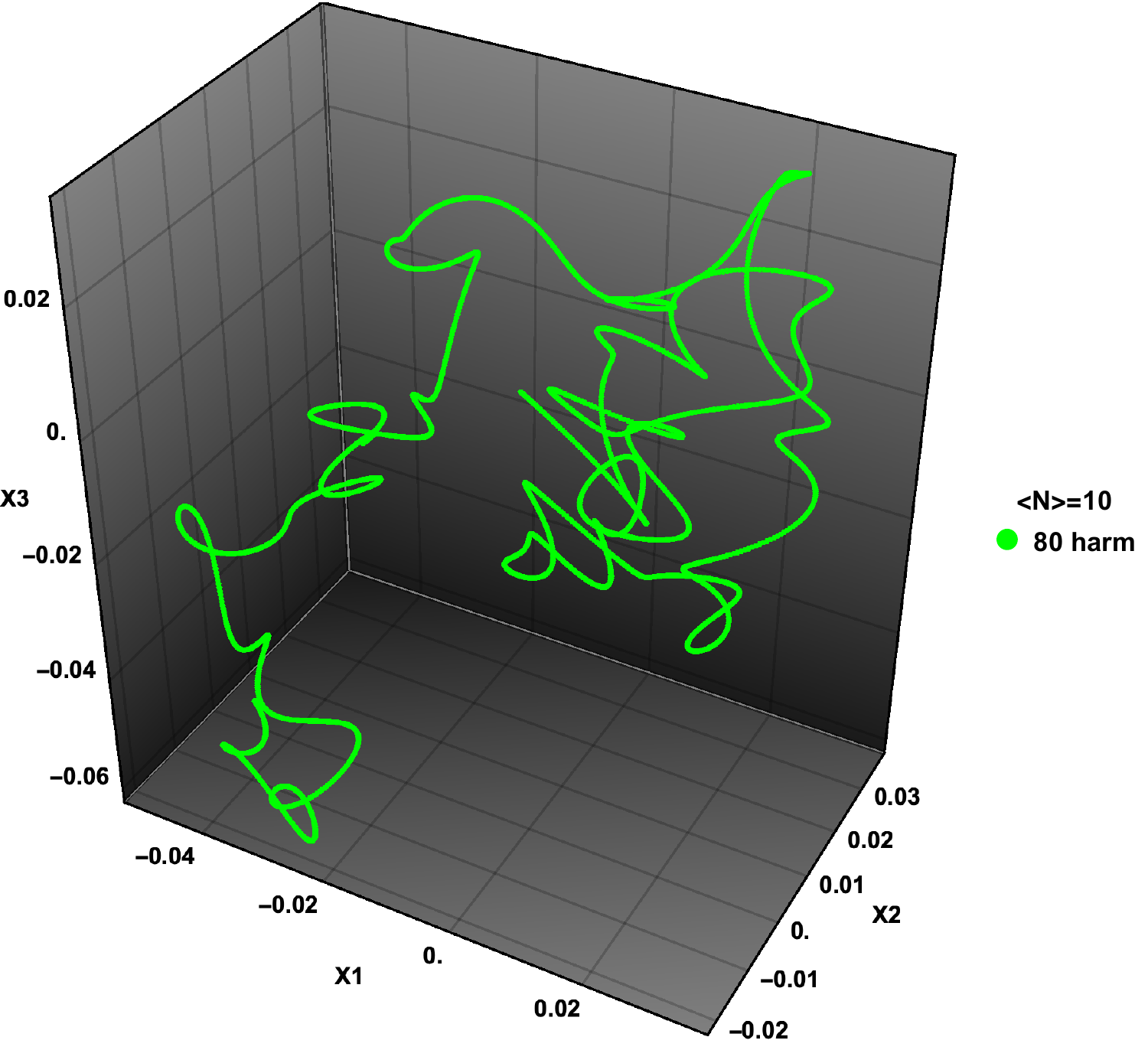}
\end{center}
\caption{3D open string profiles with different harmonics $\{10,20,40,80\}$ with the same average mass square $\braket{M^2}=9/\alpha'$}
\end{figure}
\begin{figure}[!h]
\begin{center}
\includegraphics[scale=0.5]{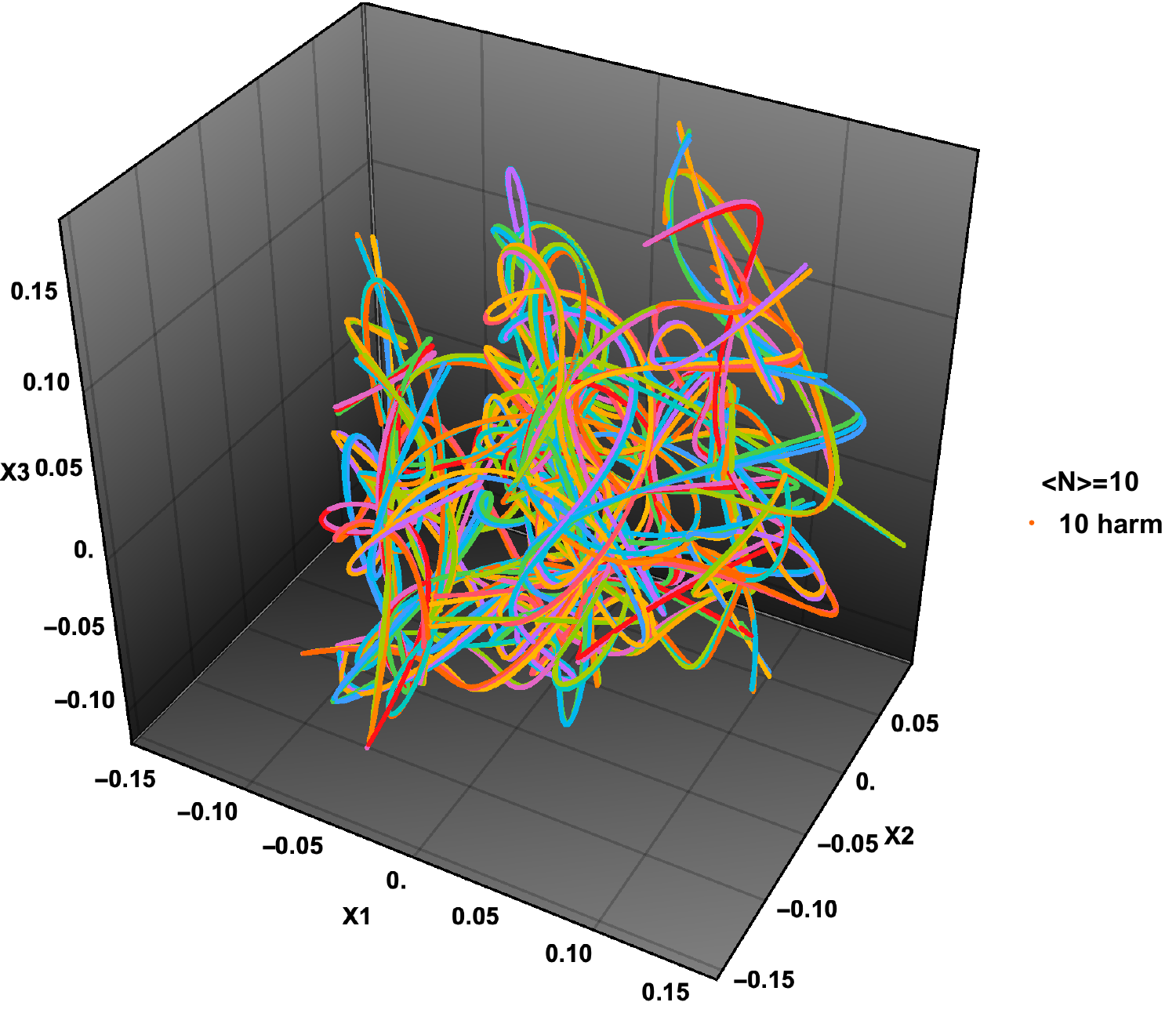} \includegraphics[scale=0.5]{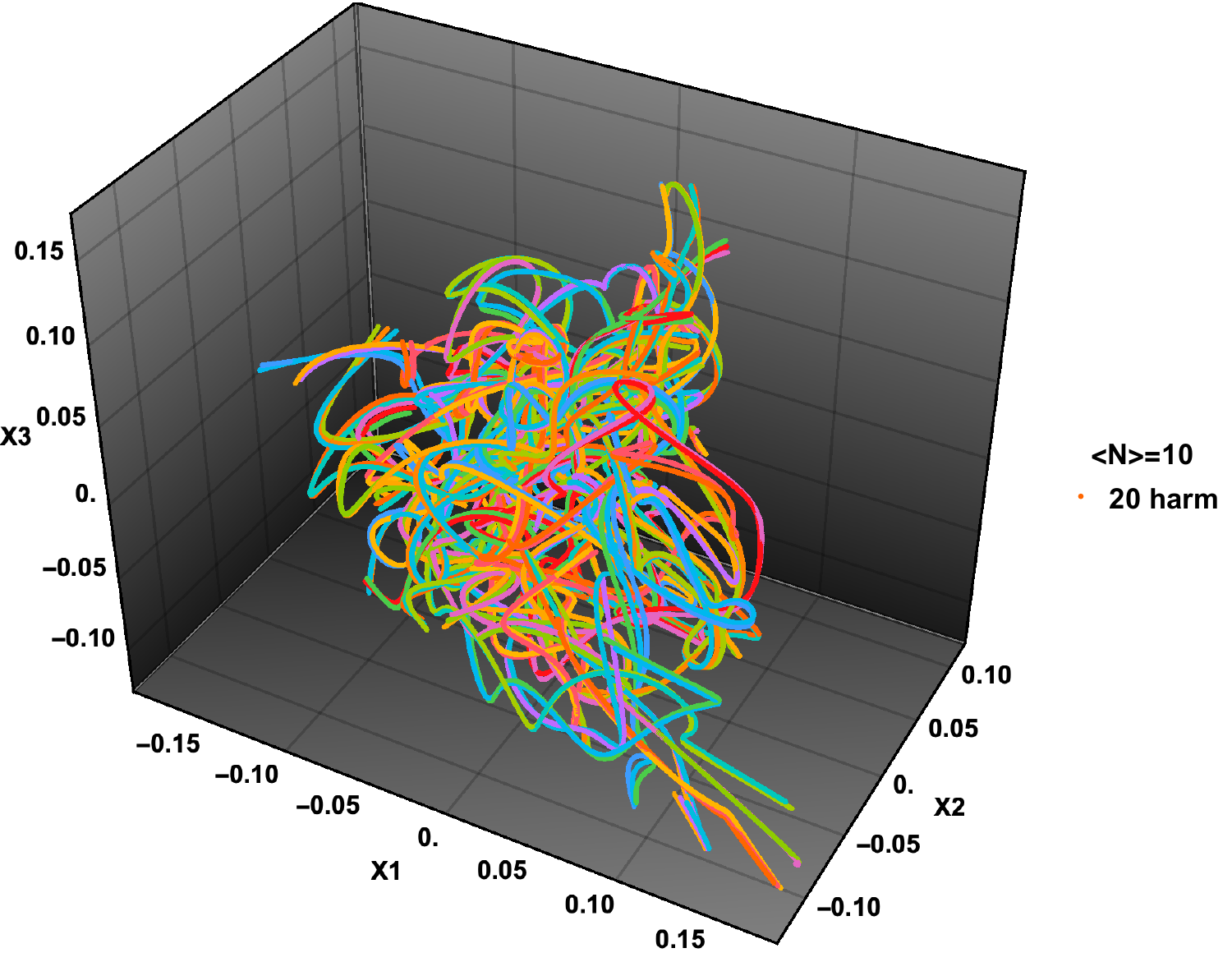}

\includegraphics[scale=0.5]{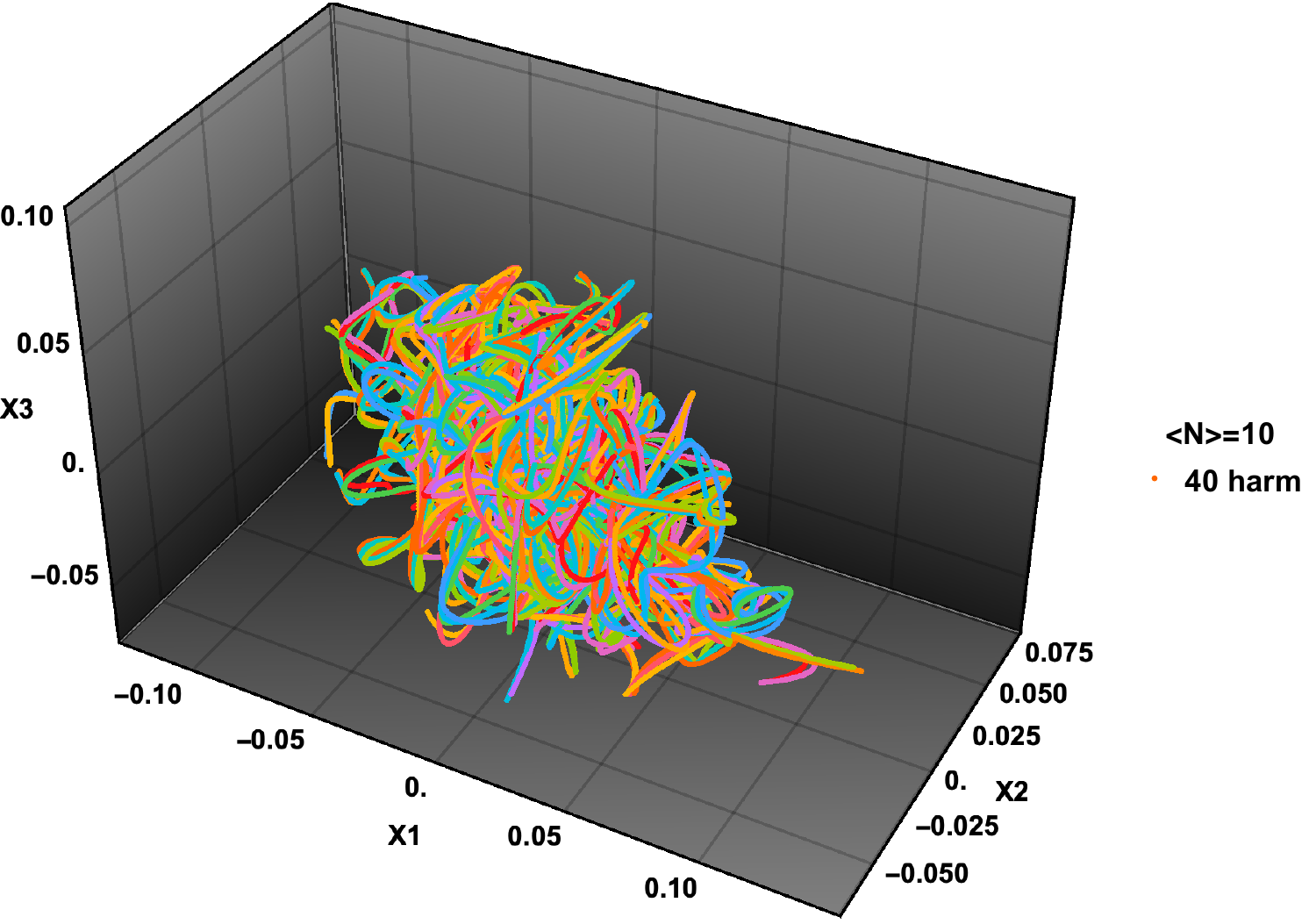}\includegraphics[scale=0.5]{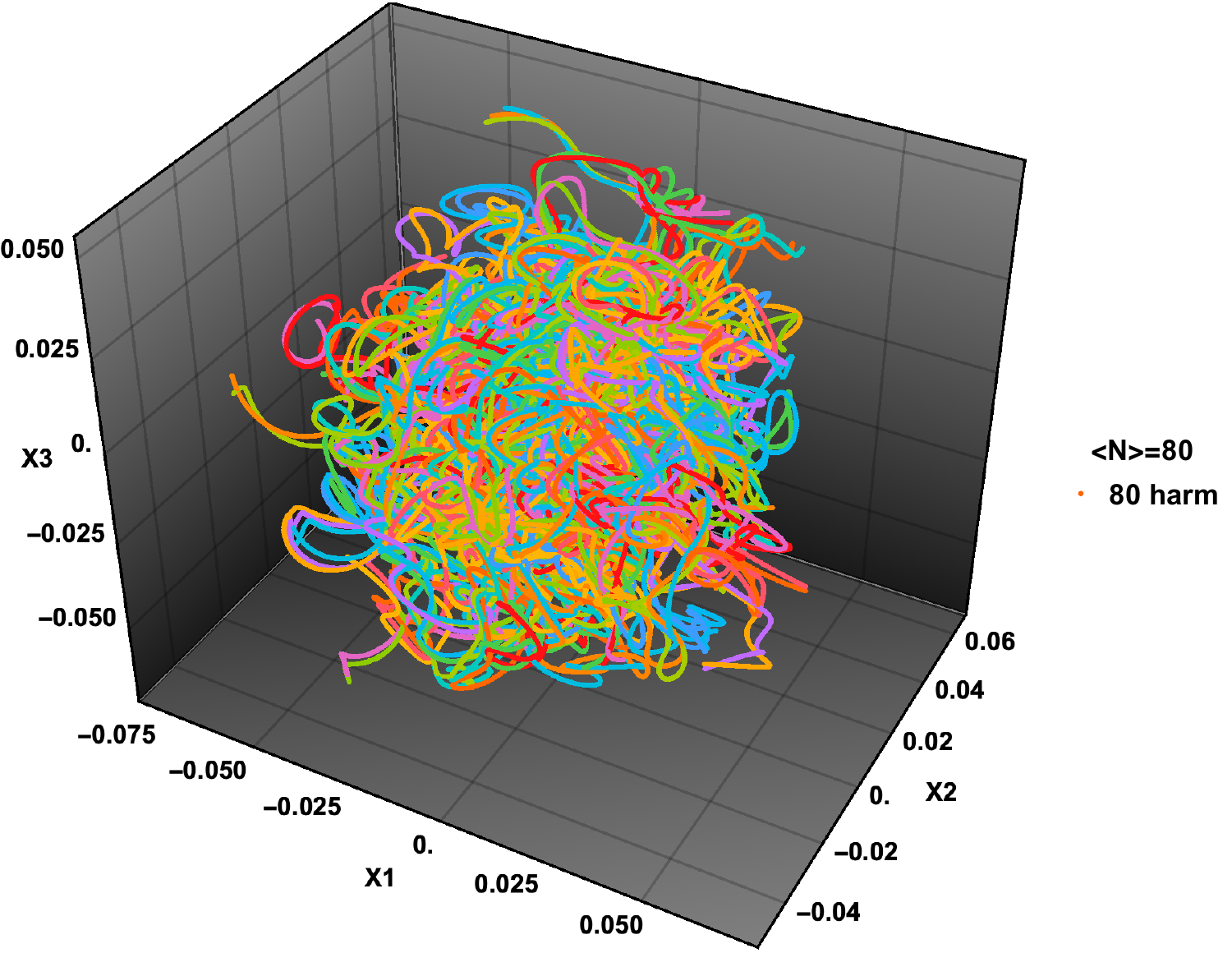}
\end{center}
\caption{Time evolution of the 3D open string profiles with different harmonics $\{10,20,40,80\}$ with the same average mass square $\braket{M^2}=9/\alpha'$, where profiles at different time are identified by different colors}
\end{figure}


At the quantum level ${{a}}_n^{\mu}$ become creation/annihilation operators that satisfy a Heisenberg algebra.  Virasoro constraints translate into the BRST condition, that is satisfied by the DDF operators, so much so that any state built acting with $A^i_{-n}$ on the tachyon vacuum $|p\rangle$ is a physical state. In particular the vertex operators (\ref{CohVertOpFin}) describe physical coherent states. It is relatively easy to fix their normalisation using standard properties of creation/annihilation operators\begin{equation}
\langle {{\cal C}}(\lambda',p')| {{\cal C}}(\lambda,p)\rangle = \lim_{\substack{z'\rightarrow \infty \\ z\rightarrow 0\,\,\,\,\,}}
\langle 0| {{V}}_{\mathcal{C}}(\lambda^*,p';z') {{V}}_{\mathcal{C}}(\lambda,p;z)|0\rangle = \langle p'|\exp\sum_{n'}{1\over n'}\lambda^*_{n'}{\cdot}A_{n'} \exp\sum_{n}{1\over n} \lambda_{n}{\cdot}A_{-n} |p\rangle \nonumber
\end{equation}   
\begin{equation}
=\langle -p'|\exp\sum_{n,n'}{n\delta_{n,n'} \delta_{ij}\over nn'}\lambda^{i*}_{n}\lambda^j_{n}|p\rangle = \delta(p-p') \exp\sum_{n}{1\over n}\lambda^{*}_{n}{\cdot}\lambda_{n}
\end{equation}

\subsection{Average values of position, momentum and angular momentum}
Replacing the standard oscillators $\alpha^i_n$ with the DDF operators $A^i_n$ -- given that they satisfy the same algebra -- and using the eigenvalue equation (\ref{EigenvEq}) in the form
\begin{equation}
A^i_n |{\mathcal{C}}(\lambda,p)\rangle = \lambda^i_n |{\mathcal{C}}(\lambda,p)\rangle 
\end{equation}
one finds 
\begin{equation}
X^i_{cl} = \langle {\mathcal{C}}(\lambda',p')|X^i | {\mathcal{C}}(\lambda,p)\rangle = \sum_n {1\over n}(\lambda^i_n e^{-in\tau} + \lambda_n^{i*} e^{+in\tau}) 
\end{equation}
for the transverse coordinates while 
\begin{equation}
X^+_{cl} = \langle {\mathcal{C}}(\lambda',p')|X^+ | {\mathcal{C}}(\lambda,p)\rangle = X^+ = x^+ + p^+ \tau
\end{equation}
and 
\begin{equation}
X^-_{cl} = \langle {\mathcal{C}}(\lambda',p')|X^- | {\mathcal{C}}(\lambda,p)\rangle = x^- + p^- \tau + \sum_{n,m} {1\over 4 n p^+} [\lambda_{n-m}{\cdot}\lambda_m e^{-in\tau} + c.c.]
\end{equation}
In the rest frame, whereby $p^i=0$, reinstating $\alpha'$, the gyration radius is given by 
\begin{equation}
\boxed{\delta{X}^2 = \langle (X-X_{cl})^2 \rangle = 2\alpha' \sum_{n}{1\over n^2} \lambda^{*}_{n}{\cdot}\lambda_{n}}
\end{equation}
The average value of the transverse momentum is given by
\begin{equation}
\langle {\mathcal{C}}(\lambda',p')|P^i | {\mathcal{C}}(\lambda,p)\rangle = \sum_n in (\lambda^i_n e^{-in\tau} - \lambda_n^{i*} e^{+in\tau})  
\end{equation}
Including the light-cone coordinates one has
\begin{equation}
\langle {\mathcal{C}}(\lambda',p')|P^\mu | {\mathcal{C}}(\lambda,p)\rangle = p^\mu - \langle N\rangle q^\mu\end{equation}
where 
\begin{equation}
\boxed{\langle N\rangle = \sum_{n}\lambda^{*}_{n}{\cdot}\lambda_{n} = \sum_{n}\zeta^{*}_{n}{\cdot}\zeta_{n}}
\end{equation}
as a result, reinstating $\alpha'$, the average mass is 
\begin{equation}
\boxed{\langle {\mathcal{C}}(\lambda',p')|{\cal M}^2 | {\mathcal{C}}(\lambda,p)\rangle = - \langle {\mathcal{C}}(\lambda',p')|P^2 | {\mathcal{C}}(\lambda,p)\rangle = {1\over \alpha'} (\langle N\rangle - 1)}
\end{equation}

\begin{figure}[!h]
\center
\includegraphics[scale=0.5]{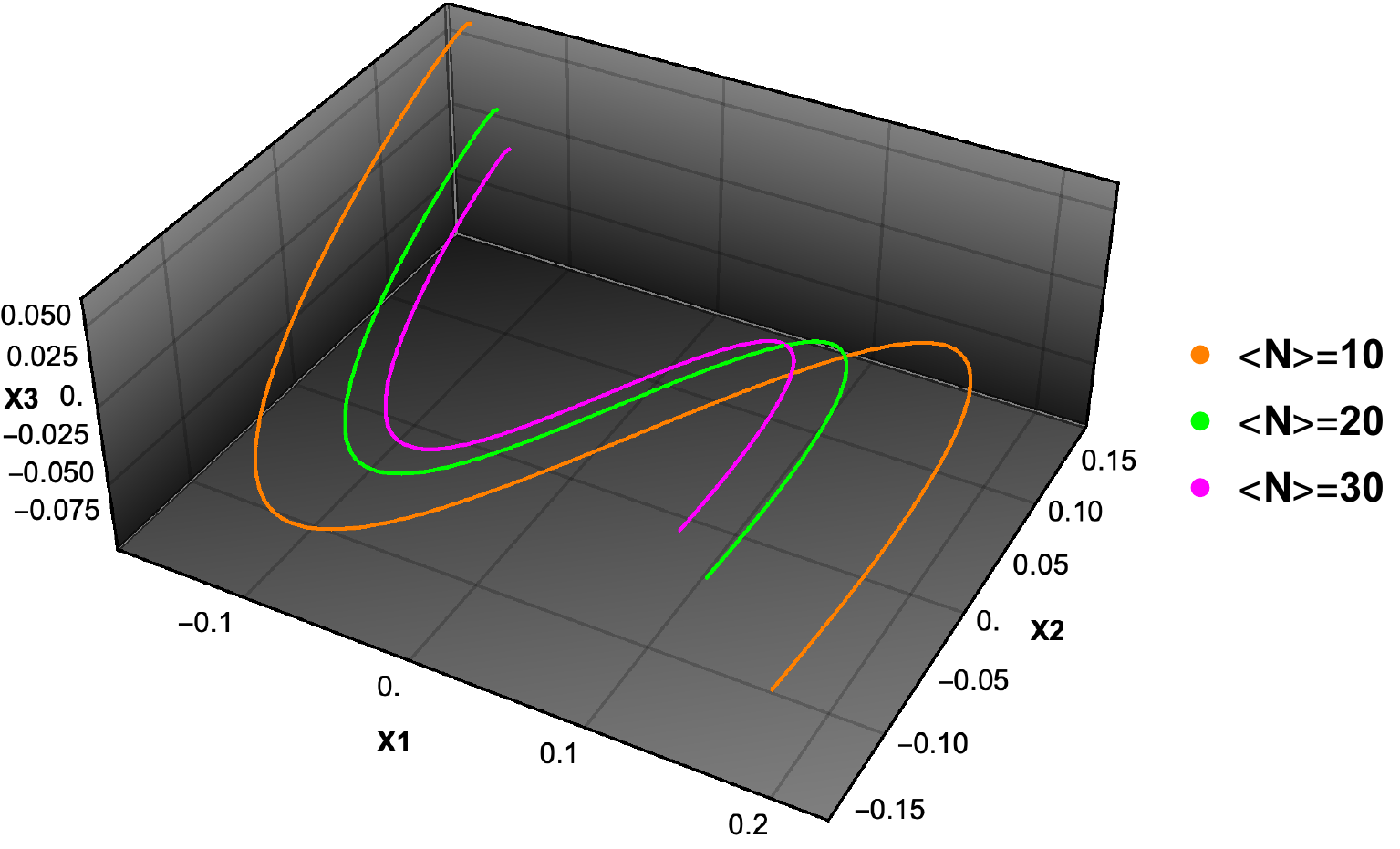} \includegraphics[scale=0.5]{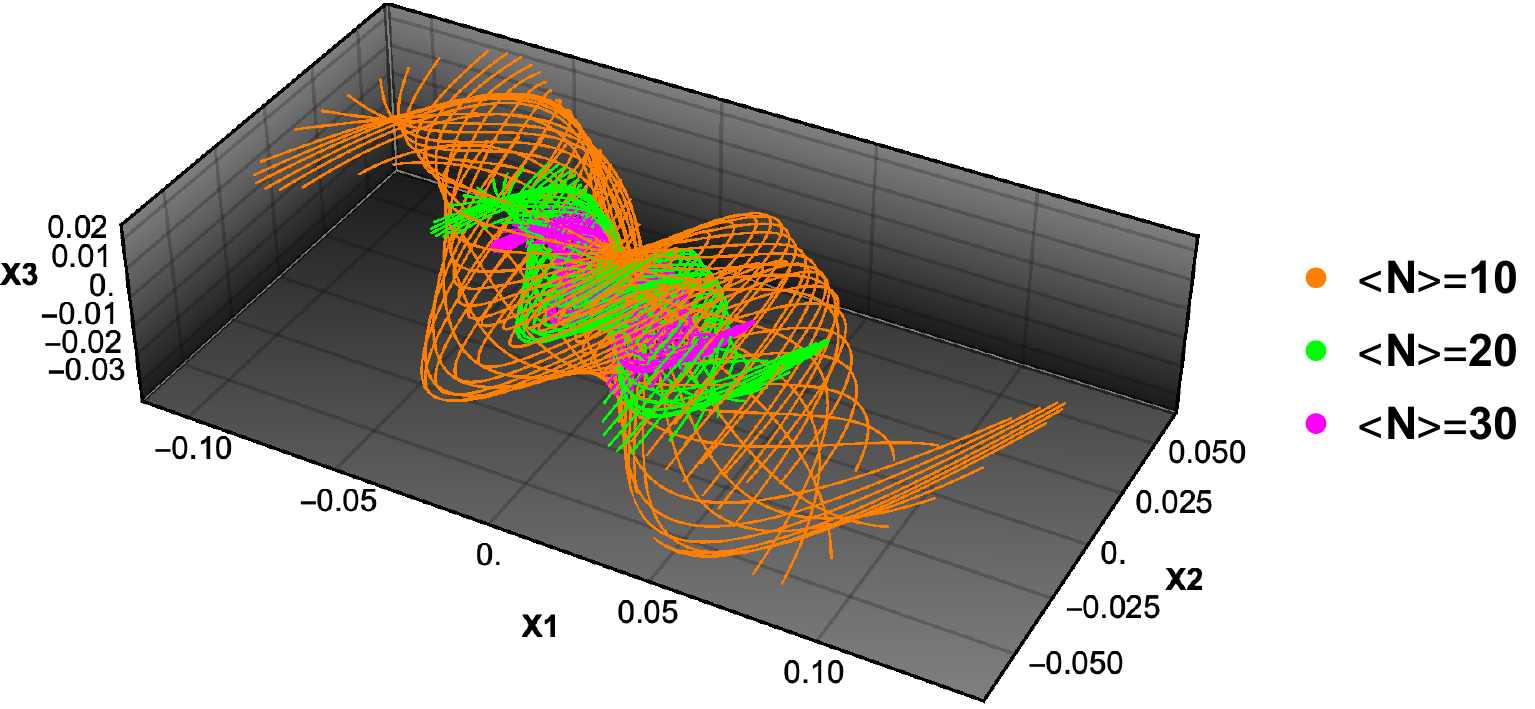} 
\caption{Open string profiles with 5 harmonics at different average mass and their time evolution}
\end{figure}

\begin{figure}[!h]
\center
\includegraphics[scale=0.5]{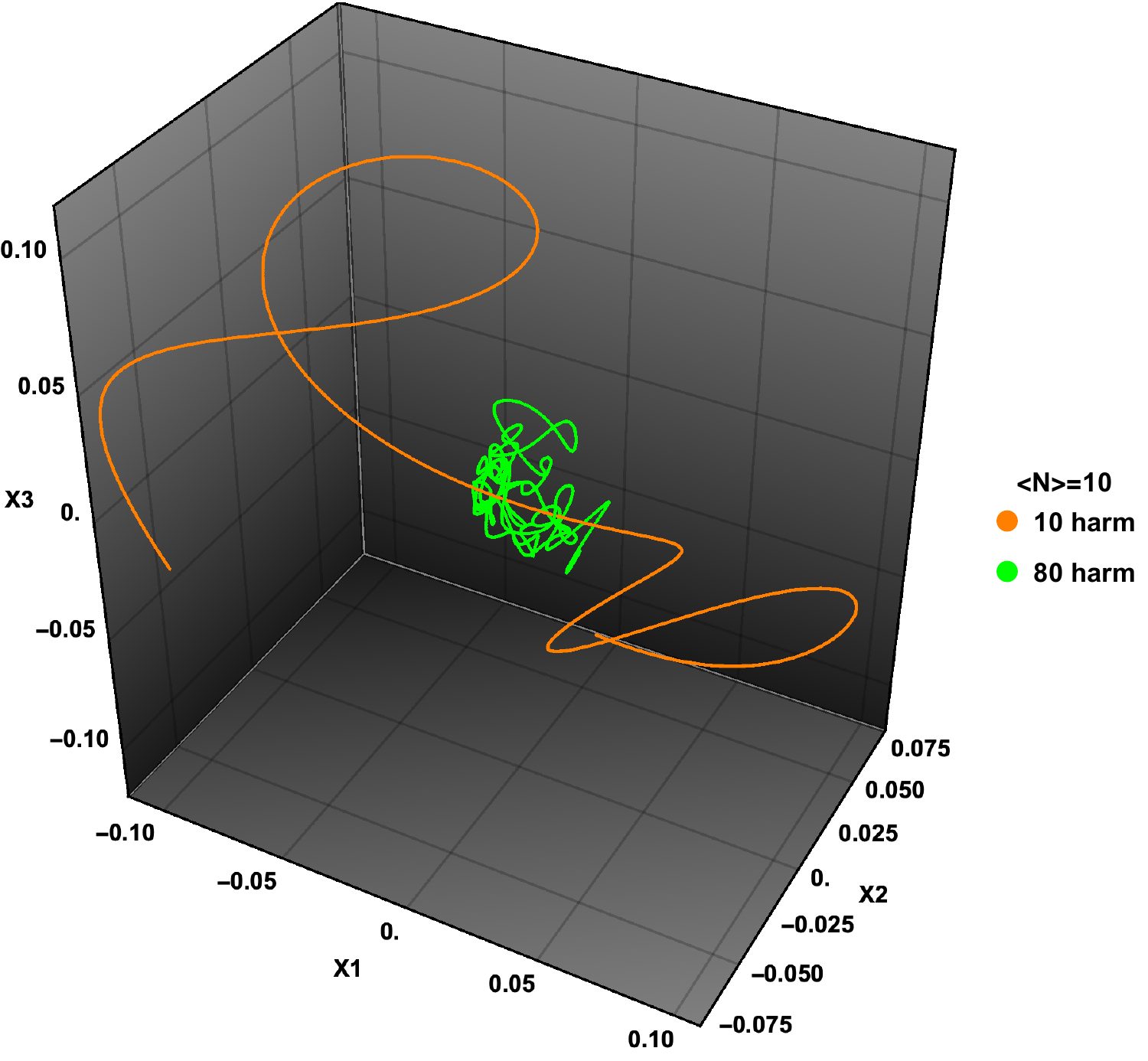} \includegraphics[scale=0.5]{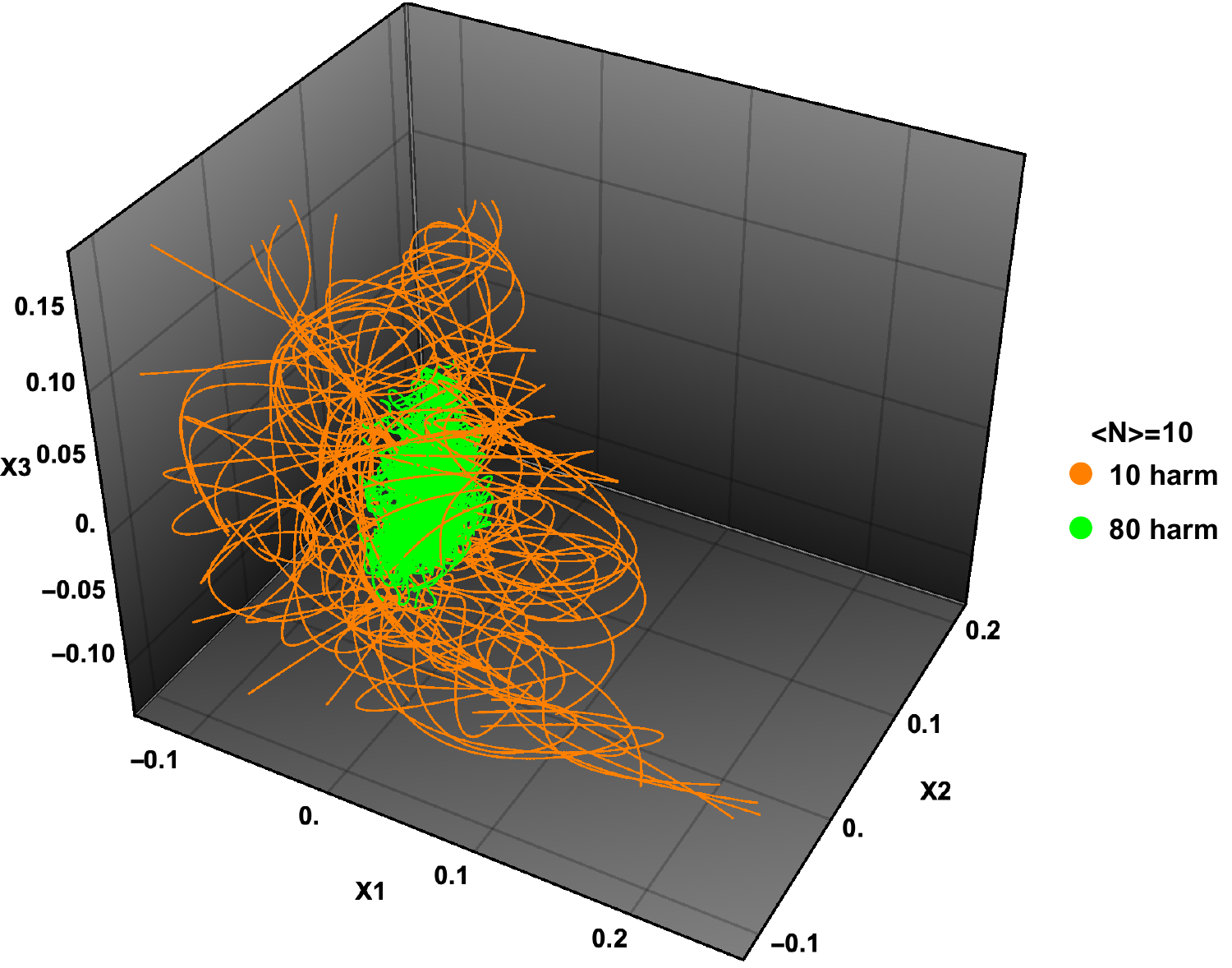} 
\caption{Open string profiles with the same average mass at different number (ie 10, 80) of harmonics  and their time evolution}
\end{figure}

Note that in the chosen frame, the only non-zero component of $q$ is $q^-$. As a result the only component of $P^\mu$ that fluctuates is $P^-$ with $\langle P^-\rangle = p^- - \langle N\rangle q^-$. 
Since all the remaining components, including $P^+$ are `classical' in that $\langle P^\mu\rangle = p^\mu$, the mass (and thus $P^2$) does not fluctuate {\it viz.}
\begin{equation}\langle {\cal M}^2\rangle = - \langle P_\mu P^\mu\rangle = \langle 2P^+P^- - P^iP_i\rangle = 2p^+\langle P^-\rangle - p_ip^i = 2p^+(p^- - \langle N\rangle q^-) - p_ip^i = 
- \langle P_\mu\rangle \langle P^\mu\rangle
\end{equation}

One can proceed similarly and compute the average value of the angular momentum $J^{\mu\nu} = 
L^{\mu\nu}+S^{\mu\nu}$. Barring the `orbital' part $L^{\mu\nu}$, let us focus on $\langle S^{\mu\nu}\rangle$. The `transverse' components are given by
\begin{equation}
\langle S^{ij}\rangle = \sum_{n>0} {2\over n} Im(\lambda^{i*}_n \lambda^j_n)
\end{equation}
After some tedious algebra one can also compute the $S^{i-}$ components ($\alpha'=1/2$)
\begin{equation}
\langle S^{i-}\rangle = {1\over p^+} \sum_{\ell>0} \sum_{m=-\infty}^{+\infty} {1\over \ell} Im(\lambda^{i*}_\ell \lambda_m{\cdot}\lambda_{\ell - m})
\end{equation}
The other components involving light-cone directions are computed along the same lines with some more effort.

\subsection{Level expansion, Chan-Paton factors, `twist' symmetry} 

Although we will mostly consider coherent states of open strings terminating on a single D-brane $N_{25}=1$, it is worth mentioning that most of the results remain valid for $N_{25}>1$, provided the correct Chan-Paton (CP) factors are included. 

For oriented strings the gauge group is $U(N_c)$ and the relevant CP factor for a `color-ordered'$n$-point amplitude ${\cal A}(1,2,..., n)$ on the disk is  $Tr(T_{1} T_{2} ... T_{n})$ with $T_a$ a generator of $U(N_c)$ in the fundamental representation. The complete amplitude obtains after summing over inequivalent `color' orderings, {\it i.e.} over non-cyclic permutation of the $n$ external legs,  
\begin{equation}
\widehat{\cal A}_n(1,2,..., n) = \sum_{\rho\in S_n/Z_n}  {\cal A}_n({\rho(1)},{\rho(2)}, ... {\rho(n)})Tr_{N_c}(T_{\rho(1)} T_{\rho(2)} ... T_{\rho(n)})
\end{equation} 

Notice that under transposition, {\it i.e.} exchange of the two ends of the string, $T\rightarrow T^t=-T=-T^*=+T^\dagger$ (for the $SU(N_c)$ generators) while $T_0=1\rightarrow T_0^t=T_0=T_0^*=T_0^\dagger=1$ (for the $U(1)$ generator) and 
\begin{eqnarray}
&&Tr(T_{\rho(1)} T_{\rho(2)} ... T_{\rho(n)})=Tr(T_{\rho(1)} T_{\rho(2)} ... T_{\rho(n)})^t = 
Tr(T^t_{\rho(n)} ... T^t_{\rho(2)}T^t_{\rho(1)})\underset{\Omega}{\rightarrow} \nonumber \\
&&Tr(T^t_{\rho(1)} T^t_{\rho(2)} ... T^t_{\rho(n)}) = (-)^{n-n_0} Tr(T_{\rho(1)} T_{\rho(2)} ... T_{\rho(n)}) = (-)^{n-n_0} Tr(T_{\rho(1)} T_{\rho(2)} ... T_{\rho(n)})
\end{eqnarray}
This has to be combined with the sign the colour-ordered amplitude for external states at level $N_i$
obtains 
\begin{equation}
{\cal A}(n,n{-}1,... 2,1)= (-)^{\sum_{i=1}^n  N_i} {\cal A}(1,2,..., n)
\end{equation}
For coherent states each level has its own symmetry and a full amplitude will involve contributions with different symmetry.

For un-oriented strings the gauge group includes $O(N_c)$ or $Usp(2N_c)$ factors and one can take for states at level $N$
$T^t = (-)^{N+\Omega_0} T$ where $\Omega_0=+1$ for $O(N_c)$ and $\Omega_0=-1$ for $Usp(2N_c)$ denote the eigen-value of the tachyonic ground state under the exchange of its ends, as implemented by world-sheet parity $\Omega$, known as `twist symmetry' in the early days.
 
When open strings end on different D-branes the situation is more intricate, boundary conditions and mode expansion get modified and one has to sum over all possible (allowed) inequivalent  orderings. 


\section{3-point open string amplitudes with Coherent states}
 
In this Section, we compute 3-point `amplitudes'  on the disk (tree level)  involving diverse numbers of open string coherent vertex operators (\ref{CohVertOpFin}), tachyons (\ref{tachVert}) and vector bosons (\ref{VectVert}). Quite remarkably but not unexpectedly, we find that the amplitudes exponentiate. 

We will not write down any CP factor and focus on a given ordering of the vertex operators inserted on the boundary of the disk. Assuming the same CP factor for all the excited levels in the coherent states and summing over the two inequivalent orderings, one finds $Tr(T_aT_bT_c)+(-)^{N_a+N_b+N_c} Tr(T_aT_cT_b)$ that produces the structure constants ($if_{abc}$) when $\sum_iN_i$ is odd or the `anomaly' coefficients ($d_{abc}$) when $\sum_iN_i$ is even. We will also drop the overall dependence on $g_s$ and the normalisation of the vertex operators that have been carefully determined in \cite{Skliros:2009cs, Hindmarsh:2010if, Skliros:2011si, SklirosOAC}. 

\subsection{${\mathcal{A}}_3({\mathcal{C}}, T,T)$}
The first and simplest computation is the 3-point amplitude with one coherent vertex and two tachyons.

The amplitude (that coincides with the 3-point correlator of `unintegrated' $cV$ type vertex operators) reads
\begin{eqnarray}
&&{\mathcal{A}}_3({\mathcal{C}}, T,T)= \langle c{{V}}_{\mathcal{C}}(z_1) \,cV_T(z_2)\,cV_T(z_3) \rangle= \langle c(z_1) c(z_2) c(z_3)\rangle  \times \nonumber \\
&&\quad \Big\langle \exp\Bigg\{\sum_{r,s=1}^{\infty} \dfrac{\zeta_{r}{{\cdot} } \zeta_s }{2rs}\, \mathcal{S}_{r,s} \,e^{-i(r+s)q{\cdot}  X} + \sum_{n=1}^{\infty} \dfrac{\zeta_n}{n}{\cdot}  \mathcal{P}_{n} \,e^{-inq{\cdot}  X}  \Bigg\} \,e^{ip{\cdot}X}(z_1) e^{ip_2{\cdot}  X}(z_2) e^{ip_3{\cdot}  X}(z_3)\Big\rangle
\end{eqnarray}
Contracting the $c$ ghosts yields
\begin{equation}
\langle c(z_1) c(z_2) c(z_3)\rangle = z_{12}z_{13}z_{23} 
\end{equation} 
In order to perform the other contractions it is crucial to recall that $e^{-i n q{\cdot}X} = e^{i n q^-X^+} $ and $\mathcal{S}_{m,n}$, that depend only 
on $q{\cdot}\pa^sX= -q^-\pa^sX^+$, can only contract with the `longitudinal' part of $e^{i p{\cdot}X}= e^{-i p^+X^-+ ...}$, while $\mathcal{P}_{n}$ that depend on both $q^-\pa^sX^+$ and $\zeta_n{\cdot}\pa^rX$ can also contract with the `transverse' part of $e^{i p{\cdot}X}= e^{... + i p_iX^i}$.

To start with we assume $\zeta_n{\cdot}\zeta_m =0$, a condition we will relax later on. In this case we have

\begin{equation}
\Big\langle \prod_{n=1}^{\infty} \sum_{g_n=0}^{\infty}\dfrac{1}{g_n!}\Bigg\{ \sum_{h=1}^{n} \dfrac{i}{n(h-1)!} {\cal Z}_{n-h}[{{\mathcal{U}}_{s}^{(n)}}] \,{\zeta_n}{\cdot}\pa^hX e^{-inqX}\Bigg\}^{g_n} e^{ip{\cdot}X}(z_1) e^{ip_2{\cdot}  X}(z_2) e^{ip_3{\cdot}  X}(z_3) \Big\rangle 
\end{equation}

For $g_n=0$ contracting the exponentials yields the Koba-Nielsen (KN) factor 
\begin{equation}
\large\langle e^{ip{\cdot}X}(z_1) e^{ip_2{\cdot}  X}(z_2) e^{ip_3{\cdot}  X}(z_3)\large\rangle = z_{12}^{p_1p_2}z_{13}^{p_1p_3}z_{23}^{p_2p_3}
\end{equation}
that for $g_n\neq 0$ combines with the contractions of $X^+$ in $e^{-i ng_n q{\cdot}X}$ that yield $z_{12}^{-ng_n qp_2}z_{13}^{-ng_n qp_3}$.  At fixed level $N=\sum_n ng_n$, after integration over the center-of-mass (CM) coordinates $X_0^\mu$, one finds 
\begin{equation}
z_{12}^{(p-Nq)p_2}z_{13}^{(p-Nq)p_3} z_{23}^{p_2p_3} \delta(p-Nq+p_2+p_3) = z_{12}^{N-1}z_{13}^{N-1} z_{23}^{-N-1} \delta(p-Nq+p_2+p_3)
\end{equation}
Contracting $i q^-\pa^sX^+$ in ${{\mathcal{U}}_{s}^{(n)}}$ yields
\begin{equation}
\begin{split}
\label{UhatToU}
 {{\mathcal{U}}_{s}^{(n)}} \rightarrow (-)^s \dfrac{n}{2} \Bigg( \dfrac{Q-1}{z_{12}^s} - \dfrac{Q+1}{z_{13}^{s}} \Bigg)
 \end{split}
\end{equation}
where 
\begin{equation}
\label{defQ}
Q=q{{\cdot} } (p_2 -p_3)
\end{equation}
while $i{\zeta_n}{\cdot}\pa^hX$ yields instead
\begin{equation}
i{\zeta_n}{\cdot}\pa^hX \rightarrow {(-)\over 2}^{h{-}1}(h-1)! 
\Bigg(\dfrac{1}{z_{12}^h} -  \dfrac{1}{z_{13}^h}\Bigg){\zeta_n}{\cdot}(p_2-p_3)
\end{equation}

 Focusing on the sum over $h$, factoring out ${\zeta_{n}}{\cdot}  p_{23}z_{12}^{-ng_nqp_2}z_{13}^{-ng_nqp_3}$ and using properties (\ref{Zproperties}) of the cycle index polynomials one finds 
 \begin{equation}
 \begin{split}
 &\sum_{h=1}^{n} {(-)^{h-1}\over 2n} \, {\cal Z}_{n-h}[{\mathcal{U}_{s}^{(n)}}] \Bigg( \dfrac{z_{13}^h -z_{12}^h}{z_{12}^h z_{13}^h} \Bigg)= \dfrac{(-)^{n+1} }{2\,n!} \dfrac{\Gamma[\frac{n}{2}(Q+1)]}{\Gamma[\frac{n}{2}(Q-1)+1]} \left(z_{23} \over z_{12} z_{13}\right)^n   \end{split}
 \end{equation}
that can be proven setting $z_1=0, z_2=1$ and $z_3=\infty$, by exploiting the BRST invariance of the construction, so much so that the complete amplitude reads
\begin{equation}
\begin{split}
&\sum_{N=\Sigma n g_n} \left[\prod_{n=1}^{\infty} \sum_{g_n=0}^{\infty}\dfrac{1}{g_n!} \Bigg\{ {{\cal R}}_{n-1}(Q)\, \zeta_{n}{\cdot}  (p_2-p_3) \left(z_{23} \over z_{12} z_{13}\right)^n \Bigg\}^{g_n}\right]_N z_{12}^{N}z_{13}^{N} z_{23}^{-N}\delta^{(D)}(p-N q +p_2+p_3)= \\
&=\int {d^{D}X_0\over(2\pi)^D}\, e^{i\, (p + p_2 + p_3){{\cdot} } X_0  }  \prod_{n=1}^{\infty}  \sum_{g_n=0}^{\infty}\dfrac{1}{g_n!} \Bigg\{{{\cal R}}_{n-1}(Q) \,   e^{-i\,n\,q{\cdot}  X_0}\,{\zeta}_{n}{{\cdot} }(p_2-p_3)  \Bigg\}^{g_n}
\end{split}
\end{equation}
where  
\begin{equation}\label{Rpoly}
\boxed{{{\cal R}}_{n-1}(Q) = 
\dfrac{(-)^{n+1} }{2\,n!} \dfrac{\Gamma[\frac{n}{2}(Q+1)]}{\Gamma[\frac{n}{2}(Q-1)+1]} }  
\end{equation}
with $Q$ defined in (\ref{defQ}) and we have used $\int{d^{D}X_0} e^{i\, (p - Nq + p_2 + p_3)} = (2\pi)^D\delta(p -Nq+ p_2 + p_3)$. Setting
\begin{equation}\label{zetaHat}
\widehat{\zeta}^\mu_n= e^{-i n q{{\cdot} }X_0} \,\zeta^\mu_n
\end{equation}
for $\zeta_n{{\cdot} }\zeta_m=0$, one finds the very compact and elegant expression
\begin{equation}
{\cal A}_3({\mathcal{C}}(p,q;\zeta_n), T(p_2), T(p_3)) =  \int {{d^{D}X_0}}\, e^{i\, (p + p_2 + p_3){{\cdot} } X_0  }\exp\sum_{n=1}^{\infty} {{\cal R}}_{n-1}(Q) \,\widehat\zeta_{n}{{\cdot} } (p_2-p_3)
\end{equation}
Quite remarkably the expression has exponentiated in a rather subtle and interesting way. Although this was to be expected thanks to the exponential form of the vertex operators for coherent states in the DDF basis, exposing this property required a decomposition in the level number $N$ as well as the redefinition of the harmonics as in (\ref{zetaHat}). 

In order to extend the amplitude to the case in which $\zeta_m{{\cdot} } \zeta_n \neq 0$, there is an additional factor to consider:
\begin{eqnarray}
&&\Big\langle \prod_{n=1}^{\infty} \sum_{g_n=0}^{\infty}\dfrac{1}{g_n!}\Bigg\{ e^{-inqX} {\zeta_n}{\cdot}\sum_{h=1}^{n} \dfrac{i}{n(h-1)!} {\cal Z}_{n-h}[{{\mathcal{U}}_{s}^{(n)}}] \,\pa^hX \Bigg\}^{g_n}  \\
&&\quad \prod_{r,s=1}^{\infty} \sum_{h_{rs}=0}^{\infty}\dfrac{1}{h_{rs}!}\left\{ {{\zeta_r}{\cdot}\zeta_s \over 2rs}e^{-i(r+s)qX} \sum_{f=1}^{s} f {\cal Z}_{r+f}[{{\mathcal{U}}_{\ell}^{(r)}}] \, {\cal Z}_{s-f}[{{\mathcal{U}}_{\ell}^{(s)}}]\right\}^{h_{rs}}
e^{ip{\cdot}X}(z_1) e^{ip_2{\cdot}  X}(z_2) e^{ip_3{\cdot}  X}(z_3) \Big\rangle \nonumber
\end{eqnarray}

Contracting $i q^-\pa^\ell X^+$ in ${{\mathcal{U}}_{\ell}^{(r/s)}}$ yields ${{\mathcal{U}}_{\ell}^{(r/s)}}$ in (\ref{UhatToU}).  Using $SL(2,R)$ invariance to set $z_1=0, z_2=1$ and $z_3=\infty$ and barring the common factor $e^{-i(r+s)qX} {\zeta_r}{\cdot}{\zeta_s}/2rs$, for given $r$ and $s$ one finds 
\begin{equation}
\sum_{f=1}^{s} f\, {\cal Z}_{r+f}[\mathcal{U}_\ell^{(r)}]\, {\cal Z}_{s-f}[\mathcal{U}_\ell^{(s)}] \, \Bigg( \dfrac{z_{23}}{z_{12} z_{13}}\Bigg)^{r+s} = \dfrac{(rs)^2 (Q^2-1)  }{(r+s)}{{\cal R}}_{r-1}(Q) {{\cal R}}_{s-1}(Q) 
\end{equation}
\begin{equation}
=  \dfrac{rs(-)^{r+s}}{(r+s)} \dfrac{(Q+1)}{2(r-1)!} \dfrac{(Q-1)}{2(s-1)!} \dfrac{\Gamma\Big[ \frac{r}{2}(Q+1) \Big] \, \Gamma\Big[ \frac{s}{2}(Q+1) \Big]}{\Gamma\Big[ \frac{r}{2}(Q-1) + 1 \Big] \, \Gamma\Big[ \frac{s}{2}(Q-1) +1 \Big]}
\end{equation}

These terms combine with the ones from the first exponential to shift the level from $N=\sum_n ng_n$ to $N=\sum_n ng_n + \sum_{r,s} (r+s)h_{rs}$. As before this can be taken track of by switching from $\zeta$ to $\widehat{\zeta}$. Taking in due account the kinematic constraints
\begin{equation}
k_1{{\cdot} }k_2=k_1{{\cdot} }k_3=ng+ (m+h)l-1\,;\quad k_2{{\cdot} }k_3=1-ng-(m+h)l
\end{equation}
the final and most general (in that $\zeta_m{{\cdot} } \zeta_n \neq 0$) yet quite compact result reads
\begin{equation}\label{A3CTT}
\boxed{\begin{split}
&{\cal A}_3({\mathcal{C}}(p,q;\zeta_n), T(p_2), T(p_3)) =\\
&\int {{d^D X_0}} \,e^{i(p{+}p_2{+}p_3){\cdot}  X_0} \,\exp\Bigg\{ \sum_{n=1}^{\infty} {{\cal R}}_{n{-}1}(Q)\widehat{\zeta}_{n}{{\cdot} } (p_2{-}p_3)+\sum_{r,s}^{1,\infty}\widehat{\zeta}_r{\cdot}  \widehat{\zeta}_s\dfrac{rs(Q^2{-}1)}{2(r{+}s)}{{\cal R}}_{r{-}1}(Q) {{\cal R}}_{s{-}1}(Q)\Bigg\}
\end{split}}
\end{equation}
where $Q$, ${{\cal R}}_{n-1}(Q)$ and $\widehat{\zeta}_m$ are defined in (\ref{defQ}), (\ref{Rpoly}) and (\ref{zetaHat}), respectively. 
Recall that this is a color-ordered amplitude and should be combined with the corresponding CP factors and summed over the two non-cyclic permutations of the external legs. (\ref{A3CTT}) determines the decay rate of a generic coherent states into a tachyon pair and encodes the tri-linear couplings of an arbitrary excited open string state to two tachyons. The latter can be extracted taking derivatives w.r.t. $\zeta_n$.

\subsection{${\mathcal{A}}_3({\mathcal{C}},V,T)$}

It is quite interesting to replace one or both tachyons with massless vector bosons. Let us start with one. Without losing generality, it is convenient to decompose the null momentum $k$ of the vector as $k=p'-q'$ with $(p')^2=2$ and $p'q'=1$ and $(q')^2=0$ with $q'$ along the same light ray as the $q$ used to express the momenta of the coherent state, so much so that $qq'=0$. As a result the polarisation ${\cal A}_\mu = {{\varepsilon}}_i (\delta^i_\mu - (p')^i q'_\mu)$ turns out to be transverse to both $q'\sim q$ and $p'$ and thus to their combination $k=p'-q'$.

The relevant correlation function is
\begin{eqnarray}
&&{\mathcal{A}}_3({\mathcal{C}},V,T) = \braket{c{{V}}_{\mathcal{C}}(z_1) \, cV_A(z_2)\, cV_T(z_3)} = 
\braket{c(z_1) \, c(z_2)\, c(z_3)} \times \\
&&\Big\langle \exp\Bigg\{\sum_{r,s=1}^{\infty} \dfrac{\zeta_{r}{{\cdot} } \zeta_s }{2rs}\, \mathcal{S}_{r,s} \,e^{-i(r+s)q{\cdot}  X} + \sum_{n=1}^{\infty} \dfrac{\zeta_n}{n}{\cdot}  \mathcal{P}_{n} \,e^{-inq{\cdot}  X}  \Bigg\} \,e^{ip{\cdot}X}(z_1)iA{\cdot}\pa{X}e^{ik_2{\cdot}  X}(z_2) e^{ip_3{\cdot}  X}(z_3) \Big\rangle \nonumber
\end{eqnarray}

With respect to the previous case ${\zeta_n}{\cdot}{\cal P}_n(z_1)$ can also contract with $A{\cdot}\pa{X}(z_2)$ and the latter can contract with  $e^{ip{\cdot}X}(z_1)$ and $e^{ip_3{\cdot}  X}(z_3)$. 

The first new kind of contractions produces terms $\zeta_n{{\cdot} }A$ of the form
\begin{equation}\label{EQQQ}
\sum_{h=1}^{n} \dfrac{(-)}{n}^{h-1} \dfrac{h}{z_{12}^{h+1}}\, {\cal Z}_{n{-}h}[\mathcal{U}_s^{(n)}] \,\Bigg( \dfrac{z_{12}z_{13}}{z_{23}} \Bigg)^{n} \,\, \Bigg(\dfrac{z_{12}z_{23}}{z_{13}}\Bigg) = 
\dfrac{(-)}{n!}^{n+1}\, \dfrac{\Gamma[\frac{n}{2}(1+Q)+1]}{\Gamma[2-\frac{n}{2}(1-Q)]} 
\end{equation}
where it is convenient to introduce the degree $n{-}1$ polynomials  
\begin{equation}\label{Mpoly}
\boxed{ {{\cal M}}_{n-1}(Q)
 = \dfrac{(-)^{n+1}}{n!}\, \dfrac{\Gamma[\frac{n}{2}(1+Q)+1]}{\Gamma[2-\frac{n}{2}(1-Q)]}}
\end{equation}
 
 The second new kind of contractions yields terms $A{{\cdot} }p_{1/3}$ of the form 
\begin{equation}\label{QQQE}
 {A{{\cdot} }p_{1}\over z_{21}} + {A{{\cdot} }p_{3}\over z_{23}} = A{{\cdot} }(p_{1}-p_3)  {z_{13} \over  2 z_{21}z_{23}}
\end{equation}

So the final quite compact result reads
\begin{equation}
\boxed{\begin{split}
&{\mathcal{A}}_3({\mathcal{C}}(p_1,q;\zeta_n),V(A, k_2),T(p_3)) =\int {{d^D X_0}} \,e^{i(p_1+k_2+p_3){\cdot}  X_0} \,\Bigg[ A{\cdot}  p_3 + \sum_{m=1}^{\infty}{{\cal M}}_{m-1}(Q) \, \widehat\zeta_m{{\cdot} }A \, \Bigg] \times  \\
&\qquad \exp\Bigg\{ \sum_{n=1}^{\infty} {{\cal R}}_{n-1}(Q)\,   \widehat{\zeta}_{n}{{\cdot} }(k_2-p_3)\,+\, \sum_{r,s}^{1,\infty}  \widehat{\zeta}_r{\cdot}  \widehat{\zeta}_s\,   \dfrac{rs (Q^2-1) }{2(r+s)}{{\cal R}}_{r-1}(Q) {{\cal R}}_{s-1}(Q)\Bigg\}
\end{split}}
\end{equation}
where $Q$, ${{\cal R}}_{n-1}(Q)$, $\widehat{\zeta}_m$ and ${{\cal M}}_{n-1}(Q)$ are defined in (\ref{defQ}), (\ref{Rpoly}), (\ref{zetaHat}), and (\ref{Mpoly}), respectively. Once again, the result has an exponential form up to the linear terms in the polarisation of the vector boson. 

\subsection{${\mathcal{A}}_3({\mathcal{C}},V,V)$}
When both tachyons are replaced by massless vector bosons, the relevant correlation function is
\begin{eqnarray}
&&{\mathcal{A}}_3({\mathcal{C}},V,V) = \braket{c{{V}}_{\mathcal{C}}(z_1) \, cV_A(z_2)\, cV_A(z_3)} = 
\braket{c(z_1) \, c(z_2)\, c(z_3)} \times \\
&&\Big\langle \exp\Bigg\{\sum_{r,s=1}^{\infty} \dfrac{\zeta_{r}{{\cdot} } \zeta_s }{2rs}\, \mathcal{S}_{r,s} \,e^{-i(r+s)q{\cdot}  X} + \sum_{n=1}^{\infty} \dfrac{\zeta_n}{n}{\cdot}  \mathcal{P}_{n} \,e^{-inq{\cdot}  X}  \Bigg\} \,e^{ip{\cdot}X}(z_1) iA_2{\cdot}\pa{X}e^{ik_2{\cdot}  X}(z_2) iA_3{\cdot}\pa{X}e^{ik_3{\cdot}  X}(z_3)  \Big\rangle \nonumber
\end{eqnarray}
Including a second vector boson, one has to take into account the additional contractions of $iA_2{\cdot}\pa{X}$ with $iA_3{\cdot}\pa{X}$, producing ${A_2{\cdot}A_3/z_{23}^2}$,
as well as the separate contractions of both of them with ${\zeta_n}{\cdot}  \mathcal{P}_{n}$ in the exponent, producing terms of the same form as the ones in (\ref{EQQQ}), and with $e^{ipX}$ producing terms of the same form as the ones in (\ref{QQQE}).

Combining the various contributions the amplitude takes the final form 
\begin{equation}
\boxed{\begin{split}
&{\mathcal{A}}_3({\mathcal{C}}(p_1,q;\zeta_n),V(A_2, k_2),V(A_3, k_2)) =\int {{d^D X_0}}  
\Bigg\{\sum_{n=1}^{\infty} (\widehat{\zeta}_n{{\cdot} }A_2) {{\cal M}}_{n-1}(Q) \sum_{m=1}^{\infty} (\widehat{\zeta}_m{{\cdot} }A_3){{\cal M}}_{m-1}(Q) \\
&+ (A_2{{\cdot} }A_3) {-} (A_2{{\cdot} }k_3)(A_3{{\cdot} }k_2) {+} \sum_{n=1}^{\infty} [(A_2{{\cdot} }k_3) (\widehat{\zeta}_n{{\cdot} }A_3) - (A_3{{\cdot} }k_2)(\widehat{\zeta}_n{{\cdot} }A_2)] {{\cal M}}_{n-1}(Q)\Bigg\} \,e^{i(p+k_2+k_3){\cdot}  X_0} \\
& \times \exp\Bigg\{ \sum_{n=1}^{\infty} {{\cal R}}_{n-1}(Q)\,   \widehat{\zeta}_{n}{{\cdot} }(k_2-k_3) + \sum_{r,s}^{1,\infty}  \widehat{\zeta}_r{\cdot}  \widehat{\zeta}_s \dfrac{rs (Q^2-1) }{2(r+s)}{{\cal R}}_{r-1}(Q) {{\cal R}}_{s-1}(Q)\Bigg\} 
\end{split}}
\end{equation}
where $Q$, ${{\cal R}}_{n-1}(Q)$, $\widehat{\zeta}_m$ and ${{\cal M}}_{n-1}(Q)$ are defined in (\ref{defQ}), (\ref{Rpoly}), (\ref{zetaHat}), and (\ref{Mpoly}), respectively.
As a first almost trivial check, setting $\zeta_n=0$ and noticing that $2\alpha' k_2 k_3 = -\alpha' M^2_T = +1$ one finds
\begin{equation}
{\mathcal{A}}_3(T_1 \,V_2\,V_3)= (A_2{{\cdot} }A_3) {-} (A_2{{\cdot} }k_3)(A_3{{\cdot} }k_2) = k_2{\cdot}k_3 (A_2{{\cdot} }A_3) {-} (A_2{{\cdot} }k_3)(A_3{{\cdot} }k_2)
\end{equation}
which is the correct tachyon-to-two-photon amplitude, which survives even for a single D25-brane, since it is symmetric under the exchange of the two photons.

Another less trivial check follows from setting $n=1$ and $g=1$ (having in mind formula (2.47)),
 one has
\begin{equation}
\begin{split}
&{\mathcal{A}}_3(V_1\,V_2\,V_3)=(A_2{{\cdot} }k_3)(A_3{{\cdot} }p_1)(\zeta_1{{\cdot} }k_2) + (A_2{{\cdot} }A_3)(\zeta_1{{\cdot} }k_2)+(A_3{{\cdot} }p_1)(\zeta_1{{\cdot} }A_2)+(A_2{{\cdot} }k_3)(\zeta_1{{\cdot} }A_3) \\
&\qquad=(A_2{{\cdot} }\dfrac{k_{31}}{2})(A_3{{\cdot} }\dfrac{k_{12}}{2})(\zeta_1{{\cdot} }\dfrac{k_{23}}{2}) + (\zeta_1{{\cdot} }\dfrac{k_{23}}{2}) (A_2{{\cdot} }A_3)+(A_3{{\cdot} }\dfrac{k_{12}}{2})(\zeta_1{{\cdot} }A_2)+(A_2{{\cdot} }\dfrac{k_{31}}{2})(\zeta_1{{\cdot} }A_3)
\end{split}
\end{equation}
where $k_{ij}=k_i-k_j$, which is -- not surprisingly -- the (totally anti-symmetric!) 3-vector boson vertex in the bosonic string, that requires multiple D25-branes to be non-zero. In particular one can recognise a higher-derivative $F^3$ interaction (the very first term), suppressed by powers of $\alpha'$, in addition to the Yang-Mills vertex (the three last terms).

\subsection{${\mathcal{A}}_3({\mathcal{C}}, {\mathcal{C}}, T)$}

A more laborious amplitude to tackle is the one with two coherent states and a tachyon. 

The relevant correlation function is
\begin{eqnarray}
&&{\mathcal{A}}_3({\mathcal{C}},{\mathcal{C}},T) = \braket{c{{V}}_{\mathcal{C}}(z_1) \, cV_{\mathcal{C}}(z_2)\, cV_T(z_3)} = 
\braket{c(z_1) \, c(z_2)\, c(z_3)}\times \nonumber \\ &&\quad \Big\langle \exp\Bigg\{\sum_{r_1,s_1=1}^{\infty} \dfrac{\zeta_{r_1}^{(1)}{{\cdot} } \zeta_{s_1}^{(1)} }{2r_1s_1}\, \mathcal{S}_{r_1,s_1} \,e^{-i(r_1+s_1)q_1{\cdot}  X} + \sum_{n_1=1}^{\infty} \dfrac{\zeta^{(1)}_{n_1}}{n_1}{\cdot}  \mathcal{P}_{n_1} \,e^{-in_1q_1{\cdot}  X}  \Bigg\} \,e^{ip_1{\cdot}X}(z_1) \\
&&\qquad\exp\Bigg\{\sum_{r_2,s_2=1}^{\infty} \dfrac{\zeta_{r_2}^{(2)}{{\cdot} } \zeta_{s_2}^{(2)} }{2r_2s_2}\, \mathcal{S}_{r_2,s_2} \,e^{-i(r_2+s_2)q_2{\cdot}  X} + \sum_{n_2=1}^{\infty} \dfrac{\zeta^{(2)}_{n_2}}{n_2}{\cdot}  \mathcal{P}_{n_2} \,e^{-in_2q_2{\cdot}  X}  \Bigg\} \,e^{ip_2{\cdot}X}(z_2) 
e^{ip_3{\cdot}  X}(z_3)  \Big\rangle \nonumber
\end{eqnarray}

Without loss of generality one can take $q_2 \sim q_1$ so that $q_1{\cdot}q_2=0$ \footnote{Indeed $q_i{\cdot}q_j=\omega_i \omega_j (1-\vec{n}_i{\cdot}\vec{n}_j) = 0$ iff $\vec{n}_i{\cdot}\vec{n}_j=1$ i.e. $\vec{n}_i =\vec{n}_j$ and $q_i = {\omega_i\over \omega_j} q_j $.}. This entails
$$\zeta_n^{(i)}{{\cdot} }q_{j}=0 \quad {\rm for} \quad i,j=1,2$$
thus preventing proliferation of contractions.

Anyway a new type of terms arises from the contractions of $\zeta^{(1)}_{n_1}{\cdot}\mathcal{P}_{n_1}$ with 
$\zeta^{(2)}_{n_2}{\cdot}\mathcal{P}_{n_2}$. For fixed $n_1$ and $n_2$  one finds 

$$\dfrac{{\zeta}_{n_{1}}^{(1)}{{\cdot} }{\zeta}_{n_{2}}^{(2)}}{n_{1}\,n_{2}} \Bigg(\dfrac{z_{12}z_{13}}{z_{23}} \Bigg)^{n_1}\Bigg(\dfrac{z_{12}z_{23}}{z_{13}} \Bigg)^{n_2}\sum_{{{u}}_{1}=1}^{n_{1}}\sum_{{{u}}_{2}=1}^{n_{2}} \dfrac{(-)^{{{u}}_{1}+{{u}}_{2}+1}}{z_{12}^{{{u}}_1+{{u}}_2}} \dfrac{\Gamma(1-{{u}}_{1}) {\cal Z}_{n_{1}{-}{{u}}_{1}}[{\cal U}_s^{(n_1)}]{\cal Z}_{n_{2}{-}{{u}}_{2}} [{\cal U}_s^{(n_2)}] }{\Gamma({{u}}_{2})\,\Gamma(1-{{u}}_{1}-{{u}}_{2})} 
$$
where 
\begin{equation}
 {\mathcal{U}}_{s}^{(n_1)}={(-)^s} \dfrac{n_{1}}{2}\Bigg(\dfrac{Q_1 - 1}{z_{12}^s} - \dfrac{Q_1 + 1}{z_{13}^s} \Bigg) \quad , \quad {\mathcal{U}}_{s}^{(n_2)} = {(-)^s} \dfrac{n_{2}}{2}\Bigg(\dfrac{(Q_2 - 1)(-)^s}{z_{12}^s} - \dfrac{Q_2 + 1}{z_{23}^s} \Bigg)  
 \end{equation}
with
\begin{equation}\label{defQ1Q2}
Q_1\equiv q_{1}{{\cdot} }(p_2{-}p_3) \quad , \quad Q_2\equiv q_{2}{{\cdot} }(p_1{-}p_3)
 \end{equation}
arises from the contractions of $q^-_i\pa^sX^+$ with the exponentials $\exp{ip_jX}$.

Barring the overall factor ${{\zeta}_{n_{1}}^{(1)}{{\cdot} }{\zeta}_{n_{2}}^{(2)}}$,
the above expression is a polynomial in $Q_1$ and $Q_2$ of degree $n_1{-}1$ and $n_2{-}1$, respectively, that is symmetric up to a sign $(-)^{n_1{+}n_2}$ under the simultaneous exchange of $n_1,Q_1$ with $n_2,Q_2$.
We will henceforth indicate the polynomial as ${\cal L}_{(n_{1}{-}1,n_{2}{-}1)}(Q_1,Q_2)$. For low values of $n_1$ and $n_2$ one finds
\begin{itemize}
\item $(1,1)$:  
${\cal L}_{(0,0)}= 1$
\item $(1,2)$:
${\cal L}_{(0,1)}= \frac{1}{2} (Q_2+1)$
\item$(2,2)$:
${\cal L}_{(1,1)} =
 -\frac{1}{4} (Q_2 Q_1+Q_1+Q_2+3)  $
 \item$(2,3)$:
${\cal L}_{(1,2)} = -\frac{1}{16} (Q_2+1) (3 Q_2 Q_1+Q_1+3 Q_2+9) $
\item$(3,3)$:
${\cal L}_{(2,2)} = \frac{1}{192} \left(27 Q_2^2 Q_1^2{+}36 Q_2 Q_1^2{+}9 Q_1^2{+}36 Q_2^2
   Q_1{+}144 Q_2 Q_1{+}108 Q_1{+}9 Q_2^2{+}108 Q_2{+}163\right) $
\end{itemize}

Thanks to $SL(2,\mathbb{R})$ invariance the result is independent of the choice of the $z$'s. In particular one can set $z_1=0$, $z_2=1$ and $z_{3}=\infty$ and the double sum produces
\begin{equation}\label{Lpoly}
\boxed{ {\cal L}_{(n_{1}{-}1,n_{2}{-}1)}(Q_1,Q_2) = {(-)^{n_1+1}\over n_{1}n_{2}} \sum_{{{u}}_{1}=1}^{n_{1}}\sum_{{{u}}_{2}=1}^{n_{2}} \dfrac{\Gamma({{u}}_{1}+{{u}}_{2})}{\Gamma({{u}}_{2})\,\Gamma({{u}}_{1})} {\cal Z}^{(1)}_{n_{1}{-}{{u}}_{1}}\Big[\dfrac{n_1}{2} (Q_1-1)  \Big] {\cal Z}^{(2)}_{n_{2}{-}{{u}}_{2}}\Big[ \dfrac{n_2}{2 }(Q_2-1) \Big] }
\end{equation}

Combining with by-now-familiar contractions, that already appeared in previous computations, yields{\small \begin{equation}
\boxed{ \begin{split}
&{\mathcal{A}}_3({\mathcal{C}}(p_1,q;\zeta^{(1)}_n),{\mathcal{C}}(p_2,q_2;\zeta^{(2)}_n),T(p_3)) = \int {{d^D X_0}}\, e^{i\,(p_1+p_2+p_3){{\cdot} }X_{0}} \exp\Bigg\{\sum_{n_{1}=1}^{\infty} \sum_{n_{2}=1}^{\infty}  \widehat{\zeta}_{n_{1}}^{(1)}{\cdot} \widehat{\zeta}_{n_{2}}^{(2)} {\cal L}_{n_1{-}1, n_2{-}1} (Q_1,Q_2) \Bigg\} \\
&\exp\Bigg\{ \sum_{r_1,s_1}^{1,\infty}  \widehat{\zeta}_{r_1}^{(1)}{\cdot}  \widehat{\zeta}_{s_1}^{(1)} \dfrac{r_1s_1(Q_1^2{-}1)  }{2(r_1{+}s_1)}{{\cal R}}_{r_1{-}1}(Q_1) {{\cal R}}_{s_1{-}1}(Q_1) +
\sum_{r_2,s_2}^{1,\infty} (-)^{r_{2}{+}s_{2}} \widehat{\zeta}_{r_2}^{(2)}{\cdot}  \widehat{\zeta}_{s_2}^{(2)} \dfrac{r_2s_2(Q_2^2{-}1)  }{2(r_2{+}s_2)}{{\cal R}}_{r_2{-}1}(Q_2) {{\cal R}}_{s_2{-}1}(Q_2)
\Bigg\}\,\\
&\exp \Bigg\{ \sum_{n_{1}=1}^{\infty} \widehat{\zeta}^{(1)}_{n_{1}}{{\cdot} }(p_{2}-p_3) {{\cal R}}_{n_1{-}1}(Q_1)+ \sum_{n_{2}=1}^{\infty} (-)^{n_{2}}\widehat{\zeta}^{(2)}_{n_{2}}{{\cdot} }(p_{3}-p_1) {{\cal R}}_{n_2{-}1}(Q_2) \Bigg\}
 \end{split}}
 \end{equation}}
where $Q_1$, $Q_2$, ${{\cal R}}_{n-1}(Q)$, $\widehat{\zeta}_m$ and ${{\cal L}}_{n_1-1,n_2-1}(Q_1,Q_2)$ are defined in (\ref{defQ1Q2}), (\ref{Rpoly}), (\ref{zetaHat}), and (\ref{Lpoly}), respectively. 
 It is not hard to check that the signs are precisely those needed for the amplitude to be (anti)symmetric under the exchange of the two coherent states. More precisely, after decomposing the coherent states in bases of states at fixed levels $N_1$ and $N_2$ the `partial' amplitude gets a sign $(-)^{N_1{+}N_2}$ under exchange.
 

\subsection{${\mathcal{A}}_3({\mathcal{C}}, {\mathcal{C}}, V)$}
The case in which the third state is a vector is very similar to the previous computation except for the KN factor and other couplings related to the vector polarization.

The relevant correlation function is
\begin{eqnarray}
&&{\mathcal{A}}_3({\mathcal{C}},{\mathcal{C}},V) = \braket{c{{V}}_{\mathcal{C}}(z_1) \, cV_{\mathcal{C}}(z_2)\, cV_A(z_3)} = 
\braket{c(z_1) \, c(z_2)\, c(z_3)}\times \nonumber \\ &&\Big\langle \exp\Bigg\{\sum_{r_1,s_1=1}^{\infty} \dfrac{\zeta_{r_1}^{(1)}{{\cdot} } \zeta_{s_1}^{(1)} }{2r_1s_1}\, \mathcal{S}_{r_1,s_1} \,e^{-i(r_1+s_1)q_1{\cdot}  X} + \sum_{n_1=1}^{\infty} \dfrac{\zeta^{(1)}_{n_1}}{n_1}{\cdot}  \mathcal{P}_{n_1} \,e^{-in_1q_1{\cdot}  X}  \Bigg\} \,e^{ip_1{\cdot}X}(z_1) \\
&&\exp\Bigg\{\sum_{r_2,s_2=1}^{\infty} \dfrac{\zeta_{r_2}^{(2)}{{\cdot} } \zeta_{s_2}^{(2)} }{2r_2s_2}\, \mathcal{S}_{r_2,s_2} \,e^{-i(r_2+s_2)q_2{\cdot}  X} + \sum_{n_2=1}^{\infty} \dfrac{\zeta^{(2)}_{n_2}}{n_2}{\cdot}  \mathcal{P}_{n_2} \,e^{-in_2q_2{\cdot}  X}  \Bigg\} \,e^{ip_2{\cdot}X}(z_2) \,
iA_3{\cdot}\pa{X}e^{ik_3{\cdot}  X}(z_3)  \Big\rangle \nonumber
\end{eqnarray}

Following the same steps as in previous computations, in particular choosing $q_1$, $q_2$ and $q_3$ to be collinear without loss of generality, no new types of contractions appear and one smoothly arrives at the final form of the amplitude that reads 
{\small\begin{equation}
\boxed{ \begin{split}
&{\mathcal{A}}_3({\mathcal{C}}, {\mathcal{C}}, V) = \int {{d^D X_0}}\, e^{i\,(p_1+p_2+p_3){{\cdot} }X_{0}} \Bigg({A}_{3}{{\cdot} }p_{2} + \sum_{n_{1}=1}^{\infty} {\cal M}_{n_1{-}1}(Q_{1}) 
\widehat{\zeta}^{(1)}_{n_{1}}{{\cdot} }{A}_{3} +
\sum_{n_{2}=1}^{\infty} {\cal M}_{n_2{-}1}(Q_{2})\widehat{\zeta}^{(2)}_{n_{2}}{{\cdot} }{A}_{3}\Bigg) \\
&\exp\Bigg\{ \sum_{r_1,s_1}^{1,\infty}  \widehat{\zeta}_{r_1}^{(1)}{\cdot}  \widehat{\zeta}_{s_1}^{(1)} \dfrac{r_1s_1(Q_1^2{-}1)  }{2(r_1{+}s_1)}{{\cal R}}_{r_1{-}1}(Q_1) {{\cal R}}_{s_1{-}1}(Q_1) +
\sum_{r_2,s_2}^{1,\infty} (-)^{r_{2}{+}s_{2}} \widehat{\zeta}_{r_2}^{(2)}{\cdot}  \widehat{\zeta}_{s_2}^{(2)} \dfrac{r_2s_2(Q_2^2{-}1)  }{2(r_2{+}s_2)}{{\cal R}}_{r_2{-}1}(Q_2) {{\cal R}}_{s_2{-}1}(Q_2)
\Bigg\}\,\\
&\exp \Bigg\{ \sum_{n_{1}=1}^{\infty} \widehat{\zeta}^{(1)}_{n_{1}}{{\cdot} }(p_{2}-k_3) {{\cal R}}_{n_1{-}1}(Q_1) \,{-} \sum_{n_{2}=1}^{\infty} (-)^{n_{2}}\widehat{\zeta}^{(2)}_{n_{2}}{{\cdot} }(k_{3}-p_1) {{\cal R}}_{n_2{-}1}(Q_2)\, {+} \sum_{n_{1}=1}^{\infty} \sum_{n_{2}=1}^{\infty}  \widehat{\zeta}_{n_{1}}^{(1)}{\cdot} \widehat{\zeta}_{n_{2}}^{(2)} {\cal L}_{n_1{-}1, n_2{-}1} (Q_1,Q_2)  \Bigg\}
 \end{split}}
 \end{equation}}
where $\widehat{\zeta}_m$, $Q_1$, $Q_2$, ${{\cal R}}_{n-1}(Q)$, ${{\cal M}}_{n-1}(Q)$, and ${{\cal L}}_{n_1-1,n_2-1}(Q_1,Q_2)$ are defined in (\ref{zetaHat}), (\ref{defQ1Q2}), (\ref{Rpoly}),  (\ref{Mpoly}), and (\ref{Lpoly}), respectively. One can check that the subtle signs that appear are precisely the ones necessary to implement the correct Bose symmetry of the amplitude.

\subsection{${\mathcal{A}}_3({\mathcal{C}}, {\mathcal{C}}, {\mathcal{C}})$}
 The most laborious 3-point amplitude corresponds to taking three coherent states as external legs. The relevant correlation function is
\begin{eqnarray}
&&{\mathcal{A}}_3({\mathcal{C}},{\mathcal{C}},{\mathcal{C}}) = \braket{c{{V}}_{\mathcal{C}}(z_1) \, cV_{\mathcal{C}}(z_2)\, cV_{\mathcal{C}}(z_3)} = 
\braket{c(z_1) \, c(z_2)\, c(z_3)}\times \nonumber \\ &&\Big\langle \exp\Bigg\{\sum_{r_1,s_1=1}^{\infty} \dfrac{\zeta_{r_1}^{(1)}{{\cdot} } \zeta_{s_1}^{(1)} }{2r_1s_1}\, \mathcal{S}_{r_1,s_1} \,e^{-i(r_1+s_1)q_1{\cdot}  X} + \sum_{n_1=1}^{\infty} \dfrac{\zeta^{(1)}_{n_1}}{n_1}{\cdot}  \mathcal{P}_{n_1} \,e^{-in_1q_1{\cdot}  X}  \Bigg\} \,e^{ip_1{\cdot}X}(z_1) \nonumber \\
&&\exp\Bigg\{\sum_{r_2,s_2=1}^{\infty} \dfrac{\zeta_{r_2}^{(2)}{{\cdot} } \zeta_{s_2}^{(2)} }{2r_2s_2}\, \mathcal{S}_{r_2,s_2} \,e^{-i(r_2+s_2)q_2{\cdot}  X} + \sum_{n_2=1}^{\infty} \dfrac{\zeta^{(2)}_{n_2}}{n_2}{\cdot}  \mathcal{P}_{n_2} \,e^{-in_2q_2{\cdot}  X}  \Bigg\} \,e^{ip_2{\cdot}X}(z_2) \\
&&\exp\Bigg\{\sum_{r_3,s_3=1}^{\infty} \dfrac{\zeta_{r_3}^{(3)}{{\cdot} } \zeta_{s_3}^{(3)} }{2r_3s_3}\, \mathcal{S}_{r_3,s_3} \,e^{-i(r_3+s_3)q_3{\cdot}  X} + \sum_{n_3=1}^{\infty} \dfrac{\zeta^{(3)}_{n_3}}{n_3}{\cdot}  \mathcal{P}_{n_3} \,e^{-in_3q_3{\cdot}  X}  \Bigg\} \,e^{ip_3{\cdot}X}(z_3)
  \Big\rangle \nonumber
\end{eqnarray}
 
Setting 
\begin{equation}\label{defQ1Q2Q3}
Q_{1}\equiv q_{1}{{\cdot} }(p_{2}-p_{3}) \quad , \quad Q_{2}\equiv q_{2}{{\cdot} }(p_{1}-p_{3}) 
\quad , \quad  Q_{3}\equiv q_{3}{{\cdot} }(p_{1}-p_{2})
\end{equation}
the contractions of $q\pa^s{X}$ with $\exp{ipX}$ yield
 \begin{equation}
\mathcal{U}^{(n_{1})}_{s}=\dfrac{n_{1}}{2}\Bigg( \dfrac{Q_{1}{-}1}{z_{21}^{s}}-\dfrac{Q_{1}{+}1}{z_{31}^{s}} \Bigg) \:\: , \quad
 \mathcal{U}^{(n_{2})}_{s} =\dfrac{n_{2}}{2}\Bigg( \dfrac{Q_{2}{-}1}{z_{12}^{s}}-\dfrac{Q_{2}{+}1}{z_{32}^{s}} \Bigg) \:\: , \quad \mathcal{U}^{(n_{3})}_{s}=\dfrac{n_{3}}{2}\Bigg( \dfrac{Q_{3}{-}1}{z_{13}^{s}}-\dfrac{Q_{3}{+}1}{z_{23}^{s}} \Bigg)
 \end{equation}
Despite appearance, no new types of contractions are necessary to compute the above amplitude. Following the same steps as in previous computations, one finally gets to 
{\small \begin{equation}\label{3pointCCC}
\boxed{ \begin{split}
&{\mathcal{A}}_3({\mathcal{C}}, {\mathcal{C}}, \mathcal{C}) = \int {{d^D X_0}}\, e^{i\,(p_1+p_2+p_3){{\cdot} }X_{0}} \exp\Bigg\{ \sum_{r_1, s_1}^{1,\infty} \widehat{\zeta}^{(1)}_{r_{1}}{\cdot}\widehat{\zeta}^{(1)}_{s_{1}} \,\dfrac{r_1 s_1(Q_1^2{-}1)  }{2(r_1{+}s_1)}  {\cal R}_{r_1{-}1}(Q_1) {\cal R}_{s_1{-}1}(Q_1) + \\
&\sum_{r_2,s_2}^{1,\infty} (-)^{r_2+s_2}\widehat{\zeta}^{(2)}_{r_{2}}{{\cdot} } \widehat{\zeta}^{(2)}_{s_{2}} \dfrac{r_2 s_2(Q_2^2{-}1)  }{2(r_2{+}s_2)}  {\cal R}_{r_2{-}1}(Q_2) {\cal R}_{s_2{-}1}(Q_2) + \sum_{r_3,s_3}^{1,\infty} \widehat{\zeta}^{(3)}_{r_{3}}{{\cdot} } \widehat{\zeta}^{(3)}_{s_{3}}  \dfrac{r_3s_3(Q_3^2{-}1)  }{2(r_3{+}s_3)}   {\cal R}_{r_3{-}1}(Q_3) {\cal R}_{s_3{-}1}(Q_3) 
\Bigg\} \\
&\exp\Bigg\{\sum_{n_{1}=1}^{\infty} \sum_{n_{2}=1}^{\infty}  \widehat{\zeta}_{n_{1}}^{(1)}{\cdot} \widehat{\zeta}_{n_{2}}^{(2)}{\cal L}_{n_1{-}1, n_{2}{-}1}(Q_1,Q_2) +
\sum_{n_{2}=1}^{\infty} \sum_{n_{3}=1}^{\infty} \widehat{\zeta}_{n_{2}}^{(2)}{\cdot} \widehat{\zeta}_{n_{3}}^{(3)}{\cal L}_{n_2{-}1, n_{3}{-}1}(-Q_2,-Q_3) +\\
&\sum_{n_{3}=1}^{\infty} \sum_{n_{1}=1}^{\infty} (-)^{n_1{+}n_{3}} \widehat{\zeta}_{n_{3}}^{(3)}{\cdot} \widehat{\zeta}_{n_{1}}^{(1)}{\cal L}_{n_1{-}1, n_{3}{-}1}(-Q_1,Q_3) 
\Bigg\}\\
&\exp \Bigg\{\sum_{n_{1}=1}^{\infty}\widehat{\zeta}^{(1)}_{n_{1}}{{\cdot} }p_{23}{\cal R}_{n_1{-}1}(Q_1)- \sum_{n_{2}=1}^{\infty} (-)^{n_2} \widehat{\zeta}^{(2)}_{n_{2}}{{\cdot} }p_{31} 
{\cal R}_{n_2{-}1}(Q_2) +\sum_{n_{3}=1}^{\infty} \widehat{\zeta}^{(3)}_{n_{3}}{{\cdot} }p_{12} 
{\cal R}_{n_3{-}1}(Q_3) \Bigg\}
 \end{split}}
 \end{equation}}
where $\widehat{\zeta}_m$, $Q_1$, $Q_2$, $Q_3$, ${{\cal R}}_{n-1}(Q)$, and ${{\cal L}}_{n_1-1,n_2-1}(Q_1,Q_2)$ are defined in (\ref{zetaHat}), (\ref{defQ1Q2Q3}), (\ref{Rpoly}), and (\ref{Lpoly}), respectively. One can check that the signs are those necessary to implement the correct symmetry properties under exchange of any pair of coherent states. 
 
 Before passing to consider 4-point amplitude let us pause and discuss the significance of the above results for 3-point amplitudes with coherent states. Expanding in the levels $N_1$, $N_2$ and $N_3$ (\ref{3pointCCC}) encodes all the physical amplitudes of arbitrarily excited open string states. These can be simply extracted taking derivatives with respect to the parameters $\zeta^{(i)}_n$. Moreover, using unitarity and completeness of the coherent states, one can in principle glue such 3-point amplitudes to construct any higher point and even higher loop amplitude. It would be interesting to compare our compact expression with the result that would obtain after saturating the three-Reggeon vertex \cite{Ademollo:1974kz, DiVecchia:1986jv} with coherent states\footnote{We thank P.~Di Vecchia to suggest this interesting check, that we leave to the future.}.
 
 More interestingly in our views is the possibility of deriving quantitative informations on the interactions of highly excited very massive open string states, even with large spin. As mentioned in the introduction, generalisation to stacks of coincident or intersecting D-branes looks quite natural and feasible, including the appropriate CP factors and taking into account of the correct boundary conditions. Moreover, using KLT relations or otherwise, one can combine such open string 3-point physical amplitudes for left- and right-movers and get closed string 3-point physical amplitudes for arbitrary excited states.
 

\section{4-point open string amplitudes with Coherent states}

In this Section, we compute 4-point `amplitudes' involving diverse numbers of coherent states as well as tachyons and vector bosons. In preparation for future applications to closed strings, we also study the soft limit at the open string level as the momentum $k$ of a massless vector is taken to zero.

Specifically, we turn our attention on 4-point amplitudes on the disk. As before we will consider processes with one, two or three coherent states, leaving the case of four coherent states to the future. Although we focus on a particular ordering of the external legs to start with, results for different orderings obtain easily after exchanging the external legs. We will comment on this at the end of the computations. One can also `dress' the amplitude with CP factors corresponding to multiple D-branes and/or different stacks of D-branes. We will briefly mention how to do so later on.      

\subsection{${\mathcal{A}}_4({ T, T, T,\mathcal{C}})$}

Let us consider the amplitude of one coherent state and three tachyons. Thanks to the symmetry under the exchange of any pair of tachyons (for a single D25-brane!), one can choose any ordering of the external legs. For definiteness we choose the following
\begin{eqnarray}
&&{\mathcal{A}}_4({ T, T, T,\mathcal{C}}) = \int_{z_2}^{z_4} dz_3 \, \braket{  cV_T(z_1) \, \, cV_T(z_2) \, V_T(z_3)\, c{{V}}_{\mathcal{C}}(z_4) }= \int_{z_2}^{z_4} dz_3 \, \langle c(z_1)c(z_2)c(z_4)\rangle \\ 
&&\Big\langle e^{ip_1{\cdot}X}(z_{1})e^{ip_2{\cdot}X}(z_{2})e^{ip_3{\cdot}X}(z_{3}) \exp\Bigg\{\sum_{r,s=1}^{\infty} \dfrac{\zeta_{r}{{\cdot} } \zeta_{s} }{2rs}\, \mathcal{S}_{r,s} \,e^{-i(r+s)q_4{\cdot}  X} + \sum_{n=1}^{\infty} \dfrac{\zeta_{n}}{n}{\cdot}  \mathcal{P}_{n} \,e^{-inq_4{\cdot}  X}  \Bigg\} \,e^{ip_4{\cdot}X}(z_4) \Big\rangle \nonumber 
\end{eqnarray}
where $z_1> z_2>z_3>z_4$. Expanding the exponentials one has
\begin{eqnarray}
&&\Big\langle e^{ip_1{\cdot}X}(z_{1})e^{ip_2{\cdot}X}(z_{2})e^{ip_3{\cdot}X}(z_{3}) \\ 
&&\qquad\prod_{r,s=1}^{\infty} \sum_{\ell_{r,s}=0}^{\infty}\dfrac{1}{\ell_{r,s}!} \Bigg(\dfrac{\zeta_{r}{{\cdot} } \zeta_{s} }{2rs}\, \mathcal{S}_{r,s} \,e^{-i(r+s)q_4{\cdot}  X}\Bigg)^{\ell_{r,s}}\prod_{n=1}^{\infty} \sum_{g_{n}=0}^{\infty}\dfrac{1}{g_{n}!}\Bigg( \dfrac{\zeta_{n}}{n}{\cdot}  \mathcal{P}_{n} e^{-inq_4{\cdot}  X}\Bigg)^{g_{n}}e^{ip_4{\cdot}X}(z_4) \Big\rangle \nonumber
\end{eqnarray}

For fixed level $N = \sum_{r,s}^{1,\infty}\ell_{r,s}(r+s) + \sum_{n=1}^{\infty} n g_{n}$ the KN factor is given by
\begin{eqnarray}
&&\langle e^{ip_1{\cdot}X}(z_{1})e^{ip_2{\cdot}X}(z_{2})e^{ip_3{\cdot}X}(z_{3}) e^{i(p_4-Nq_4){\cdot}X}(z_{4}) \rangle= \\ &&\qquad \Bigg(\dfrac{z_{12} z_{34}}{z_{13}z_{24}} \Bigg)^{{-}\frac{s}{2}{-}2} \Bigg(\dfrac{z_{14} z_{23}}{z_{13}z_{24}} \Bigg)^{{-}\frac{t}{2}{-}2} \Bigg(\dfrac{z_{34}z_{14}}{z_{13}}\Bigg)^{N}(z_{13}z_{24})^{-2} \,\delta(p_1+p_2+p_3+p_4-Nq_4)\nonumber
\end{eqnarray}
where the standard 4-point kinematics has been implemented, more precisely:
\begin{eqnarray}
&& s= -4 - 2p_{1}{\cdot}p_{2}= 2N_4 - 4 - 2p_{3}{\cdot}(p_{4}{-}N_4q_4) \\
&& t =  - 4 - 2p_{3}{\cdot}p_{2} = 2N_4 - 4 - 2p_{1}{\cdot}(p_{4}{-}N_4q_4) \\
&& u =-4 -2 p_{1}{\cdot}p_{3} =2N_4 -4 -2p_{2}{\cdot}(p_{4}{-}N_4q_4)
\end{eqnarray}

Performing the contractions of the light-cone coordinate $X^+$ present in $\mathcal{S}_{r,s}$ (in the form $q{\cdot}\pa^sX= q^-\pa^sX^+$), in ${\cal P}_n$ and in the exponentials $\exp i\ell q{\cdot}X = \exp i\ell q^-X^+$ and of the transverse coordinates $X^i$ present in ${\cal P}_n$ and in $\exp ip{\cdot}X$,  including the ghost contribution, using the $SL(2,\mathbb{R})$ invariance to fix $z_1=\infty\,,z_2=1\,,z_4=0$ and renaming $z_3=z$ the amplitude looks like
\begin{eqnarray}
&&{\mathcal{A}}_4({ T, T, T,\mathcal{C}}) = \int {{d^D X_0}}\, e^{i(p_1+p_2+p_3+p_4){\cdot}X_{0}} \int_{0}^{1} dz\, z^{-\frac{s}{2}-2} (1-z)^{-\frac{t}{2}-2} \\
&&\qquad\exp \Bigg\{ \sum_{r,s=1}^{\infty} \dfrac{\widehat{\zeta}_{r}{\cdot}\widehat{\zeta}_{s}}{2rs} {\cal S}_{r,s}\big[\mathcal{U}^{(r)}_{k} ; \mathcal{U}^{(s)}_{k}\big] z^{r+s} - \sum_{n=1}^{\infty} {z^{n}\over n}\sum_{h=1}^{n} {\cal Z}_{n-h}\big[ \mathcal{U}_{k}^{(n)}\big]  \Bigg( \widehat{\zeta}_{n}{\cdot}p_2  {+} \dfrac{\widehat{\zeta}_{n}{\cdot}p_3}{z^h}\Bigg)   \Bigg\} \nonumber
\end{eqnarray}
with
\begin{equation}
 \mathcal{U}^{(n)}_{k}=  n\Bigg( q_4{\cdot}p_2  + \dfrac{q_4 {{\cdot} } p_3}{ z^k}   \Bigg)
 \end{equation} 
Now relying on the chain of identities\footnote{We thank Dimitri Skliros for his crucial help in their derivation.}:
\begin{equation}
\exp\Big\{\sum_{s=1}^{\infty}{\cal Y}_{s}\frac{d^{s}}{d\beta^{s}}\Big\} \,e^{\beta z}\,\Big|_{\beta=0}=\sum_{f=0}^{\infty}{\cal Z}_{f}\Big( s {\cal Y}_{s} \Big)\frac{d^{f}}{d\beta^{f}}\,e^{\beta z}\Big|_{\beta=0}=\sum_{f=0}^{\infty}{\cal Z}_{f}\Big( s {\cal Y}_{s}z^s\Big)=\exp\Big\{ \sum_{s=1}^{\infty}{\cal Y}_{s}\,z^{s} \Big\}
\end{equation}
one can represent the $z$ integral in the following way:
\begin{equation}
\begin{split}
\int_{0}^{1} dz \,z^{{A}} (1-z)^{{B}}\,\exp\Big\{ \sum_{s=1}^{\infty}{\cal Y}_{s}\,z^{s} \Big\}= \exp\Big\{ \sum_{s=1}^{\infty} {\cal Y}_{s} \frac{d^{s}}{d\beta^{s}}\Big\}   \int_{0}^{1} dz \,z^{{A}} (1-z)^{{B}}\,\exp\{ \beta z\} \Big|_{\beta=0}
\end{split}
\end{equation}
more precisely one has to massage the terms inside the exponential to get
\begin{eqnarray}
&&-\sum_{n=1}^{\infty} {z^{n}\over n} \sum_{h=1}^{n}  {\cal Z}_{n-h}\big[ \mathcal{U}_{k}^{(n)}\big] \Bigg( \widehat{\zeta}_{n}{\cdot}p_2  {+} \dfrac{\widehat{\zeta}_{n}{\cdot}p_3}{z^h}\Bigg) =  \sum_{f=0}^{\infty} \big({\cal Y}_{f}^{\zeta{\cdot}p_2} +  {\cal Y}_{f}^{\zeta{\cdot}p_3} \big) z^f \\
&&\sum_{r,s=1}^{\infty} \dfrac{\widehat{\zeta}_{r}{\cdot}\widehat{\zeta}_{s}}{2rs} {\cal S}_{r,s}\big[\mathcal{U}^{(r)}_{k} ; \mathcal{U}^{(s)}_{k}\big] z^{r+s}= \sum_{f=0}^{\infty} {\cal Y}_{f}^{\zeta{\cdot}\zeta} z^f
\end{eqnarray}
where
\begin{equation}
\label{Yfzp2}
{\cal Y}_{f}^{\zeta{\cdot}p_2}=-\sum_{n=1}^\infty {\widehat\zeta_n{\cdot}p_2\over n}
\sum_{h=1}^n  {\cal Z}_{n-f}(nq_4{\cdot}p_3){\cal Z}_{f-h}(nq_4{\cdot}p_2)\, 
\end{equation}
and
\begin{equation}
\label{Yfzp3}
{\cal Y}_{f}^{\zeta{\cdot}p_3}=- \sum_{n=1}^\infty { \widehat\zeta_n{\cdot}p_3\over n} \sum_{h=1}^n  
{\cal Z}_{n-h-f}(nq_4{\cdot}p_3){\cal Z}_{f}(nq_4{\cdot}p_2)\, 
\end{equation}
and 
\begin{equation}
\label{Yfzz}
{\cal Y}_{f}^{\zeta{\cdot}\zeta}=\sum_{r,s}^{1,\infty} {\widehat\zeta_r{\cdot}\widehat\zeta_s\over 2rs}  \sum_{h=1}^r h \sum_{m=0}^{r+h} 
{\cal Z}_{s+h-m}(sq_4{\cdot}p_3){\cal Z}_{m}(sq_4{\cdot}p_2){\cal Z}_{r-h-f+m}(sq_4{\cdot}p_3)
{\cal Z}_{f-m}(rq_4{\cdot}p_2)
\end{equation}
and using
\begin{equation}
\int_{0}^{1}dz\,z^{a-1} (1-z)^{b-a-1} e^{\beta z}=\dfrac{\Gamma(a)\Gamma(b-a)}{\Gamma(b)} \,_1F_{1}(a,b,\beta)
\end{equation}

where $_1F_{1}(a,b;\beta)$ is the confluent hyper-geometric function 
\begin{equation}
_1F_{1}(a,b;z)=\sum_{h=0}^{\infty} \dfrac{(a)_{h}}{(b)_{h}} \dfrac{z^h}{h!}\,;\quad (c)_{h}=\dfrac{\Gamma(c+h)}{\Gamma(c)}
\end{equation}
and performing the following steps
\begin{equation}
\begin{split}
&\exp\Big\{ \sum_{\ell=1}^{\infty} {\cal Y}_\ell \frac{d^{\ell}}{d\beta^{\ell}}\Big\}\, _{1}F_{1}\Big( A, B; \beta \Big)\Big|_{\beta=0}=\sum_{\ell=0}^{\infty}{\cal Z}_{\ell}({\cal Y}_f) \dfrac{d^{\ell}}{d\beta^{\ell}}\, _{1}F_{1}\Big( A, B; \beta \Big)\Big|_{\beta=0}=\sum_{\ell=0}^{\infty}{\cal Z}_{\ell}({\cal Y}_f) \dfrac{(A)_{\ell}}{(B)_{\ell}}
\end{split}
\end{equation}
one finally finds 
\begin{equation}
\boxed{\mathcal{A}_4({ T, T, T,\mathcal{C}}) = \int {{d^D X_0}}\, e^{i(p_1+p_2+p_3+p_4){\cdot}X_{0}}e^{{\cal Y}_0} {\Gamma(p_1p_2{+}1) \Gamma(p_2p_3{+}1)\over \Gamma(p_2(p_1{+}p_3){+}2)} 
\sum_{\ell=0}^{\infty}{\cal Z}_{\ell}({f\cal Y}_f){(p_1p_2{+}1)_\ell \over (p_2(p_1{+}p_3){+}2)_\ell}}
\end{equation}
where ${\cal Y}_0 = {\cal Y}_{f=0}$ with
\begin{equation}
{\cal Y}_f = {\cal Y}^{\widehat\zeta{\cdot}p_2}_f + {\cal Y}^{\widehat\zeta{\cdot}p_3}_f+{\cal Y}^{\widehat\zeta{\cdot}\widehat\zeta}_f
\end{equation}
The final expression is quite compact and elegant but somewhat deceiving. Most of the complication is hidden in the definitions of ${\cal Y}^{\widehat\zeta{\cdot}p_2}_f$ in (\ref{Yfzp2}), 
${\cal Y}^{\widehat\zeta{\cdot}p_3}$ in (\ref{Yfzp3}) and ${\cal Y}^{\widehat\zeta{\cdot}\widehat\zeta}_f$ in in (\ref{Yfzz}), that we have not been able to further simplify.
\subsection{$\mathcal{A}_4( T,T,V,\mathcal{C})$}
As for 3-point amplitudes, it is quite interesting to replace one or more tachyons with massless vector bosons. Even for a single D25-brane and for a single vector boson, different orderings of the external legs lead to different 4-point amplitudes. For definiteness we consider the following ordering 
\begin{equation}
\begin{split}
&{\mathcal{A}}_4({ T, T, V,\mathcal{C}}) = \int_{z_4}^{z_2} dz_3 \, \braket{  cV_T(z_1) \, \, cV_T(z_2) \, V_T(z_3)\, c{{V}}_{\mathcal{C}}(z_4) }= \int_{z_4}^{z_2} dz_3 \, \langle c(z_1)c(z_2)c(z_4)\rangle \\ 
&\Big\langle e^{ip_1{\cdot}X(z_{1})}e^{ip_2{\cdot}X(z_{2})}\,A_3{\cdot}\pa X e^{ik_3{\cdot}X}(z_{3}) \exp\Bigg\{\sum_{r,s=1}^{\infty} \dfrac{\zeta_{r}{{\cdot} } \zeta_{s} }{2rs}\, \mathcal{S}_{r,s} \,e^{-i(r+s)q_4{\cdot}  X}{+}\sum_{n=1}^{\infty} \dfrac{\zeta_{n}}{n}{\cdot}  \mathcal{P}_{n} \,e^{-inq_4{\cdot}  X}  \Bigg\} \,e^{ip_4{\cdot}X}(z_4) \Big\rangle 
\end{split}
\end{equation}
where $z_4<z_3<z_2<z_1$ and later on comment on how to get the 4-point amplitudes corresponding to different orderings.

For fixed level $N = \sum_{r,s}^{1,\infty}\ell_{r,s}(r+s) + \sum_{n=1}^{\infty} n g_{n}$  KN factor with ghost contribution is given by
\begin{equation}
\begin{split}
&\langle c\,e^{ip_1{\cdot}X}(z_{1}) c\,e^{ip_2{\cdot}X}(z_{2})e^{ik_3{\cdot}X}(z_{3}) c\,e^{i(p_4{-}Nq_4){\cdot}X}(z_{4}) \rangle= \\
&\qquad\Bigg(\dfrac{z_{12} z_{34}}{z_{13}z_{24}} \Bigg)^{{-}\frac{s}{2}{-}1} \Bigg(\dfrac{z_{14} z_{23}}{z_{13}z_{24}} \Bigg)^{{-}\frac{t}{2}{-}1} \Bigg(\dfrac{z_{14}z_{34}}{z_{13}}\Bigg)^{N}\,\delta(p_1{+}p_2{+}k_3{+}p_4{-}Nq_4)
\end{split}
\end{equation}
where the standard 4-point kinematics has been implemented, more precisely:

\begin{eqnarray}
&& s= -4 - 2p_{1}{\cdot}p_{2}= 2N_4 - 2 - 2k_{3}{\cdot}(p_{4}{-}N_4q_4) \\
&& t =  - 2 - 2k_{3}{\cdot}p_{2} = 2N_4 - 4 - 2p_{1}{\cdot}(p_{4}{-}N_4q_4) \\
&& u =-2 -2 p_{1}{\cdot}k_{3} =2N_4 -4 -2p_{2}{\cdot}(p_{4}{-}N_4q_4)
\end{eqnarray}

Using the same strategy as in the previous computation, with the addition of a new type of contractions, related to the presence of $A_3{\cdot}\pa X (z_3)$, $SL(2,\mathbb{R})$ invariance allows to fix $z_4=0\,,z_2=1\,,z_1=\infty$. Renaming $z_3=z$, the amplitude reads

\begin{equation}
\begin{split}
&{\mathcal{A}}_4({ T, T, V,\mathcal{C}}) = \int {{d^D X_0}}\, e^{i(p_1+p_2+k_3+p_4){\cdot}X_{0}} \int_{0}^{1} dz\, z^{-\frac{s}{2}-1} (1{-}z)^{-\frac{t}{2}-1} \Bigg(\sum_{n=1}^{\infty} \widehat{\zeta}_n{\cdot}A_3 \sum_{h=1}^{n} {\cal Z}_{n-h}\big[ \mathcal{U}_{k}^{(n)}\big]\dfrac{h z^n}{n z^{h{+}1}} \\
&{+}\dfrac{A_3{\cdot}p_4}{z} {-}\dfrac{A_3{\cdot}p_2}{1{-}z}\Bigg)\exp\Bigg\{ \sum_{r,s=1}^{\infty} \dfrac{\widehat{\zeta}_{r}{\cdot}\widehat{\zeta}_{s}}{2rs} {\cal S}_{r,s}\big[\mathcal{U}^{(r)}_{k} ; \mathcal{U}^{(s)}_{k}\big] z^{r+s}{-}\sum_{n=1}^{\infty} \dfrac{z^{n}}{n}\sum_{h=1}^{n} {\cal Z}_{n-h}\big[ \mathcal{U}_{k}^{(n)}\big] \Bigg( \widehat{\zeta}_{n}{\cdot}p_2  {+} \dfrac{\widehat{\zeta}_{n}{\cdot}p_3}{z^h}\Bigg)\Bigg\}
\end{split}
\end{equation}
with
\begin{equation}
\mathcal{U}_{k}^{(n)}= n \Bigg(q_4{\cdot}p_2 +  \dfrac{q_4{\cdot}k_3}{z^k}\Bigg)
\end{equation}
with the help of the following identifications:
\begin{eqnarray}
&& \sum_{n=1}^{\infty}\dfrac{z}{n}^n\sum_{h=1}^{n} {\cal Z}_{n-h}\big[ \mathcal{U}_{k}^{(n)}\big]\widehat{\zeta}_n{\cdot}A \dfrac{h}{z^{h{+}1}} =\sum_{k=0}^{\infty} \mathcal{Y}_{k}^{\widehat{\zeta}{\cdot}A} \,z^{k-1}\\
&& \sum_{n=1}^{\infty}(-)\dfrac{z}{n}^n\sum_{h=1}^{n} {\cal Z}_{n-h}\big[ \mathcal{U}_{k}^{(n)}\big]\Bigg(\widehat{\zeta}_n{\cdot}p_2 + \dfrac{\widehat{\zeta}_n{\cdot}k_3}{z^h}\Bigg)=\sum_{k=0}^{\infty} \Big( \mathcal{Y}_{k}^{\widehat{\zeta}{\cdot}p_2}+\mathcal{Y}_{k}^{\widehat{\zeta}{\cdot}k_3}\Big) \,z^{k}\\
&& \sum_{r,s=1}^{\infty} \dfrac{\widehat{\zeta}_{r}{\cdot}\widehat{\zeta}_{s}}{2rs} {\cal S}_{r,s}\big[\mathcal{U}^{(r)}_{k} ; \mathcal{U}^{(s)}_{k}\big] z^{r+s}= \sum_{k=0}^{\infty} {\cal Y}_{k}^{\widehat{\zeta}{\cdot}\widehat{\zeta}} \,z^k
\end{eqnarray}
where
\begin{eqnarray}
&&\mathcal{Y}_{k}^{\widehat{\zeta}{\cdot}A_3}=\sum_{n=1}^{\infty}{1\over n} \widehat{\zeta}_{n}{\cdot}{A_3} \sum_{h=1}^{n}h\, {\cal Z}_{n-h-k}(nq_4{\cdot}k_3) {\cal Z}_k(nq_4{\cdot}p_2) \\
&&\mathcal{Y}_{k}^{\widehat{\zeta}{\cdot}k_3}=-\sum_{n=1}^{\infty}{1\over n} \widehat{\zeta}_{n}{\cdot}k_3 \sum_{h=1}^{n}\, {\cal Z}_{n-h-k}(nq_4{\cdot}k_3) {\cal Z}_k(nq_4{\cdot}p_2) \\
&&\mathcal{Y}_{k}^{\widehat{\zeta}{\cdot}p_2}=-\sum_{n=1}^{\infty}{1\over n} \widehat{\zeta}_{n}{\cdot}p_2 \sum_{h=1}^{n}\, {\cal Z}_{n-k}(nq_4{\cdot}k_3) {\cal Z}_{k-h}(nq_4{\cdot}p_2) \\
&&\mathcal{Y}_{k}^{\widehat{\zeta}{\cdot}\widehat{\zeta}}=\sum_{r,s}^{1,\infty}   {\widehat{\zeta}_{r}\widehat{\zeta}_{s}\over 2rs}  \sum_{h=1}^{r}\sum_{m=0}^{r-h}\, {\cal Z}_{s+h-k+m}(sq_4{\cdot}k_3) {\cal Z}_{k-m}(sq_4{\cdot}p_2) {\cal Z}_{r-h-m}(rq_4{\cdot}k_3) {\cal Z}_{m}(rq_4{\cdot}p_2) 
\end{eqnarray}
using the same strategy as in the previous computation the full amplitudes reads:
\begin{equation}
\boxed{\begin{split}
&{\mathcal{A}}_4({ T, T, V,\mathcal{C}}) = \int {{d^D X_0}}\, e^{i(p_1+p_2+k_3+p_4){\cdot}X_{0}}\,e^{{\cal Y}_0}  \\
&\sum_N \Big\{A_3{\cdot}(p_4{-}Nq_4){\Gamma\big[k_3{\cdot}(p_4{-}Nq_4){-}N\big] \Gamma\big[k_3{\cdot}p_2+1\big]\over \Gamma\big[k_3{\cdot}(p_4{-}Nq_4{+}p_2){+}1{-}N\big]} 
    \sum_{\ell=0}^{\infty}{\cal Z}_{\ell}(f{\cal Y}_f){[k_3{\cdot}(p_4{-}Nq_4){-}N\big] _\ell \over [k_3{\cdot}(p_4{-}Nq_4{+}p_2){+}1{-}N\big]_\ell} +  \\
&-A_3{\cdot}p_2{\Gamma\big[k_3{\cdot}(p_4{-}Nq_4){-}N{+}1\big] \Gamma\big[k_3{\cdot}p_2\big]\over \Gamma\big[k_3{\cdot}(p_4{-}Nq_4{+}p_2){+}1{-}N\big]} 
    \sum_{\ell=0}^{\infty}{\cal Z}_{\ell}(f{\cal Y}_f){[k_3{\cdot}(p_4{-}Nq_4){-}N{+}1\big] _\ell \over [k_3{\cdot}(p_4{-}Nq_4{+}p_2){+}1{-}N\big]_\ell} +\\
&+ \sum_{k=0}^{\infty}\mathcal{Y}_{k}^{\zeta{\cdot}A_3} {\Gamma\big[k_3{\cdot}(p_4{-}Nq_4){-}N{+}k\big] \Gamma\big[k_3{\cdot}p_2{+}1\big]\over \Gamma\big[k_3{\cdot}(p_4{-}Nq_4{+}p_2){+}1{-}N{+}k\big]} 
    \sum_{\ell=0}^{\infty}{\cal Z}_{\ell}(f{\cal Y}_f){[k_3{\cdot}(p_4{-}Nq_4){-}N{+}k\big] _\ell \over [k_3{\cdot}(p_4{-}Nq_4{+}p_2){+}1{-}N{+}k\big]_\ell} \Big\} \\
\end{split}}
\end{equation}

where ${\cal Y}_0 = {\cal Y}_{f=0}$ with
\begin{equation}
{\cal Y}_f = {\cal Y}^{\widehat\zeta{\cdot}p_2}_f + {\cal Y}^{\widehat\zeta{\cdot}k_3}_f+{\cal Y}^{\widehat\zeta{\cdot}\widehat\zeta}_f
\end{equation}

For a single D25-brane, the only other inequivalent ordering of the external legs corresponds to the 4-point amplitude $\mathcal{A}_4(T, V, T,\mathcal{C})$. We will not display the final result but it is easy to convince oneself that its form is actually simpler in that no singularity in $k_3{\cdot}(p_4-Nq_4)$ can be exposed, due to planarity of disk amplitudes. 

\subsection{Soft behaviour} 

We now pass to investigate the soft behavior of the 4-point amplitude we computed, that has poles in $k_3{\cdot}(p_4-Nq_4)$. To leading order only the first two terms of the amplitude contribute because $\zeta{\cdot}A_{3}$ is sub-leading in the $k_{3}\rightarrow 0$ limit.

Starting from: 
\begin{equation}
e^{{\cal Y}_{0}} A_3{\cdot}p_2{\Gamma\big[k_3{\cdot}(p_4{-}Nq_4){-}N{+}1\big] \Gamma\big[k_3{\cdot}p_2\big]\over \Gamma\big[k_3{\cdot}(p_4{-}Nq_4{+}p_2){+}1{-}N\big]} 
    \sum_{\ell=0}^{\infty}{\cal Z}_{\ell}(f{\cal Y}_f){[k_3{\cdot}(p_4{-}Nq_4){-}N{+}1\big] _\ell \over [k_3{\cdot}(p_4{-}Nq_4{+}p_2){+}1{-}N\big]_\ell}
\end{equation}

and using the explicit form of the Pochhammer symbols as ratios of $\Gamma$ functions,
one can rewrite the above as
\begin{equation}
e^{{\cal Y}_{0}}A_3{\cdot}p_2\, \Gamma\left(k_3{\cdot}p_2\right)
   \sum_{N=0}^{\infty} \sum_{\ell=0}^{N}{\cal Z}_{\ell}(f{\cal Y}_f)    {[k_3{\cdot}(p_4{-}Nq_4){-}N{+}\ell] \,\Gamma\left[k_3{\cdot}(p_4{-}Nq_4){-}N{+}\ell\right] \over [k_3{\cdot}(p_4{-}Nq_4{+}p_2){-}N{+}\ell]\, \Gamma\left[k_3{\cdot}(p_4{-}Nq_4{+}p_2){-}N{+}\ell\right]}
\end{equation}
where the upper bound $\ell\leq N$ is made manifest. Using $\Gamma(k_3{\cdot}p_2)\approx 1/k_3{\cdot}p_2$ ad similar for the other $\Gamma$ functions, one then finds
\begin{equation}
e^{{\cal Y}_{0}}{A_3{\cdot}p_2 \over k_{3}{\cdot}p_{2}}\, 
   \sum_{N=0}^{\infty} \sum_{\ell=0}^{N}{\cal Z}_{\ell}(f{\cal Y}_f)   {k_3{\cdot}(p_4{-}Nq_4){-}N{+}\ell \over k_3{\cdot}(p_4{-}Nq_4) } {k_3{\cdot}(p_4{-}Nq_4{+}p_2)\over k_3{\cdot}(p_4{-}Nq_4{+}p_2){-}N{+}\ell}                                         
\end{equation}
 The leading terms, as $k_3\rightarrow 0$, correspond to $\ell=N$ and read
 \begin{equation}
 e^{{\cal Y}_{0}}{A_3{\cdot}p_2 \over k_{3}{\cdot}p_{2}}\, 
   \sum_{N=0}^{\infty} {\cal Z}_{\ell}(f{\cal Y}_f) = {A_3{\cdot}p_2 \over k_{3}{\cdot}p_{2}}\exp\Bigg\{\sum_{k=0}^{\infty} {\cal Y}^{\widehat\zeta{\cdot}p_2}_k + {\cal Y}^{\widehat\zeta{\cdot}k_3}_k+{\cal Y}^{\widehat\zeta{\cdot}\widehat\zeta}_k\Bigg\}
 \end{equation}
The second term in the exponential is sub-dominant for $k_3\rightarrow 0$, while the other terms produce \begin{equation}
 \begin{split}
&\sum_{k=0}^{\infty} {\cal Y}^{\widehat\zeta{\cdot}p_2}_k=
\sum_{k=0}^{\infty}\sum_{n=1}^{\infty}{(-)\over n} \widehat{\zeta}_{n}{\cdot}p_2 \sum_{h=1}^{n}\,{\Gamma(nq_4{\cdot}k_3+n-k)\over \Gamma(nq_4{\cdot}k_3)\, \Gamma(n-k+1)} {\cal Z}_{k-h}(nq_4{\cdot}p_2) \\
\end{split}
\end{equation}
the leading term correspond to $k=n$, and the expression reduces as follow
\begin{equation}
\begin{split}
\sum_{k=0}^{\infty} {\cal Y}^{\widehat\zeta{\cdot}p_2}_k\Big|_{k_{3}\rightarrow 0} =\sum_{n=1}^{\infty}{1\over 2 n} \widehat{\zeta}_{n}{\cdot}(p_1{-}p_2) \sum_{h=1}^{n}\, {\cal Z}_{n-h}\left[ -\frac{n}{2}(Q{+}1)\right]=\sum_{n=1}^{\infty} \widehat{\zeta}_{n}{\cdot}(p_1{-}p_2) R_{n{-}1}(Q)
\end{split}
\end{equation}
Similarly one finds
\begin{equation}
\begin{split}
&\sum_{k=0}^{\infty}\mathcal{Y}_{k}^{\widehat{\zeta}{\cdot}\widehat{\zeta}}=
\sum_{k=0}^{\infty}\sum_{r,s}^{1,\infty}   {\widehat{\zeta}_{r}\widehat{\zeta}_{s}\over 2rs}  \sum_{h=1}^{r}\sum_{m=0}^{r-h}\,{\Gamma\left(sq_4{\cdot}k_3{+}s{+}h{-}k{+}m\right) {\cal Z}_{k-m}(sq_4{\cdot}p_2) \over \Gamma\left(sq_4{\cdot}k_3\right) \, \Gamma\left(s{+}h{-}k{+}m{+}1\right)} {\Gamma\left(rq_4{\cdot}k_3{+}r{-}h{-}m\right) {\cal Z}_{m}(rq_4{\cdot}p_2)\over \Gamma\left(rq_4{\cdot}k_3\right) \, \Gamma\left(r{-}h{-}m{+}1\right)} 
\end{split}
\end{equation}
the leading terms correspond to $k=s+h-m$ and $m=r-h$ separately, and one has
\begin{equation}
\begin{split}
\sum_{k=0}^{\infty}\mathcal{Y}_{k}^{\widehat{\zeta}{\cdot}\widehat{\zeta}}\Big|_{k_3 \rightarrow 0}=\sum_{r,s}^{1,\infty}   {\widehat{\zeta}_{r}\widehat{\zeta}_{s}\over 2rs}  \sum_{h=1}^{r} {\cal Z}_{s{+}h}(nq_4{\cdot}p_2)  {\cal Z}_{r{-}h}(nq_4{\cdot}p_2)=\sum_{r,s}^{1,\infty}  \widehat{\zeta}_{r}\widehat{\zeta}_{s} {rs (Q+1) \over 2(r+s)} R_{s{-}1}(Q)\,R_{r{-}1}(Q)
\end{split}
\end{equation}

The same happens for the other term $A_{3}{\cdot}(p_4{-}Nq_4)$, and finally the soft leading behavior is the following:
\begin{equation}
\boxed{{\mathcal{A}}_4({ T, T, V,\mathcal{C}})\Big|_{k_3 \rightarrow 0}= \sum_N \Bigg({A_3{\cdot}(p_4{-}Nq_{4})\over k_3{\cdot}(p_4{-}Nq_{4})} - {A_3{\cdot}p_2\over k_3{\cdot}p_2} \Bigg)\, {\mathcal{A}}_3(T, T,\mathcal{C}_N)}
\end{equation}
where we exposed the decomposition in terms of levels labelled by $N$, that matches precisely the level $N$ contribution to the 3-point physical amplitude of one coherent state into two tachyons.

The result is perfectly in line with expectations based on the old literature \cite{Gell-MannGoldberger, Low} for QED and \cite{BurnettKroll} for gauge theories, that has been recently revived, after  \cite{Cachazo:2014fwa}, in various contexts \cite{Casali:2014xpa, Schwab:2014xua, Afkhami-Jeddi:2014fia, Bern:2014vva}, including string theory \cite{Bianchi:2014gla, DiVecchia:2015oba, Bianchi:2015lnw, Bianchi:2016tju}.


\subsection{$\mathcal{A}_4( T,\mathcal{C},T,\mathcal{C})$}

The simplest 4-point amplitude with two coherent states is the one with two tachyons. For a single D25-brane there are two inequivalent orderings: $\mathcal{A}_4( T,T,\mathcal{C},\mathcal{C})$ and $\mathcal{A}_4( T,\mathcal{C},T,\mathcal{C})$. For definiteness we consider the latter and comment later on the former. The relevant amplitude is given by 
\begin{equation}
\begin{split}
&\mathcal{A}_4( T\,\mathcal{C}\,T\,\mathcal{C}) = \int_{z_4}^{z_2} dz_3 \, \braket{  cV_T(z_1) \, 
c{{V}}_{\mathcal{C}}(z_2) \, V_T(z_3)\, c{{V}}_{\mathcal{C}}(z_4) }= \int_{z_4}^{z_2} dz_3 \, \langle c(z_1)c(z_2)c(z_4)\rangle \\ 
&\Big\langle e^{ip_1{\cdot}X}(z_{1})\,\exp\Bigg\{\sum_{r_2,s_2=1}^{\infty} \dfrac{\zeta^{(2)}_{r_2}{{\cdot} } \zeta^{(2)}_{s_2} }{2r_2s_2}\, \mathcal{S}_{r_2,s_2} \,e^{-i(r_2+s_2)q_2{\cdot}  X} + \sum_{n_2=1}^{\infty} \dfrac{\zeta^{(2)}_{n_2}}{n_2}{\cdot}  \mathcal{P}_{n_2} \,e^{-in_2q_2{\cdot}  X}  \Bigg\} \,e^{ip_2{\cdot}X}(z_2)
\\
&e^{ip_3{\cdot}X}(z_{3}) \, \exp\Bigg\{\sum_{r_4,s_4=1}^{\infty} \dfrac{\zeta^{(4)}_{r_4}{{\cdot} } \zeta^{(4)}_{s_4} }{2r_4s_4}\, \mathcal{S}_{r_4,s_4} \,e^{-i(r_4+s_4)q_4{\cdot}  X} + \sum_{n_4=1}^{\infty} \dfrac{\zeta^{(4)}_{n_4}}{n_4}{\cdot}  \mathcal{P}_{n_4} \,e^{-in_4q_4{\cdot}  X}  \Bigg\} \,e^{ip_4{\cdot}X}(z_4) \Big\rangle 
\end{split}
\end{equation}

where $z_1>z_2>z_3>z_4$. 

For fixed level of both coherent vertex operators, respectively $N_2$ and $N_4$, the KN factor with ghost contribution assumes the following form:
\begin{equation}
\begin{split}
&\langle c\, e^{ip_1{\cdot}X}(z_{1}) c\, e^{i(p_2{-}N_{2}q_2){\cdot}X}(z_{2})e^{ik_3{\cdot}X}(z_{3}) c\,e^{i(p_4{-}N_{4}q_4){\cdot}X}(z_{4}) \rangle= \\
&\quad\Bigg(\dfrac{z_{12} z_{34}}{z_{13}z_{24}} \Bigg)^{{-}\frac{s}{2}{-}1} \Bigg(\dfrac{z_{14} z_{23}}{z_{13}z_{24}} \Bigg)^{{-}\frac{t}{2}{-}1}  \Bigg(\dfrac{z_{12}z_{23}}{z_{13}}\Bigg)^{N_{2}}   \Bigg(\dfrac{z_{14}z_{34}}{z_{13}}\Bigg)^{N_{4}}\,\delta(p_1{+}p_2{-}N_2 q_2{+}p_3{+}p_4{-}N_4 q_4)
\end{split}
\end{equation}
with the 4-point kinematics:

\begin{eqnarray}
&& s= 2N_2 - 4 - 2p_{1}{\cdot}(p_{2}{-}N_2q_2) = 2N_4 - 4 - 2p_{3}{\cdot}(p_{4}{-}N_4q_4) \\
&& t = 2N_2 - 4 - 2p_{3}{\cdot}(p_{2}{-}N_2q_2) = 2N_4 - 4 - 2p_{1}{\cdot}(p_{4}{-}N_4q_4) \\
&& u = -4-2 p_{1}{\cdot}p_{3} =2N_2 + 2N_4 -4 -2(p_{2}{-}N_2q_2){\cdot}(p_{4}{-}N_4q_4)
\end{eqnarray}


Using $SL(2,\mathbb{R})$ invariance in order to fix $z_1=\infty$, $z_2=1$, $z_4=0$, renaming $z_3=z$ and performing the contraction, the amplitude reads
\begin{equation}
\begin{split} 
&{\mathcal{A}}_4({ T, \mathcal{C}, T,\mathcal{C}}) = \int {{d^D X_0}}\, e^{i(p_1+p_2+p_3+p_4){\cdot}X_{0}} \sum_{N_2, N_4}^{0,\infty} \int_{0}^{2 \pi} {d\theta_2 d\theta_4 \over (2\pi)^2} e^{i N_2 \theta_2 + i N_4 \theta_4} \int_{0}^{1} dz\, z^{-\frac{s}{2}-2} (1{-}z)^{-\frac{t}{2}-2}    \\
&\exp \Bigg\{ \sum_{r_{2},s_{2}=1}^{\infty} \dfrac{\widetilde{\zeta}^{(2)}_{r_{2}}{\cdot}\widetilde{\zeta}^{(2)}_{s_{2}}}{2r_{2}s_{2}} {\cal S}_{r_{2},s_{2}}\big[\mathcal{U}^{(r_{2})}_{k} ; \mathcal{U}^{(s_{2})}_{k}\big] (1-z)^{r_{2}+s_{2}}     {+}   \sum_{r_{4},s_{4}=1}^{\infty} \dfrac{\widetilde{\zeta}^{(4)}_{r_{4}}{\cdot}\widetilde{\zeta}^{(4)}_{s_{4}}}{2r_{4}s_{4}} {\cal S}_{r_{4},s_{4}}\big[\mathcal{U}^{(r_{4})}_{k} ; \mathcal{U}^{(s_{4})}_{k}\big] z^{r_{4}+s_{4}}                                       \Bigg\}  \\
&\exp\Bigg\{ \sum_{n_{2},n_{4}}^{1,\infty}{(1-z)^{n_{2}}\over n_{2}}{z^{n_{4}}\over n_{4}}\sum_{h_{2},h_{4}}^{1;n_{2},n_{4}}{\Gamma(h_{2}{+}h_{4})\over \Gamma(h_{2})\Gamma(h_{4})} {\cal Z}_{n_{2}{-}h_{2}}\big[ \mathcal{U}_{k}^{(n_{2})}\big]\,{\cal Z}_{n_{4}{-}h_{4}}\big[ \mathcal{U}_{k}^{(n_{4})}\big] (-)^{h_{2}{+}1} \widetilde{\zeta}^{(2)}_{n_{2}}{\cdot}\widetilde{\zeta}^{(4)}_{n_{4}}  \Bigg\}\\
&\exp \Bigg\{-\sum_{n_{4}=1}^{\infty} \dfrac{z^{n_{4}}}{n_{4}}\sum_{h_{4}=1}^{n_{4}} {\cal Z}_{n_{4}{-}h_{4}}\big[ \mathcal{U}_{k}^{(n_{4})}\big] \Bigg( \widetilde{\zeta}^{(4)}_{n_{4}}{\cdot}p_2  {+} {\widetilde{\zeta}^{(4)}_{n_{4}}{\cdot}p_3 \over z^{h_{4}}}\Bigg)   \Bigg\}\\
&\exp\Bigg\{\sum_{n_{2}=1}^{\infty} \dfrac{(1-z)^{n_{2}}}{n_{2}}\sum_{h_{2}=1}^{n_{2}}  (-)^{h_{2}{+}1}{\cal Z}_{n_{2}-h_{2}}\big[ \mathcal{U}_{k}^{(n_{2})}\big]\Bigg( {\widetilde{\zeta}^{(2)}_{n_{2}}{\cdot}p_3\over (1-z)^{h_{2}}}  {+} \widetilde{\zeta}^{(2)}_{n_{2}}{\cdot}p_4\Bigg)   \Bigg\} 
\end{split}
\end{equation}
with
\begin{equation}
\widetilde{\zeta}_n = \widehat\zeta_n e^{-in\theta} =  \zeta_n e^{-in(qX_0 +\theta)}
\end{equation}
for each coherent state, that combined with the integration over $\theta$ and the factor $\exp iN\theta$ projects onto fixed level $N$, and 
\begin{equation}
\mathcal{U}_{k}^{(n_{2})}= n_{2}\, (-)^{k}  \Bigg(q_2{\cdot}p_4 +  \dfrac{q_2{\cdot}p_3}{(1-z)^k}\Bigg)\,,\quad \mathcal{U}_{k}^{(n_{4})}= n_{4} \Bigg(q_4{\cdot}p_2 +  \dfrac{q_4{\cdot}p_3}{z^k}\Bigg)
\end{equation}
As in the previous computation one can completely encode the $z$ dependence introducing some ${\cal Y}$ coefficient functions
\begin{equation}
\sum_{r_{2},s_{2}}^{1,\infty} \dfrac{\widetilde{\zeta}^{(2)}_{r_{2}}{\cdot}\widetilde{\zeta}^{(2)}_{s_{2}}}{2r_{2}s_{2}} {\cal S}_{r_{2},s_{2}}\big[\mathcal{U}^{(r_{2})}_{k} ; \mathcal{U}^{(s_{2})}_{k}\big] (1-z)^{r_{2}+s_{2}}=\sum_{k=0}^{\infty} {\cal Y}^{\widetilde{\zeta}^{(2)}{\cdot}\widetilde{\zeta}^{(2)}}_{k} z^k
\end{equation}
\begin{equation}
\sum_{r_{4},s_{4}}^{1,\infty} \dfrac{\widetilde{\zeta}^{(4)}_{r_{4}}{\cdot}\widetilde{\zeta}^{(4)}_{s_{4}}}{2r_{4}s_{4}} {\cal S}_{r_{4},s_{4}}\big[\mathcal{U}^{(r_{4})}_{k} ; \mathcal{U}^{(s_{4})}_{k}\big] z^{r_{4}+s_{4}}                                      =\sum_{k=0}^{\infty} {\cal Y}^{\widetilde{\zeta}^{(4)}{\cdot}\widetilde{\zeta}^{(4)}}_{k} z^k \end{equation}
\begin{equation}
-\sum_{n_{4}=1}^{\infty} \dfrac{z^{n_{4}}}{n_{4}}\sum_{h_{4}=1}^{n_{4}} {\cal Z}_{n_{4}{-}h_{4}}\big[ \mathcal{U}_{k}^{(n_{4})}\big]  \Bigg( \widetilde{\zeta}^{(4)}_{n_{4}}{\cdot}p_2  {+} {\widetilde{\zeta}^{(4)}_{n_{4}}{\cdot}p_3 \over z^{h_{4}}}\Bigg) =\sum_{k=0}^{\infty}\Big( {\cal Y}^{\widetilde{\zeta}^{(4)}{\cdot}p_2}_{k}{+} {\cal Y}^{\widetilde{\zeta}^{(4)}{\cdot}p_3}_{k}\Big) z^k\end{equation}
\begin{equation}
\sum_{n_{2}=1}^{\infty} \dfrac{(1-z)^{n_{2}}}{n_{2}}\sum_{h_{2}=1}^{n_{2}} (-)^{h_{2}{+}1}{\cal Z}_{n_{2}-h_{2}}\big[ \mathcal{U}_{k}^{(n_{2})}\big]\Bigg( {\widetilde{\zeta}^{(2)}_{n_{2}}{\cdot}p_3\over (1-z)^{h_{2}}}  {+} \widetilde{\zeta}^{(2)}_{n_{2}}{\cdot}p_4\Bigg) =\sum_{k=0}^{\infty}\Big( {\cal Y}^{\widetilde{\zeta}^{(2)}{\cdot}p_3}_{k}{+} {\cal Y}^{\widetilde{\zeta}^{(2)}{\cdot}p_4}_{k}\Big) z^k 
\end{equation}
\begin{equation}
\sum_{n_{2},n_{4}}^{1,\infty}{(1-z)^{n_{2}}z^{n_{4}}\over n_{2} n_{4}}\sum_{h_{2},h_{4}}^{1;n_{2},n_{4}}{\Gamma(h_{2}{+}h_{4})\over \Gamma(h_{2})\Gamma(h_{4})} {\cal Z}_{n_{2}{-}h_{2}}\big[ \mathcal{U}_{k}^{(n_{2})}\big]\,{\cal Z}_{n_{4}{-}h_{4}}\big[ \mathcal{U}_{k}^{(n_{4})}\big] (-)^{h_{2}{+}1} \widetilde{\zeta}^{(2)}_{n_{2}}{\cdot}\widetilde{\zeta}^{(4)}_{n_{4}} =\sum_{k=0}^{\infty} {\cal Y}^{\widetilde{\zeta}^{(2)}{\cdot}\widetilde{\zeta}^{(4)}}_{k} z^k 
\end{equation}
where
\begin{equation}
\label{Yz4z4}
{\cal Y}^{\widetilde{\zeta}^{(4)}{\cdot}\widetilde{\zeta}^{(4)}}_{k}=\sum_{{r}_{4},{s}_{4}}^{1,\infty}\dfrac{\widetilde{\zeta}^{(4)}_{{r}_{4}}{\cdot}\widetilde{\zeta}^{(4)}_{{s}_{4}}}{2{r}_{4}{s}_{4}}  \sum_{{h}_{4}=1}^{{r}_{4}}{h}_{4}\sum_{{m}_{4}=0}^{{r}_{4}-{h}_{4}} {\cal Z}_{{s}_{4}{-}{h}_{4}{+}{m}_{4}{-}k}({s}_{4}{q}_{4}{\cdot}p_{3}) {\cal Z}_{k{-}{m}_{4}}({s}_{4}{q}_{4}{\cdot}{p}_{2}){\cal Z}_{{r}_{4}{-}{h}_{4}{-}{m}_{4}}({r}_{4}{q}_{4}{\cdot}p_{3}){\cal Z}_{{m}_{4}}({r}_{4}{q}_{4}{\cdot}{p}_{2}) 
\end{equation}
\begin{align}
\label{Yz2z2}
{\cal Y}^{\widetilde{\zeta}^{(2)}{\cdot}\widetilde{\zeta}^{(2)}}_{k}=\sum_{{r}_{2},{s}_{2}}^{1,\infty}(-)^{r_2{+}s_2}\dfrac{\widetilde{\zeta}^{(2)}_{{r}_{2}}{\cdot}\widetilde{\zeta}^{(2)}_{{s}_{2}}}{2{r}_{2}{s}_{2}}\sum_{{h}_{2}=1}^{{r}_{2}}{h}_{2}\sum_{{m}_{2}=0}^{{r}_{2}-{h}_{2}}\sum_{{t}_{2}=0}^{{s}_{2}+{h}_{2}}\, &\dfrac{({t}_{2}{+}{m}_{2})!\, (-)^k}{k!\,({t}_{2}{+}{m}_{2}{-}k)! } \, {\cal Z}_{{s}_{2}{+}{h}_{2}{-}{t}_{2}}({s}_{2}{q}_{2}{\cdot}p_{3}) \\ &{\cal Z}_{{t}_{2}}({s}_{2}{q}_{2}{\cdot}{p}_{4})\,{\cal Z}_{{r}_{2}{-}{h}_{2}{-}{m}_{2}}({r}_{2}{q}_{2}{\cdot}p_{3})\,{\cal Z}_{{m}_{2}}({r}_{2}{q}_{2}{\cdot}{p}_{4}) \nonumber 
\end{align}
\begin{eqnarray}
\label{Yz4p3}
&&{\cal Y}^{\widetilde{\zeta}^{(4)}{\cdot}p_3}_{k}=-\sum_{{n}_{4}}^{1,\infty}{\widetilde{\zeta}^{(4)}_{{n}_{4}}{\cdot}p_{3} \over {n}_{4}} \sum_{{h}_{4}=1}^{{n}_{4}}{\cal Z}_{{n}_{4}{-}{h}_{4}{-}k}({n}_{4}{q}_{4}{\cdot}p_{3})\,{\cal Z}_{k}({n}_{4}{q}_{4}{\cdot}{p}_{2})\, \\
\label{Yz4p2}
&&{\cal Y}^{\widetilde{\zeta}^{(4)}{\cdot}p_2}_{k}=-\sum_{{n}_{4}}^{1,\infty}{\widetilde{\zeta}^{(4)}_{{n}_{4}}{\cdot}{p}_{2}\over {n}_{4}} \sum_{{h}_{4}=1}^{{n}_{4}}{\cal Z}_{{n}_{4}{-}k}({n}_{4}{q}_{4}{\cdot}p_{3})\,{\cal Z}_{k{-}{h}_{4}}({n}_{4}{q}_{4}{\cdot}{p}_{2})
\end{eqnarray}
\begin{eqnarray}
\label{Yz2p4}
&&{\cal Y}^{\widetilde{\zeta}^{(2)}{\cdot}p_4}_{k}=\sum_{{n}_{2}}^{1,\infty}{(-)\over {n}_{2}}^{n_{2+1} }\widetilde{\zeta}^{(2)}_{{n}_{2}}{\cdot}{p}_{4}\sum_{{h}_{2}=1}^{{n}_{2}}\sum_{{m}_{2}=0}^{{n}_{2}{-}{h}_{2}}\,\dfrac{({h}_{2}{+}{m}_{2})!\,(-)^{k}}{k!\,({h}_{2}{+}{m}_{2}{-}k)!}{\cal Z}_{{n}_{2}{-}{h}_{2}{-}{m}_{2}}({n}_{2}{q}_{2}{\cdot}{p}_{4})\,{\cal Z}_{{m}_{2}}({n}_{2}{q}_{2}{\cdot}p_{3})\, \nonumber \\
\\
\label{Yz2p3}
&&{\cal Y}^{\widetilde{\zeta}^{(2)}{\cdot}p_3}_{k}=\sum_{{n}_{2}}^{1,\infty}{(-)\over {n}_{2}}^{n_{2+1} }\widetilde{\zeta}^{(2)}_{{n}_{2}}{\cdot}p_{3}\sum_{{h}_{2}=1}^{{n}_{2}}\sum_{{m}_{2}=0}^{{n}_{2}{-}{h}_{2}}\,\dfrac{{m}_{2}!\,(-)^{k}}{k!\,({m}_{2}{-}k)!}{\cal Z}_{{n}_{2}{-}{h}_{2}{-}{m}_{2}}({n}_{2}{q}_{2}{\cdot}{p}_{4})\,{\cal Z}_{{m}_{2}}({n}_{2}{q}_{2}{\cdot}p_{3}) \nonumber \\
\end{eqnarray}
\begin{equation}
\label{Yz2z4}
\begin{split}
{\cal Y}^{\widetilde{\zeta}^{(2)}{\cdot}\widetilde{\zeta}^{(4)}}_{k}=\sum_{{n}_{4},{n}_{2}}^{1,\infty}  &{(-)^{{n}_{2}}\over {n}_{4}{n}_{2}}\widetilde{\zeta}^{(2)}_{{n}_{2}}{\cdot}\widetilde{\zeta}^{(4)}_{{n}_{4}}     \sum_{{h}_{4},{h}_{2}}^{1;{n}_{4}{n}_{2}}\sum_{{m}_{4}=0}^{{n}_{4}{-}{h}_{4}}\sum_{{m}_{2}=0}^{{n}_{2}{-}{h}_{2}}\dfrac{({h}_{2}{+}{m}_{2})!\,(-)^{k}}{(k{-}{m}_{4}{-}{h}_{4})!} \dfrac{(-)^{{m}_{4}{+}h_4{+}1}}{({m}_{2}{+}{h}_{2}{+}{m}_{4}{+}{h}_{4}{-}k)!}\\
&{\Gamma({h}_{4}+{h}_{2})\over \Gamma({h}_{4})\Gamma({h}_{2})}   {\cal Z}_{{n}_{4}{-}{h}_{4}{-}{m}_{4}}({n}_{4}{q}_{4}{\cdot}p_{3}) {\cal Z}_{{m}_{4}}({n}_{4}{q}_{4}{\cdot}{p}_{2}){\cal Z}_{{n}_{2}{-}{h}_{2}{-}{m}_{2}}({n}_{4}{q}_{2}{\cdot}p_{3}){\cal Z}_{{m}_{2}}({n}_{2}{q}_{2}{\cdot}{p}_{4}) 
\end{split}
\end{equation}

After computing the integral one finds 
\begin{equation}\boxed{
\begin{split}
&\mathcal{A}_4({ T, \mathcal{C}, T,\mathcal{C}}) = \int {{d^D X_0}}\, e^{i(p_1+p_2+p_3+p_4){\cdot}X_{0}}  \sum_{N_2, N_4}^{0,\infty} \int_{0}^{2\pi} {d\theta_2 d\theta_4 \over (2\pi)^2} e^{i N_2 \theta_2 + i N_4 \theta_4}  \,e^{{\cal Y}_0}   \\ &{\Gamma[p_1{\cdot}(p_2{-}N_2q_2) {+}1{-}N_2] \Gamma[(p_2{-}N_2q_2){\cdot}p_3+1{-}N_2]\over \Gamma[(p_2{-}N_2q_2){\cdot}(p_1{+}p_3) {+}2(1{-}N_2)]}   
 \sum_{\ell=0}^{\infty}{\cal Z}_{\ell}(k{\cal Y}_k){[p_1{\cdot}(p_2{-}N_2q_2) {+}1{-}N_2]_\ell \over [(p_2{-}N_2q_2){\cdot}(p_1{+}p_3) {+}2(1{-}N_2)]_\ell}
\end{split}
}
\end{equation}
where ${\cal Y}_0 = {\cal Y}_{k=0}$ with ${\cal Y}_{k}$, encoding most of the intricacy of the result, given by 
\begin{equation}\label{2cohe coeff}
{\cal Y}_k = {\cal Y}^{\widetilde{\zeta}^{(2)}{\cdot}p_3}_k + {\cal Y}^{\widetilde{\zeta}^{(2)}{\cdot}p_4}_k+ {\cal Y}^{\widetilde{\zeta}^{(4)}{\cdot}p_3}_k+{\cal Y}^{\widetilde{\zeta}^{(4)}{\cdot}p_2}_k+{\cal Y}^{\widetilde{\zeta}^{(4)}{\cdot}\widetilde{\zeta}^{(4)}}_k+{\cal Y}^{\widetilde{\zeta}^{(2)}{\cdot}\widetilde{\zeta}^{(2)}}_k+{\cal Y}^{\widetilde{\zeta}^{(2)}{\cdot}\widetilde{\zeta}^{(4)}}_k
\end{equation}
with the various ${\cal Y}$'s defined in 
Eqs. (\ref{Yz2z2}), (\ref{Yz4z4}), (\ref{Yz2z4}), (\ref{Yz2p3}), (\ref{Yz2p4}), (\ref{Yz4p3}), (\ref{Yz4p2}).

\subsection{$\mathcal{A}_{4}( T,\mathcal{C},V,\mathcal{C})$}

The next-simplest 4-point amplitude with two coherent states requires one tachyon and one vector boson. For a single D25-brane, there are two inequivalent orderings $\mathcal{A}( T, V,\mathcal{C},\mathcal{C})$ and $\mathcal{A}( T\,\mathcal{C}\,V\,\mathcal{C})$. We consider only the latter. The relevant amplitude reads 
\begin{equation}
\begin{split}
&\mathcal{A}_4( T\,\mathcal{C}\,V\,\mathcal{C}) = \int_{z_4}^{z_2} dz_3 \, \braket{  cV_T(z_1) \, 
c{{V}}_{\mathcal{C}}(z_2) \, V_T(z_3)\, c{{V}}_{\mathcal{C}}(z_4) }= \int_{z_2}^{z_4} dz_3 \, \langle c(z_1)c(z_2)c(z_4)\rangle \\ 
&\Big\langle e^{ip_1{\cdot}X}(z_{1})\,\exp\Bigg\{\sum_{r_2,s_2=1}^{\infty} \dfrac{\zeta^{(2)}_{r_2}{{\cdot} } \zeta^{(2)}_{s_2} }{2r_2s_2}\, \mathcal{S}_{r_2,s_2} \,e^{-i(r_2+s_2)q_2{\cdot}  X} + \sum_{n_2=1}^{\infty} \dfrac{\zeta^{(2)}_{n_2}}{n_2}{\cdot}  \mathcal{P}_{n_2} \,e^{-in_2q_2{\cdot}  X}  \Bigg\} \,e^{ip_2{\cdot}X}(z_2)
\\
&iA_3\pa X e^{ik_3{\cdot}X}(z_{3}) \, \exp\Bigg\{\sum_{r_4,s_4=1}^{\infty} \dfrac{\zeta^{(4)}_{r_4}{{\cdot} } \zeta^{(4)}_{s_4} }{2r_4s_4}\, \mathcal{S}_{r_4,s_4} \,e^{-i(r_4+s_4)q_4{\cdot}  X} + \sum_{n_4=1}^{\infty} \dfrac{\zeta^{(4)}_{n_4}}{n_4}{\cdot}  \mathcal{P}_{n_4} \,e^{-in_4q_4{\cdot}  X}  \Bigg\} \,e^{ip_4{\cdot}X}(z_4) \Big\rangle 
\end{split}
\end{equation}
where $z_1>z_2>z_3>z_4$.

For fixed level of both coherent vertex operators, respectively $N_2$ and $N_4$, the KN factor with ghost contribution takes the following form:
\begin{equation}
\begin{split}
&\langle c\, e^{ip_1{\cdot}X}(z_{1}) c\,e^{i(p_2{-}N_{2}q_2){\cdot}X}(z_{2})e^{ik_3{\cdot}X}(z_{3}) c\,e^{i(p_4{-}N_{4}q_4){\cdot}X}(z_{4}) \rangle= \\
&\Bigg(\dfrac{z_{12} z_{34}}{z_{13}z_{24}} \Bigg)^{{-}\frac{s}{2}{-}1} \Bigg(\dfrac{z_{14} z_{23}}{z_{13}z_{24}} \Bigg)^{{-}\frac{t}{2}{-}1} z_{24}\,\,\, \Bigg(\dfrac{z_{12}z_{23}}{z_{13}}\Bigg)^{N_{2}}   \Bigg(\dfrac{z_{14}z_{34}}{z_{13}}\Bigg)^{N_{4}}\,\delta(p_1{+}p_2{-}N_2 q_2{+}k_3{+}p_4{-}N_4 q_4)
\end{split}
\end{equation}
with the 4-point kinematics:
\begin{eqnarray}
&& s= 2N_2 - 4 - 2p_{1}{\cdot}(p_{2}{-}N_2q_2) = 2N_4 - 2 - k_{3}{\cdot}(p_{4}{-}N_4q_4) \\
&& t = 2N_2 - 2 - 2k_{3}{\cdot}(p_{2}{-}N_2q_2) = 2N_4 - 4 - 2p_{1}{\cdot}(p_{4}{-}N_4q_4) \\
&& u = -2-2 p_{1}{\cdot}k_{3} =2N_2+2N_4 -4 -2(p_{2}{-}N_2q_2){\cdot}(p_{4}{-}N_4q_4)
\end{eqnarray}

Using $SL(2,\mathbb{R})$ invariance in order to fix $z_1=\infty\,,z_2=1\,,z_4=0$, renaming $z_3=z$ and performing the contractions, the amplitude reads
\begin{equation}
\begin{split}
\mathcal{A}_4( T\,\mathcal{C}\,V\,\mathcal{C})=\int &{{d^D X_0}} e^{i(p_{1}{+}p_2{+}k_3{+}p_4){\cdot}X_{0}}  \sum_{N_2, N_4}^{0,\infty} \int_{0}^{2 \pi} {d\theta_2 d\theta_4 \over (2\pi)^2} e^{i N_2 \theta_2 + i N_4 \theta_4}   \int_{0}^{1} d z\,z^{-\frac{s}{2}-1} \, (1-z)^{-\frac{t}{2}-1} \\
 \exp\Bigg\{ \sum_{k=0}^{\infty} {\cal Y}_{k}\,z^{k}\Bigg\}
&\Bigg( \dfrac{A_3{\cdot}p_{2}}{1-z} - \dfrac{A_3{\cdot}p_{4}}{z} +\sum_{m_{2}=0}^{\infty}{\cal Y}^{\widetilde{\zeta}^{(2)}{\cdot}A_3}_{m_{2}} (1-z)^{m_{2}-1}+ \sum_{m_{4}=0}^{\infty}{\cal Y}^{\widetilde{\zeta}^{(4)}{\cdot}A_3}_{m_{4}} z^{m_{4}-1}\Bigg)
\end{split}
\end{equation}
where
\begin{eqnarray}
&&{\cal Y}^{\widetilde{\zeta}^{(2)}{\cdot}A_3}_{m_{2}}=\sum_{n_{2}=1}^{\infty}{(-)\over n_{2}}^{n_{2}+1}
\widetilde{\zeta}_{n_{2}}^{(2)}{\cdot}A_3\sum_{h_{2}=1}^{n_{2}}h_{2} \, {\cal Z}_{n_{2}{-}h_{2}{-}m_{2}}(n_{2}q_{2}{\cdot}k_{3}){\cal Z}_{m_{2}}(n_{2}q_{2}{\cdot}p_{4}) \\
&&{\cal Y}^{\widetilde{\zeta}^{(4)}{\cdot}A_3}_{m_{4}}=\sum_{n_{4}=1}^{\infty} {1\over n_{4}} \widetilde{\zeta}_{n_{4}}^{(4)}{\cdot}A_3\sum_{h_{4}=1}^{n_{4}}h_{4}\, {\cal Z}_{n_{4}{-}h_{4}{-}m_{4}}(n_{4}q_{4}{\cdot}k_{3}){\cal Z}_{m_{4}}(n_{4}q_{4}{\cdot}p_{2})
\end{eqnarray}
and 
\begin{equation}
{\cal Y}_{k}={\cal Y}^{\widetilde{\zeta}^{(2)}{\cdot}\widetilde{\zeta}^{(2)}}_{k}+{\cal Y}^{\widetilde{\zeta}^{(4)}{\cdot}\widetilde{\zeta}^{(4)}}_{k}+{\cal Y}^{\widetilde{\zeta}^{(2)}{\cdot}\widetilde{\zeta}^{(4)}}_{k}+{\cal Y}^{\widetilde{\zeta}^{(2)}{\cdot}p_3}_{k}+{\cal Y}^{\widetilde{\zeta}^{(2)}{\cdot}p_4}_{k}+{\cal Y}^{\widetilde{\zeta}^{(4)}{\cdot}p_3}_{k}+{\cal Y}^{\widetilde{\zeta}^{(4)}{\cdot}p_2}_{k}
\end{equation}
with 
the various ${\cal Y}$'s
given in 
Eqs. (\ref{Yz2z2}), (\ref{Yz4z4}), (\ref{Yz2z4}), (\ref{Yz2p3}), (\ref{Yz2p4}), (\ref{Yz4p3}), (\ref{Yz4p2}).

Performing the integrals, the final result takes the following form
{\small \begin{equation}
\boxed{\begin{split}
&\mathcal{A}_4( T\,\mathcal{C}\,V\,\mathcal{C})=\int {{d^D X_0}} \,e^{i(p_{1}{+}p_2{+}k_3{+}p_4){\cdot}X_{0}}   \sum_{N_2, N_4}^{0,\infty} \int_{0}^{2 \pi} {d\theta_2 d\theta_4 \over (2\pi)^2} e^{i N_2 \theta_2 + i N_4 \theta_4}  \,e^{{\cal Y}_{0}}\,\\
&\sum_{N_2,N_4} \Bigg\{A_3{\cdot}p_{4}\,\dfrac{\Gamma\Big(k_3{\cdot}(p_4{-}N_4q_4){-}N_4 \Big)\,\Gamma\Big(k_3{\cdot}(p_2{-}N_2q_2){+}1{-}N_2 \Big)}{\Gamma\Big(k_3{\cdot}(p_4{-}N_4q_4{+}p_2{-}N_2q_2){+}1{-}N_2{-}N_4\Big)} \sum_{j=0}^{\infty} \dfrac{{\cal Z}_{j}\Big(k{\cal Y}_{k}\Big)  \Big(k_3{\cdot}(p_4{-}N_4q_4){-}N_4\Big)_{j}}{\Big(k_3{\cdot}(p_4{-}N_4q_4{+}p_2{-}N_2q_2){+}1{-}N_2{-}N_4 \Big)_{j}}\,+ \\
&-\, A_3{\cdot}p_{2}\,\dfrac{\Gamma\Big(k_3{\cdot}(p_4{-}N_4q_4){+}1-N_4 \Big)\,\Gamma\Big(k_3{\cdot}(p_2{-}N_2q_2){-}N_2\Big)}{\Gamma\Big(k_3{\cdot}(p_4{-}N_4q_4{+}p_2{-}N_2q_2){+}1{-}N_2{-}N_4\Big)} \sum_{j=0}^{\infty} \dfrac{{\cal Z}_{j}\Big(k{\cal Y}_{k}\Big) \Big(k_3{\cdot}(p_4{-}N_4q_4){+}1-N_4 \Big)_{j}}{\Big(k_3{\cdot}(p_4{-}N_4q_4{+}p_2{-}N_2q_2){+}1{-}N_2{-}N_4\Big)_{j}}\,+\\
&{+}\sum_{m_{4}=0}^{\infty}\dfrac{\Gamma\Big(k_3{\cdot}(p_4{-}N_4q_4){-}N_4+m_4 \Big)\,\Gamma\Big(k_3{\cdot}(p_2{-}N_2q_2){+}1{-}N_2 \Big)}{\Gamma\Big(k_3{\cdot}(p_4{-}N_4q_4{+}p_2{-}N_2q_2){+}1{-}N_2{-}N_4{+}m_4 \Big)} \,\sum_{j=0}^{\infty} \dfrac{{\cal Y}_{m_{4}}^{\widetilde{\zeta}^{(4)}{\cdot}A_3}{\cal Z}_{j}\Big(k{\cal Y}_{k}\Big) \Big(k_3{\cdot}(p_4{-}N_4q_4){-}N_4+m_4 \Big)_{j}}{\Big(k_3{\cdot}(p_4{-}N_4q_4{+}p_2{-}N_2q_2){+}1{-}N_2{-}N_4{+}m_4\Big)_{j}}\\
&{+}\sum_{m_{2}=0}^{\infty}\dfrac{\Gamma\Big(k_3{\cdot}(p_4{-}N_4q_4){+}1-N_4\Big)\,\Gamma\Big(k_3{\cdot}(p_2{-}N_2q_2){-}N_2+m_2\Big)}{\Gamma\Big(k_3{\cdot}(p_4{-}N_4q_4{+}p_2{-}N_2q_2){+}1{-}N_2{-}N_4{+}m_2 \Big)} \sum_{j=0}^{\infty} \dfrac{{\cal Y}_{m_{2}}^{\widetilde{\zeta}^{(2)}{\cdot}A_3}{\cal Z}_{j}\Big(k{\cal Y}_{k}\Big)\Big(k_3{\cdot}(p_4{-}N_4q_4){+}1-N_4 \Big)_{j}}{\Big(k_3{\cdot}(p_4{-}N_4q_4{+}p_2{-}N_2q_2){+}1{-}N_2{-}N_4{+}m_{2} \Big)_{j}} \Bigg\}
\end{split}}
\end{equation}}
where the sum over levels $N_2$ and $N_4$ of the two coherent states has been explicitly indicated.

\subsection{$\mathcal{A}_4( T,\mathcal{C},\mathcal{C},\mathcal{C})$}
The last 4-point amplitude we consider involves three coherent states and one tachyon\footnote{The amplitude with four coherent states presents some subtleties as how to send one of the insertion points to infinity. We thank G.~Veneziano for suggesting this interesting case, whose analysis we defer to the future.}  
\begin{equation}
\begin{split}
&\mathcal{A}_4( T\,\mathcal{C}\,\mathcal{C}\,\mathcal{C}) = \int_{z_4}^{z_2} dz_3 \, \braket{  cV_T(z_1) \, 
c{{V}}_{\mathcal{C}}(z_2) \, V_T(z_3)\, c{{V}}_{\mathcal{C}}(z_4) }= \int_{z_2}^{z_4} dz_3 \, \langle c(z_1)c(z_2)c(z_4)\rangle \\ 
&\Big\langle e^{ip_1{\cdot}X}(z_{1})\,\exp\Bigg\{\sum_{r_2,s_2=1}^{\infty} \dfrac{\zeta^{(2)}_{r_2}{{\cdot} } \zeta^{(2)}_{s_2} }{2r_2s_2}\, \mathcal{S}_{r_2,s_2} \,e^{-i(r_2+s_2)q_2{\cdot}  X} + \sum_{n_2=1}^{\infty} \dfrac{\zeta^{(2)}_{n_2}}{n_2}{\cdot}  \mathcal{P}_{n_2} \,e^{-in_2q_2{\cdot}  X}  \Bigg\} \,e^{ip_2{\cdot}X}(z_2)
\\
&\hspace*{19 mm}\exp\Bigg\{\sum_{r_3,s_3=1}^{\infty} \dfrac{\zeta^{(3)}_{r_3}{{\cdot} } \zeta^{(3)}_{s_3} }{2r_3s_3}\, \mathcal{S}_{r_3,s_3} \,e^{-i(r_3+s_3)q_3{\cdot}  X} + \sum_{n_3=1}^{\infty} \dfrac{\zeta^{(3)}_{n_3}}{n_3}{\cdot}  \mathcal{P}_{n_3} \,e^{-in_3q_3{\cdot}  X}  \Bigg\} \,e^{ip_3{\cdot}X}(z_3)\\
&\hspace*{19 mm} \exp\Bigg\{\sum_{r_4,s_4=1}^{\infty} \dfrac{\zeta^{(4)}_{r_4}{{\cdot} } \zeta^{(4)}_{s_4} }{2r_4s_4}\, \mathcal{S}_{r_4,s_4} \,e^{-i(r_4+s_4)q_4{\cdot}  X} + \sum_{n_4=1}^{\infty} \dfrac{\zeta^{(4)}_{n_4}}{n_4}{\cdot}  \mathcal{P}_{n_4} \,e^{-in_4q_4{\cdot}  X}  \Bigg\} \,e^{ip_4{\cdot}X}(z_4) \Big\rangle 
\end{split}
\end{equation}

where $z_1>z_2>z_3>z_4$. Thanks to Bose symmetry, the other orderings gives essentially identical amplitudes, up to relabelling of the external legs.

For fixed levels $N_2$, $N_3$ and $N_4$ of the coherent vertex operators the KN factor combined with the ghost contribution takes the following form:
\begin{equation}
\begin{split}
&\langle c\,e^{ip_1{\cdot}X}(z_{1}) c\,e^{i(p_2{-}N_{2}q_2){\cdot}X}(z_{2})e^{i(p_3-N_3q_3){\cdot}X}(z_{3}) c\,e^{i(p_4{-}N_{4}q_4){\cdot}X}(z_{4}) \rangle= \\
&\Bigg(\dfrac{z_{12} z_{34}}{z_{13}z_{24}} \Bigg)^{{-}\frac{s}{2}{-}1} \Bigg(\dfrac{z_{14} z_{23}}{z_{13}z_{24}} \Bigg)^{{-}\frac{t}{2}{-}1} {z_{24}\over z_{34} z_{23}}\,\,\, \Bigg(\dfrac{z_{12}z_{23}}{z_{13}}\Bigg)^{N_{2}}  
\Bigg(\dfrac{z_{34}z_{23}}{z_{24}}\Bigg)^{N_{3}}   \Bigg(\dfrac{z_{14}z_{34}}{z_{13}}\Bigg)^{N_{4}}\,\delta\left(p_1{+}\sum_{\ell=2}^{4} (p_{\ell}-N_{\ell}q_{\ell})\right)
\end{split}
\end{equation}
with the standard 4-point kinematics:
\begin{eqnarray}
&& s= 2N_2 - 4 - 2p_{1}{\cdot}(p_{2}{-}N_2q_2) = 2N_3+2N_4 - 4 - (p_{3}{-}N_3q_3){\cdot}(p_{4}{-}N_4q_4) \\
&& t = 2N_2+2N_3 - 4 - 2(p_{3}{-}N_3q_3){\cdot}(p_{2}{-}N_2q_2) = 2N_4 - 4 - 2p_{1}{\cdot}(p_{4}{-}N_4q_4) \\
&& u = -4-2 p_{1}{\cdot}(p_{3}-N_3q_3) =2N_2 + 2N_4 -4 -2(p_{2}{-}N_2q_2){\cdot}(p_{4}{-}N_4q_4)
\end{eqnarray}
Using $SL(2,\mathbb{R})$ invariance to fix $z_1=\infty\,,z_2=1\,,z_4=0$, renaming $z_3=z$ and performing the contraction, the amplitude reads
\begin{eqnarray}
&&{\mathcal{A}}_4({ T\, \mathcal{C}\, \mathcal{C}\,\mathcal{C}}) = \\
&&\int {d^{D}X_{0}}\, e^{i(p_1+p_2+p_3+p_4){\cdot}X_{0}} \sum_{N_2,N_3, N_4}^{0,\infty} \int_{0}^{2\pi} {d\theta_2 d\theta_3 d\theta_4 \over (2\pi)^3} e^{i N_2 \theta_2{+}i N_3 \theta_3 {+} i N_4 \theta_4} \int_{0}^{1} dz\, z^{-\frac{s}{2}-2} (1{-}z)^{-\frac{t}{2}-2}  \nonumber  \\
&&\exp \Bigg\{ \sum_{r_{2},s_{2}}^{1,\infty} \dfrac{\widetilde{\zeta}^{(2)}_{r_{2}}{\cdot}\widetilde{\zeta}^{(2)}_{s_{2}}}{2r_{2}s_{2}} {\cal S}_{r_{2},s_{2}}\big[\mathcal{U}^{(r_{2})}_{k} ; \mathcal{U}^{(s_{2})}_{k}\big] (1{-}z)^{r_{2}{+}s_{2}}     {+}  \sum_{r_{4},s_{4}}^{1,\infty} \dfrac{\widetilde{\zeta}^{(4)}_{r_{4}}{\cdot}\widetilde{\zeta}^{(4)}_{s_{4}}}{2r_{4}s_{4}} {\cal S}_{r_{4},s_{4}}\big[\mathcal{U}^{(r_{4})}_{k} ; \mathcal{U}^{(s_{4})}_{k}\big] z^{r_{4}{+}s_{4}}                                       \Bigg\} \nonumber\\
&&\exp\Bigg\{  \sum_{r_{3},s_{3}}^{1,\infty} \dfrac{\widetilde{\zeta}^{(3)}_{r_{3}}{\cdot}\widetilde{\zeta}^{(3)}_{s_{3}}}{2r_{3}s_{3}} {\cal S}_{r_{3},s_{3}}\big[\mathcal{U}^{(r_{3})}_{k} ; \mathcal{U}^{(s_{3})}_{k}\big] [z(1-z)]^{r_{3}{+}s_{3}}   \Bigg\}  \nonumber \\
&&\exp\Bigg\{\sum_{n_{2}=1}^{\infty} \dfrac{(1-z)^{n_{2}}}{n_{2}}\sum_{h_{2}=1}^{n_{2}} (-)^{h_{2}{+}1}{\cal Z}_{n_{2}-h_{2}}\big[ \mathcal{U}_{k}^{(n_{2})}\big]\Bigg( {\widetilde{\zeta}^{(2)}_{n_{2}}{\cdot}p_3\over (1-z)^{h_{2}}}  {+} \widetilde{\zeta}^{(2)}_{n_{2}}{\cdot}p_4\Bigg)   \Bigg\}\nonumber\\
&&\exp \Bigg\{-\sum_{n_{3}=1}^{\infty} \dfrac{[z(1-z)]^{n_{3}}}{n_{3}}\sum_{h_{3}=1}^{n_{3}} {\cal Z}_{n_{3}{-}h_{3}}\big[ \mathcal{U}_{k}^{(n_{3})}\big]  \Bigg( {\widetilde{\zeta}^{(3)}_{n_{3}}{\cdot}p_2 \over (1-z)^{h_{3}}}  {+} {\widetilde{\zeta}^{(3)}_{n_{3}}{\cdot}p_4 \over z^{h_{3}}} (-)^{h_3}  \Bigg)   \Bigg\} \nonumber\\
&&\exp \Bigg\{-\sum_{n_{4}=1}^{\infty} \dfrac{z^{n_{4}}}{n_{4}}\sum_{h_{4}=1}^{n_{4}} {\cal Z}_{n_{4}{-}h_{4}}\big[ \mathcal{U}_{k}^{(n_{4})}\big]  \Bigg( \widetilde{\zeta}^{(4)}_{n_{4}}{\cdot}p_2  {+} {\widetilde{\zeta}^{(4)}_{n_{4}}{\cdot}p_3 \over z^{h_{4}}}\Bigg)   \Bigg\}\nonumber\\
&&\exp\Bigg\{ \sum_{n_{2},n_{4}}^{1,\infty} \widetilde{\zeta}^{(2)}_{n_{2}}{\cdot}\widetilde{\zeta}^{(4)}_{n_{4}} {(1-z)^{n_{2}}\over n_{2}}{z^{n_{4}}\over n_{4}}\sum_{h_{2},h_{4}}^{1;n_{2},n_{4}}{\Gamma(h_{2}{+}h_{4})\over \Gamma(h_{2})\Gamma(h_{4})} {\cal Z}_{n_{2}{-}h_{2}}\big[ \mathcal{U}_{k}^{(n_{2})}\big]\,{\cal Z}_{n_{4}{-}h_{4}}\big[ \mathcal{U}_{k}^{(n_{4})}\big] (-)^{h_{2}{+}1} \Bigg\}\nonumber\\
&&\exp\Bigg\{ \sum_{n_{2},n_{3}}^{1,\infty}\widetilde{\zeta}^{(2)}_{n_{2}}{\cdot}\widetilde{\zeta}^{(3)}_{n_{3}}{(1-z)^{n_{2}{+}n_{3}}\over n_{2}}{z^{n_{3}}\over n_{3}}\sum_{h_{2},h_{3}}^{1;n_{2},n_{3}}{\Gamma(h_{2}{+}h_{3})\over \Gamma(h_{2})\Gamma(h_{3})}{(-)^{h_{2}-1}\over 
(1{-}z)^{h_{2}{+}h_{3}}} {\cal Z}_{n_{2}{-}h_{2}}\big[ \mathcal{U}_{k}^{(n_{2})}\big]\,{\cal Z}_{n_{3}{-}h_{3}}\big[ \mathcal{U}_{k}^{(n_{3})}\big]    \Bigg\}\nonumber\\
&&\exp\Bigg\{ \sum_{n_{4},n_{3}}^{1,\infty}\widetilde{\zeta}^{(4)}_{n_{4}}{\cdot}\widetilde{\zeta}^{(3)}_{n_{3}}{(1-z)^{n_{3}}\over n_{3}}{z^{n_{4}{+}n_{3}}\over n_{4}}\sum_{h_{4},h_{3}}^{1;n_{4},n_{3}}{\Gamma(h_{4}{+}h_{3})\over \Gamma(h_{4})\Gamma(h_{3})}{(-)^{h_{3}-1}\over z^{h_{4}{+}h_{3}}} {\cal Z}_{n_{4}{-}h_{4}}\big[ \mathcal{U}_{k}^{(n_{4})}\big]\,{\cal Z}_{n_{3}{-}h_{3}}\big[ \mathcal{U}_{k}^{(n_{3})}\big]     \Bigg\} \nonumber
\end{eqnarray}
with

\begin{equation}
\mathcal{U}_{k}^{(n_{2})}=(-)^{k} n_{2} \Bigg(q_2{\cdot}p_4 +  \dfrac{q_2{\cdot}p_3}{(1-z)^k}\Bigg)\,,\: \mathcal{U}_{k}^{(n_{3})}= (-)^{k} n_{3} \Bigg({q_3{\cdot}p_2 (-)^{k} \over (1-z)^{k}} +  \dfrac{q_3{\cdot}p_4}{z^k}\Bigg)\,, \:  \mathcal{U}_{k}^{(n_{4})}= n_{4} \Bigg(q_4{\cdot}p_2 +  \dfrac{q_4{\cdot}p_3}{z^k}\Bigg)\,
\end{equation}

Performing the integal, the final result is:
\begin{equation}
\boxed{
\begin{split}
\mathcal{A}_4({ T\, \mathcal{C}\, \mathcal{C}\,\mathcal{C}}) =& \int {{d^D X_0}}\, e^{i(p_1+p_2+p_3+p_4){\cdot}X_{0}}
\sum_{N_2,N_3, N_4}^{0,\infty} \int_{0}^{2\pi} {d\theta_2 d\theta_3 d\theta_4 \over (2\pi)^3} e^{i N_2 \theta_2{+}i N_3 \theta_3 {+} i N_4 \theta_4} \\
&e^{{\cal Y}_0} {\Gamma[(p_3{-}N_3q_3){\cdot}(p_4{-}N_4q_4){+}1{-}N_3{-}N_4]     \Gamma[(p_2{-}N_2q_2){\cdot}(p_3{-}N_3q_3){-}N_2{-}N_3{+}1]   \over \Gamma[(p_3{-}N_3q_3){\cdot}(p_2{-}N_2q_2{+}p_4{-}N_4q_4) {+}2{-}N_2{-}N_4{-}2N_3]} \\
&\sum_{\ell=0}^{\infty}{{\cal Z}_{\ell}(k{\cal Y}_k)\,[(p_3{-}N_3q_3){\cdot}(p_4{-}N_4q_4){+}1{-}N_3{-}N_4]_\ell \over [(p_3{-}N_3q_3){\cdot}(p_2{-}N_2q_2{+}p_4{-}N_4q_4) {+}2{-}N_2{-}N_4{-}2N_3]_\ell}
\end{split}
}
\end{equation}
where ${\cal Y}_0 = {\cal Y}_{k=0}$ with
\begin{eqnarray}
{\cal Y}_k &=& {\cal Y}^{\widetilde{\zeta}^{(2)}{\cdot}p_3}_k {+} {\cal Y}^{\widetilde{\zeta}^{(2)}{\cdot}p_4}_k {+} {\cal Y}^{\widetilde{\zeta}^{(4)}{\cdot}p_3}_k  {+}   {\cal Y}^{\widetilde{\zeta}^{(4)}{\cdot}p_2}_k {+} {\cal Y}^{\widetilde{\zeta}^{(4)}{\cdot}\widetilde{\zeta}^{(4)}}_k {+} {\cal Y}^{\widetilde{\zeta}^{(2)}{\cdot}\widetilde{\zeta}^{(2)}}_k {+} {\cal Y}^{\widetilde{\zeta}^{(2)}{\cdot}\widetilde{\zeta}^{(4)}}_k {+}\nonumber\\
&{+}& {\cal Y}^{\widetilde{\zeta}^{(3)}{\cdot}p_2}_k {+} {\cal Y}^{\widetilde{\zeta}^{(3)}{\cdot}p_4}_k {+}{\cal Y}^{\widetilde{\zeta}^{(3)}{\cdot}\widetilde{\zeta}^{(3)}}_k{+} {\cal Y}^{\widetilde{\zeta}^{(2)}{\cdot}\widetilde{\zeta}^{(3)}}_k{+} {\cal Y}^{\widetilde{\zeta}^{(3)}{\cdot}\widetilde{\zeta}^{(4)}}_k
\end{eqnarray}
where the coefficients in first line are the same as in (\ref{2cohe coeff}) and the others are:
\begin{equation}
{\cal Y}_{k}^{\widetilde{\zeta}^{(3)}{\cdot}p_2}=\sum_{n_{3}=1}^{\infty}{(-1)\over n_{3}}\widetilde{\zeta}^{(3)}_{n_{3}}{\cdot}p_2 \sum_{h_{3}=1}^{n_{3}} \sum_{m_{3}=0}^{n_{3}-h_{3}}(-)^{k{+}n_{3}}  m_{3}! \,{   {\cal Z}_{n_{3}{-}h_{3}{-}m_{3}}(n_{3}q_{3}{\cdot}p_{2})\,{\cal Z}_{m_{3}}(n_{3}q_{3}{\cdot}p_{4})             \over (k{-}n_{3}{-}m_{3})!\,(n_{3}{-}k)!}  
\end{equation}
\begin{equation}
{\cal Y}_{k}^{\widetilde{\zeta}^{(3)}{\cdot}p_4}=\sum_{n_{3}=1}^{\infty}{(-1)\over n_{3}}\widetilde{\zeta}^{(3)}_{n_{3}}{\cdot}p_4 \sum_{h_{3}=1}^{n_{3}} \sum_{m_{3}=0}^{n_{3}-h_{3}}(-)^{k{+}n_{3}}  (m_{3}{+}h_{3})! \,{   {\cal Z}_{n_{3}{-}h_{3}{-}m_{3}}(n_{3}q_{3}{\cdot}p_{4})\,{\cal Z}_{m_{3}}(n_{3}q_{3}{\cdot}p_{2})             \over (n_{3}{-}k)!\,(k{-}n_3{+}h_3{+}m_3)!}
\end{equation}
\begin{equation}
\begin{split}
{\cal Y}^{\widetilde{\zeta}^{(3)}{\cdot}\widetilde{\zeta}^{(3)}}=\sum_{r_{3},s_{3}}^{1,\infty}{\widetilde{\zeta}^{(3)}_{s_{3}}{\cdot}\widetilde{\zeta}^{(3)}_{r_{3}}\over 2 s_{3} r_{3}}&\sum_{h_{3}=1}^{r_{3}}h_{3} \sum_{n_{3}=0}^{s_{3}{+}h_{3}} \sum_{m_{3}=0}^{r_{3}{-}h_{3}} {(n_{3}{+}m_{3})!\,(-)^{k{+}r_{3}{+}s_{3}}  \over (k{+}n_{3}{+}m_{3}{-}r_{3}{-}s_{3})! \, (r_{3}{+}s_{3}{-}k)!} \\
&{\cal Z}_{s_{3}{+}h_{3}{-}n_{3}}(s_{3}q_{3}{\cdot}p_{2}) {\cal Z}_{n_{3}}(s_{3}q_{3}{\cdot}p_{4})          {\cal Z}_{r_{3}{-}h_{3}{-}m_{3}}(r_{3}q_{3}{\cdot}p_{2}) {\cal Z}_{m_{3}}(r_{3}q_{3}{\cdot}p_{4})
\end{split}
\end{equation}

\begin{equation}
\begin{split}
{\cal Y}^{\widetilde{\zeta}^{(2)}{\cdot}\widetilde{\zeta}^{(3)}}=\sum_{n_{2},n_{3}}^{1,\infty}(-)^{n_2{+}n_3} {\widetilde{\zeta}^{(2)}_{n_{2}}{\cdot}\widetilde{\zeta}^{(3)}_{n_{3}}\over n_{2} n_{3}}&\sum_{h_{2},h_{3}}^{1,\infty}\sum_{m_{2}=0}^{n_{2}{-}h_{2}}\sum_{m_{3}=0}^{n_{3}{-}h_{3}} {(m_{2}{+}m_{3})! \,(-)^{k} \over (k{-}n_{3}{+}m_{3})!\,(n_{3}{+}m_{2}{-}k)!} {\Gamma(h_{2}{+}h_{3})\over \Gamma(h_{2})\,\Gamma(h_{3})}\\&{\cal Z}_{n_{2}{-}h_{2}{-}m_{2}}(n_{2}q_{2}{\cdot}p_{3}) {\cal Z}_{m_{2}}(n_{2}q_{2}{\cdot}p_{4})  {\cal Z}_{n_{3}{-}h_{3}{-}m_{3}}(n_{3}q_{3}{\cdot}p_{2}) {\cal Z}_{m_{3}}(n_{3}q_{3}{\cdot}p_{4})
\end{split}
\end{equation}

\begin{equation}
\begin{split}
{\cal Y}^{\widetilde{\zeta}^{(3)}{\cdot}\widetilde{\zeta}^{(4)}}=\sum_{n_{3},n_{4}}^{1,\infty} {\widetilde{\zeta}^{(4)}_{n_{2}}{\cdot}\widetilde{\zeta}^{(3)}_{n_{3}}\over n_{4} n_{3}}&\sum_{h_{4},h_{3}}^{1,\infty}\sum_{m_{4}=0}^{n_{4}{-}h_{4}}\sum_{m_{3}=0}^{n_{3}{-}h_{3}} {(h_{3}{+}m_{3})! \,(-)^{k{+}n_{3}{+}m_{4}{+}1} \over (k{-}n_{3}{+}m_{3}{+}h_{3}{-}m_{4})!\,(n_{3}{+}m_{4}{-}k)!} {\Gamma(h_{4}{+}h_{3})\over \Gamma(h_{4})\,\Gamma(h_{3})}\\
&{\cal Z}_{n_{4}{-}h_{4}{-}m_{4}}(n_{4}q_{4}{\cdot}p_{3}) {\cal Z}_{m_{4}}(n_{4}q_{4}{\cdot}p_{2})  {\cal Z}_{n_{3}{-}h_{3}{-}m_{3}}(n_{3}q_{3}{\cdot}p_{2}) {\cal Z}_{m_{3}}(n_{3}q_{3}{\cdot}p_{4})
\end{split}
\end{equation}

Despite its compact appearance the above 4-point amplitudes encodes interactions of arbitrary massive open string states with one tachyon. Contributions of states at fixed level can be extracted by simply expanding the amplitudes. As for 3-point amplitudes, the above formula can be used as a generating function where individual amplitudes obtain after taking derivatives with respect to the parameters $\zeta^{(i)}_n$. Generalization to different boundary conditions look straightforward. Extension to closed strings seems a bit more involved but feasible.


\section{High-energy asymptotics at fixed angle}

In this section we would like to consider the asymptotics of the scattering amplitude 
${\cal A}(T,T,T,C)$ in the fixed angle regime. For simplicity we will only consider the case $\zeta_1\neq 0$ with 
$\zeta_{n\neq 1}= 0$ and further restrict our attention on $\zeta_1{\cdot}\zeta_1=0$. Setting $\zeta_1^\mu = \zeta^\mu$ for notational simplicity, the expression of the amplitude becomes
\begin{equation}
{\cal A}(T,T,T,C) = \int dX_0 e^{i(p_1{+}p_2{+}p_3{+}p_4)X_0} \int_0^1 dz z^{-{s\over 2} -2} 
(1-z)^{-{t\over 2} -2} e^{  \hat\zeta p_3 {+} z \hat\zeta p_2}
\end{equation} 
with $\hat\zeta^\mu = \zeta^\mu e^{-iqX_0}$. 

For large $s, |t|$ and $\zeta{\cdot}p_i$\footnote{Recall that $\langle N\rangle = |\zeta|^2$ for the coherent state, which means that we are taking the mass 
$M= \sqrt{\langle N\rangle/\alpha'}$ to scale as the momenta involved in the process.} we look for a saddle-point of the exponent (neglecting subleading terms)
\begin{equation}
E(z) = - S \log z - T \log (1{-}z) + A z + B 
\end{equation}
where $S= \alpha' s$, $T=\alpha' t$, $A= 2\alpha' \hat\zeta p_2$ and $B= 2\alpha' \hat\zeta p_3$.

\begin{figure}[!h]
\hspace{1cm} \includegraphics[scale=0.5]{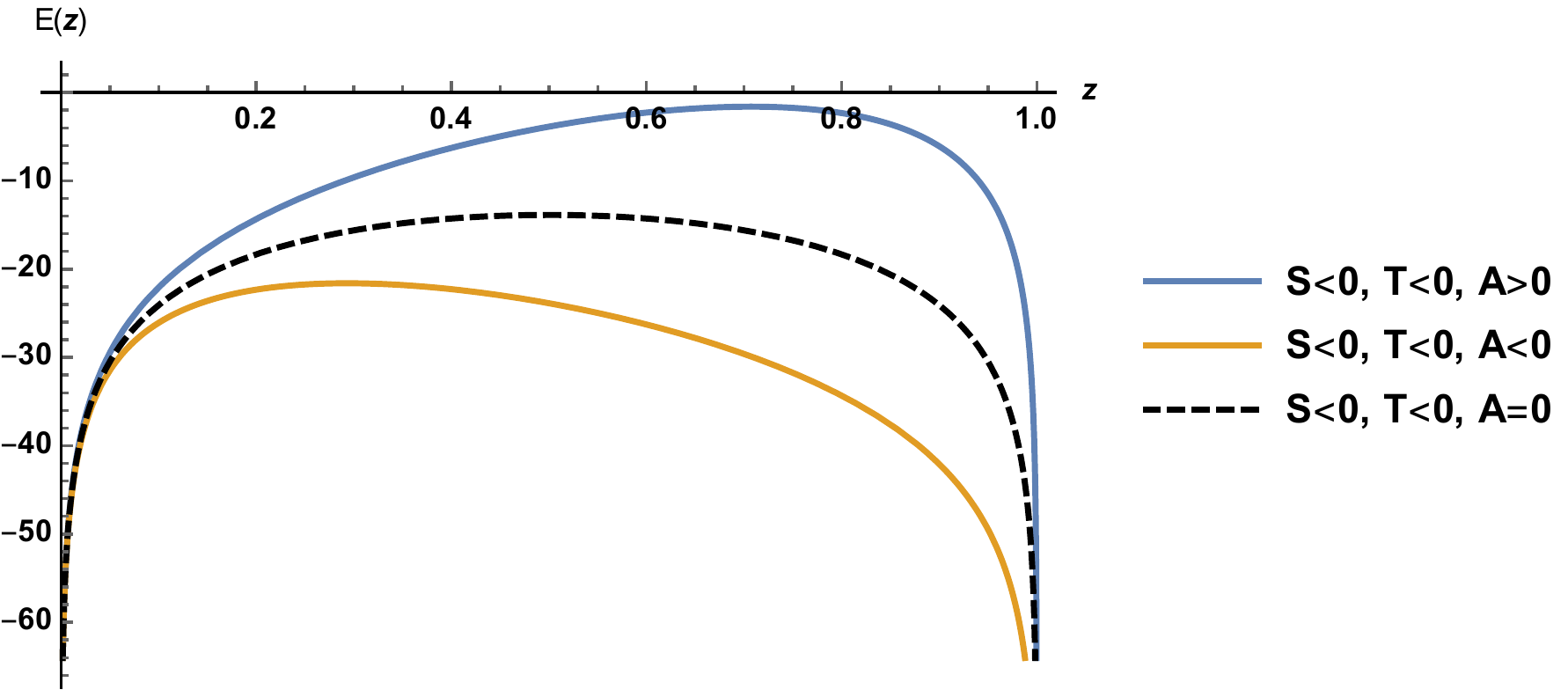}
\includegraphics[scale=0.5]{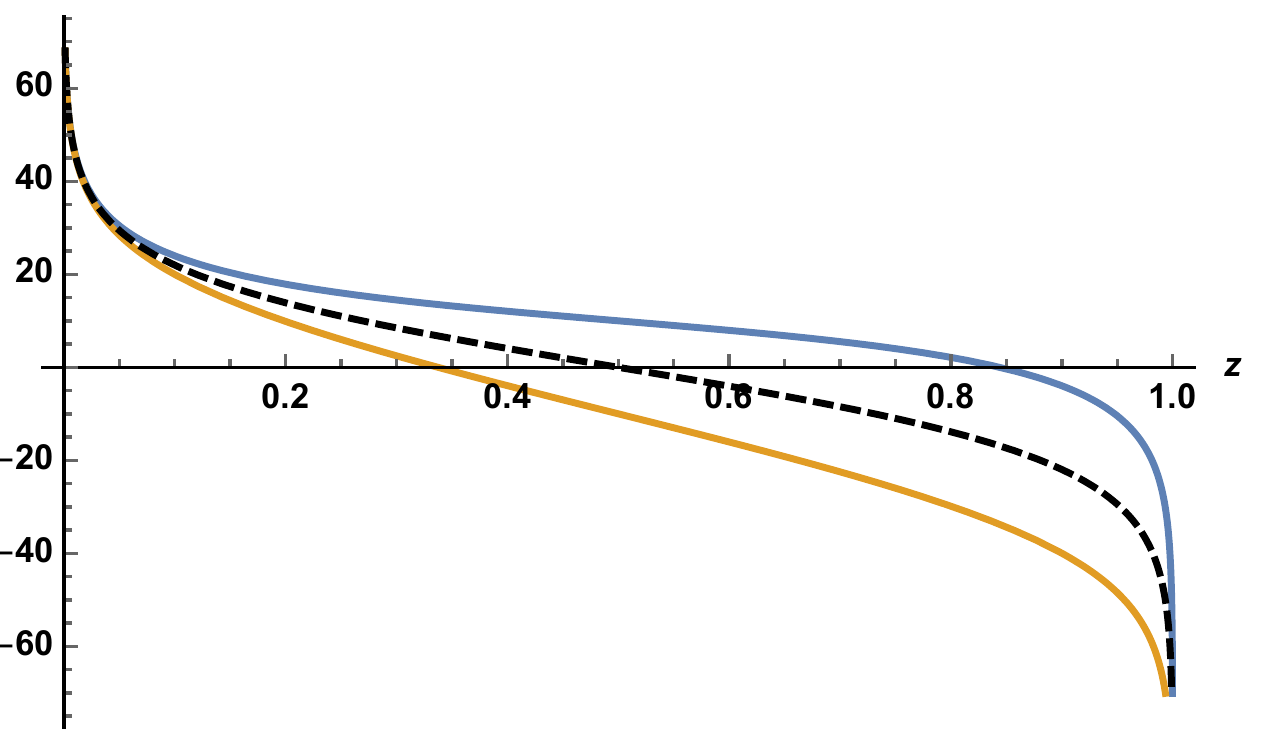}
\caption{plots with $S=T=10$ and $A=20,\, B=0$, the dashed line represents the Veneziano case.}
\end{figure}

The `mathematically' correct way to proceed it to consider ${\cal A}(T,T,T,C)$ as a function of $S$, $T$ and $A$ (the term in $B$ can be taken out of the integral) in the `unphysical' region $S,T$ large and both negative (!). One can take $A>0$ so that $E(z)>0$ in the interval $(0,1)$. Find the saddle-point approximation of the integral and then continue to the `physical' region (e.g. $S>0$ and $T<0$)\footnote{In a series of interesting papers \cite{Taiwanese} the problem for a single very massive state has been addressed with different conclusions/strategy.}. 

The saddle-point is determined by the condition $E'(z)=0$ i.e.
\begin{equation} 
-S (1-z) + T z + Az(1-z) = 0
\end{equation} 
whose solutions are
\begin{equation}
z_{\pm} = {1\over 2A} [ S+T+A \pm \sqrt{(S+T+A)^2 - 4 AS}]
\end{equation}
The solution $z_{-}$ yields the `known' saddle-point for $A=0$  i.e. $z_{-}\vert_{A=0} = S/(S+T)<1$ (in the chosen `unphysical' region), so that $z_-$ is inside the interval of integration. The other one $z_{+}$ tends to infinity for  $A=0$, outside the interval. 
It is convenient to compute 
\begin{equation}
1-z_{\pm} = {1\over 2A} [ A - (S+T) \mp \sqrt{(S+T+A)^2 - 4 AS}]
\end{equation}
and 
\begin{equation}
E''(z_{\pm}) = +{S\over z_{\pm}^2} +{T\over (1-z_{\pm})^2}  < 0
\end{equation}
in the chosen `unphysical' region. Moreover 


\begin{equation}
\begin{split}
&E(z_-)={-}S\log\left( {A{+}S{+}T {-} \sqrt{(A{+}S{+}T)^2{-}4AS}\over 2A}\right) -T\log\left( {A{-}S{-}T {+} \sqrt{(A{+}S{+}T)^2{-}4AS}\over 2A}\right)+ \\
&{+}{1\over 2}\left(A{+}S{+}T {-} \sqrt{(S{+}T{+}A)^2{-}4AS} \right){+}B =B - S\log S - T \log T + (S+T)\log(S+T) + ... (A)
\end{split}
\end{equation}

Plugging this in the expression for the amplitude one finds
\begin{equation}
{\cal A}(T,T,T,C) = \int dX_0 e^{i(p_1{+}p_2{+}p_3{+}p_4)X_0} \exp E(z_{-})
\end{equation} 

\begin{figure}[!h]
\center
 \includegraphics[scale=0.9]{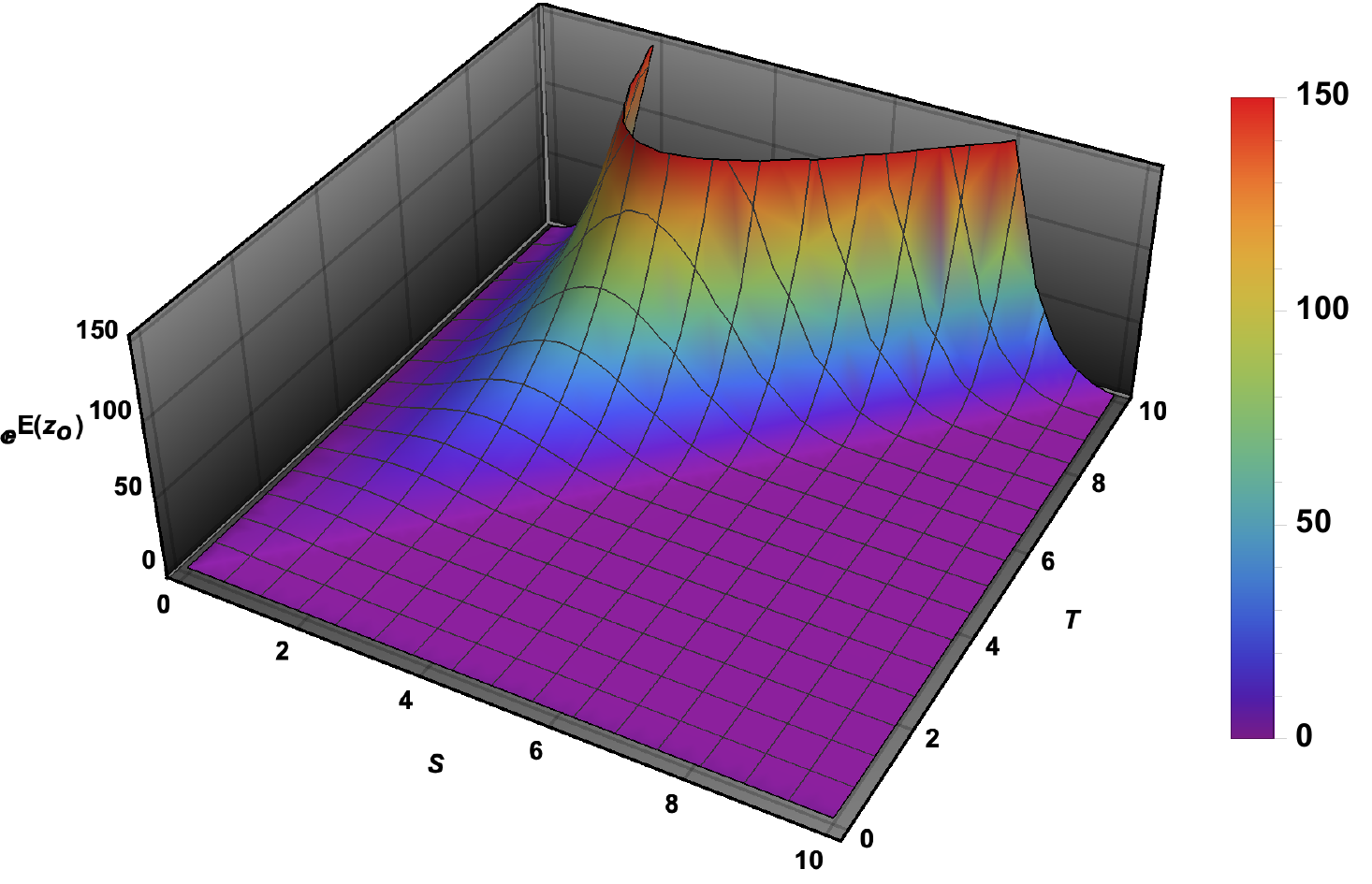}
 
\includegraphics[scale=0.9]{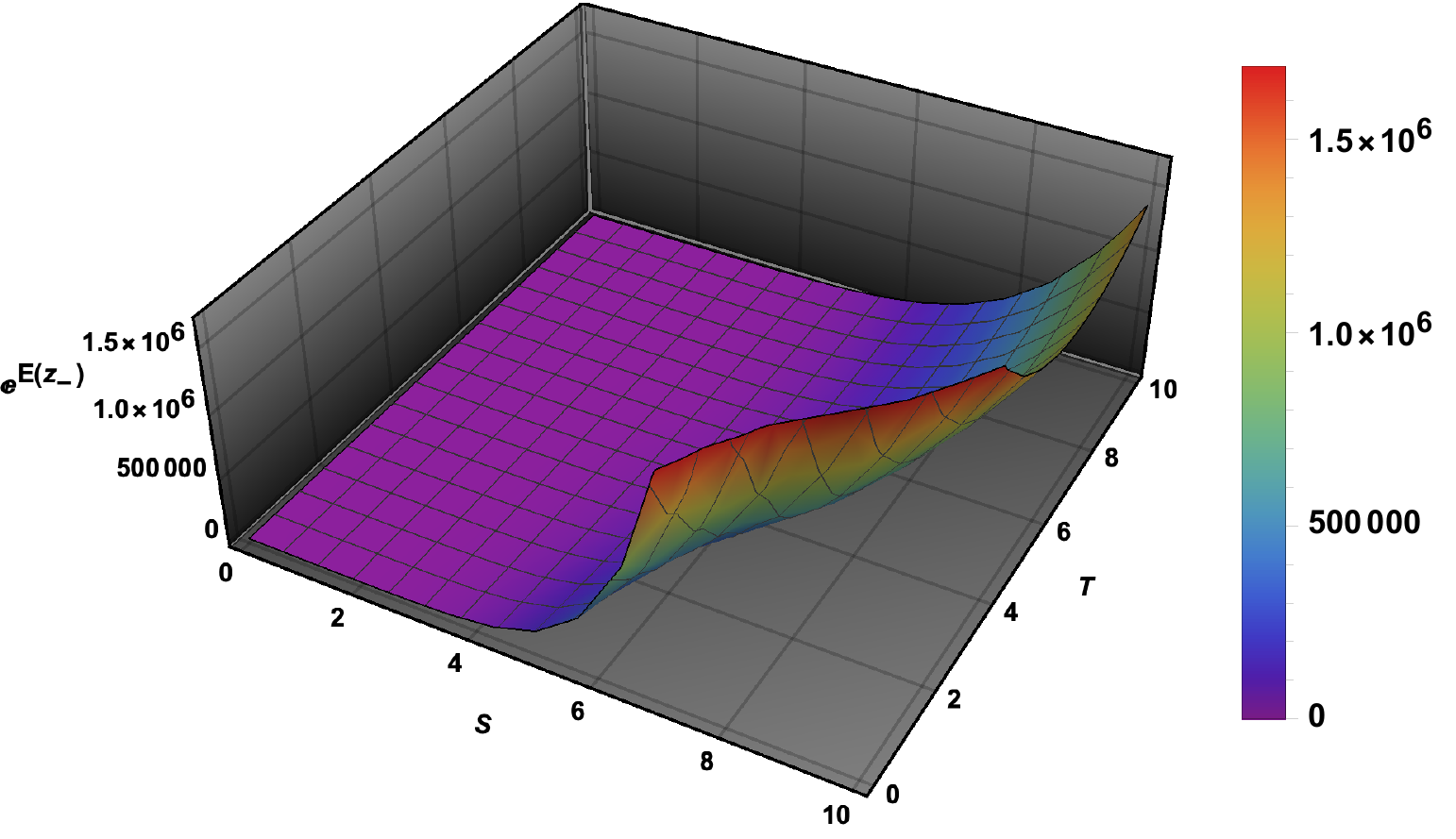}
\caption{The first plot is referred to the saddle point evaluation in the Veneziano case where $z_0=S/(S{+}T)$, and the second one is the Coherent case, both in the physical region with the choices $A=20, B=0$}
\end{figure}

There are two points that we would like to stress. First of all, the presence of $A$ (and $B$) in the exponent shifts significantly the position of the saddle-point and the final result of the approximation valid in the regime where $A\approx S, T$. Second, $A$ depends on $X_0$ and the last integral is not completely straightforward and should be performed expanding the exponent $E(z_{-})$ in powers of $A\sim \exp{-iqX_0} A'$ or equivalently as in the general case (arbitrary $s$ and $t$) using level expansion.

\section{Summary, conclusions and outlook}

We conclude with a summary of the results and some very preliminary considerations on how to extend our analysis to closed (bosonic) strings. 

Relying on the time-honored DDF approach, that has the virtue of combining the nice properties of both the light-cone and covariant approaches, we have reviewed the construction of arbitrarily massive higher-spin physical (BRST invariant) states for open bosonic strings, closely following \cite{Skliros:2009cs, Hindmarsh:2010if, Skliros:2011si} and \cite{Skliros:2016fqs, SklirosOAC}. In particular we have shown how construct coherent states for the case of a single D25-brane. After recalling some properties of the coherent states, that emerge from the computation of 2-point `amplitudes',  such as the average mass, gyration radius and spin and studied some random open string profile. 

Then we have computed 3-point physical `amplitudes'. Modulo few exceptions, they correspond to physical on-shell processes, such as decay or production of massive states. We have considered several  cases with diverse numbers of coherent states as well as tachyons and vector bosons. Quite remarkably, thanks to level expansion and a proper redefinition of the harmonics, we found that the amplitudes exponentiate. 
We have then turned our attention on the computation of 4-point `amplitudes' involving diverse numbers of coherent states as well as tachyons and vector bosons. We have also studied the high-energy fixed angle regime, dominated by a different saddle-point for a different coherent state, and the soft limit when the momentum of a massless vector boson is taken to zero. 

There is a number of generalisations and applications that one may envisage. First of all one can consider non-abelian interactions associated to multiple coincident or separated D25-branes. By T-duality the case of Dp-branes should be straightforward to analyse. Extension to intersecting or magnetised D-branes should also proceed quite easily. For superstrings, some work has been done in the NS sector \cite{SklirosOAC, Hornfeck:1987wt} but very little is known for the Ramond sector. 

It might prove more interesting however to consider closed strings, either bosonic or fermionic. Our results for 3-point or 4-point open string scattering amplitudes represent building blocks for closed string amplitudes, whereby the doubling of modes 
\begin{equation}
\label{closedsvo}
W(z,\bar{z})=V_L(z)V_R(\bar{z})
\end{equation}
requires a doubling of the DDF operators  with similar commutation relations. Physical states will share the same L and R momenta and 
level matching for coherent states can be imposed via integration over an auxiliary variable  
\cite{Skliros:2009cs, Hindmarsh:2010if, Skliros:2011si}
\begin{equation}
\begin{split}
&\mathcal{W}_{\mathcal{C}}(z, \bar{z})=\int_0^{2\pi} d\beta\:V^L_{\mathcal{C}}(p,q, \lambda, \beta, z) V^R_{\mathcal{C}} (p,q, \tilde\lambda, \beta, \bar{z}) =  \\
&\quad\int_0^{2\pi} d\beta\:\exp\Bigg\{\sum_{m,n=1}^{\infty} \dfrac{\zeta_{m}{{\cdot} } \zeta_n }{2\,mn}\, \mathcal{S}_{m,n} \,e^{-i(m+n)[q{\cdot}  X_L + \beta]} + \sum_{n=1}^{\infty} \dfrac{\zeta_n}{n}{\cdot}  \mathcal{P}_{n} \,e^{-in[q{\cdot}  X_L+\beta]}  \Bigg\}  \\
&\qquad\exp\Bigg\{\sum_{r,s=1}^{\infty} \dfrac{\tilde\zeta_{r}{{\cdot} } \tilde\zeta_n }{2\,rs}\, \widetilde{\mathcal{S}}_{r,s} \,e^{-i(r+s)[q{\cdot}  X_R - \beta]} + \sum_{s=1}^{\infty} \dfrac{\tilde\zeta_s}{s}{\cdot}  \widetilde{\mathcal{P}}_{s} \,e^{-is[q{\cdot}  X_R-\beta]}  \Bigg\} \,e^{ip{\cdot}  [X_L+X_R]} \\
\end{split}
\end{equation}
Using KLT relations or otherwise one should be able to compute scattering amplitudes of coherent states with tachyons and massless states (graviton, dilaton, Kalb-Ramond field). 

In general one can compute physical processes with highly excited closed string physical states that should help clarifying some of the issues raised in the introduction, in connection with the dynamics of BH's, the emission of GW's and the role of cosmic strings. We plan to tackle some of these problems in the near future \cite{WIP}.


\vspace{5mm} 

{\large \bf Acknowledgments}

\vspace{5mm}

We would like to thank  Andrea Addazi, Ram Brustein, Dario Consoli, Giorgio Di Russo, Paolo Di Vecchia, Antonino Marcian\`o, Gabriele Veneziano, for useful discussions and comments on the manuscript and above all Dimitri Skliros for sharing his deep insights in the subject, helping us clarifying some of the issues and collaborating at some stage of the computations.
M.~B. also acknowledges CERN for hospitality during the completion of this paper. 
 M.~B. was partially supported by the MIUR-PRIN contract 2015MP2CX4002 {\it ``Non-perturbative aspects of gauge theories and strings''}.


\end{document}